\documentclass{article}

\usepackage[preprint]{neurips_2025}

\usepackage{comment}
\usepackage{ragged2e}
\usepackage{notations}
\usepackage[utf8]{inputenc} 
\usepackage[T1]{fontenc}    
\usepackage{hyperref}       
\usepackage{url}            
\usepackage{booktabs}       
\usepackage{amsfonts}       
\usepackage{nicefrac}       
\usepackage{microtype}      
\usepackage{xcolor}         
\usepackage{amsmath}
\newcommand{\unsure}[1]{{\textcolor{black}{#1}}}

\newcommand{\rod}[1]{{\textcolor{orange}{#1}}}

\newcommand{\emm}[1]{{\textcolor{cyan}{#1}}}
\usepackage{xr-hyper}  

\usepackage{graphicx}       
\usepackage{subfigure}
\usepackage{amsmath,amsfonts,amssymb,amsthm}

\usepackage{amsmath}
\usepackage{amssymb}
\usepackage{comment}
\usepackage{graphicx}
\usepackage{amssymb}
\usepackage{mathtools}

\catcode`,\active

\catcode`\,12

\newcommand{\bw}{\mathbf{w}}
\newcommand{\defeq}{\vcentcolon=}

\usepackage{lmodern}
\usepackage[T1]{fontenc}

\let\saveaddcontentsline\addcontentsline
\renewcommand{\addcontentsline}[3]{}

\title{
Generalization performance of narrow one-hidden layer networks in the teacher-student setting}

\author{
Rodrigo P\'erez Ortiz$^*$ \\
	Alma Mater Studiorum – Università di Bologna (Unibo) \\ 
	IT-40126 Bologna, Italy \\
	\texttt{rodrigo.perezortiz2@unibo.it}
	\AND 
	Gibbs Nwemadji$^*$ \\
	International School of Advanced Studies (SISSA) \\
	Trieste, Italy \\
	\texttt{anwemadj@sissa.it}
	\AND 
	Jean Barbier  \\
	The Abdus Salam International Centre for Theoretical Physics\\
	Trieste, Italy \\
	\texttt{jbarbier@ictp.it}
	\AND Federica Gerace \\
	Department of Mathematics, University of Bologna \\ 
	Piazza di Porta San Donato 5, 40126, Bologna (BO), Italy \\
	\texttt{federica.gerace@unibo.it}
	\AND Alessandro Ingrosso \\
	Donders Institute for Brain, Cognition and Behaviour\\
	Radboud University, Nijmegen, The Netherlands \\	
	\texttt{alessingrosso@gmail.com}
	\AND Clarissa Lauditi \\
	John A. Paulson School of Engineering and Applied Sciences\\
	Harvard University \\
	\texttt{clauditi@g.harvard.edu}
	\AND Enrico M. Malatesta \\
	Department of Computing Sciences \\
	Bocconi University, 20136 Milano, Italy \\
	\texttt{enrico.malatesta@unibocconi.it}
}

\begin{document}

\maketitle

\def\thefootnote{*}\footnotetext{Equal contributions}\def\thefootnote{\arabic{footnote}}

\newpage
\begin{abstract}
Understanding the generalization properties of neural networks on simple input–output distributions is key to explaining their performance on real datasets. The classical teacher–student setting, where a network is trained on data generated by a teacher model, provides a canonical theoretical test bed. In this context, a complete theoretical characterization of fully connected one-hidden-layer networks with generic activation functions remains missing.
In this work, we develop a general framework for such networks with large width, yet much smaller than the input dimension.
Using methods from statistical physics, we derive closed-form expressions for the typical performance of both finite-temperature (Bayesian) and empirical risk minimization estimators in terms of a small number of order parameters. We uncover a transition to a specialization phase, where hidden neurons align with teacher features once the number of samples becomes sufficiently large and proportional to the number of network parameters. Our theory accurately predicts the generalization error of networks trained on regression and classification tasks using either noisy full-batch gradient descent (Langevin dynamics) or deterministic full-batch gradient descent.
\end{abstract}

\section{Introduction and related works}
Predicting theoretically the generalization abilities of complex neural architectures for generic tasks in terms of the number of parameters and training samples is a daunting task. The statistical physics approach has been successful in doing so in the past few decades~\cite{seung1992statistical,engel2001statistical,zdeborova2020understanding}, where the performance of a network is characterized by the typical generalization error over synthetic data generated by probabilistic models of increasing statistical complexity. The replica method from the physics of disordered systems~\cite{mezard1987spin,charbonneau2023spin} is particularly effective for studying feed-forward networks solving supervised learning tasks in the high-dimensional regime. The two classical, complementary settings in this area are the \textit{storage capacity problem}~\cite{gardner_optimal_1988}, where a network is tasked with memorizing a random input-output mapping, and the \textit{teacher-student} scenario~\cite{schwarze_statistical_1992}, where a rule mapping labels to input data is instantiated also in a form of a similar random network architecture.

While the analysis of the typical performance of single-layer networks (perceptrons) is -- at least in the case of continuous parameters -- relatively straightforward and classic~\cite{gardner_space_1988}, the one-hidden layer case is much more challenging. In particular, research has been mainly focused on the case of committee machines, where the second-layer weights are not trained. A variety of works (detailed in the next section) analyzed the information-theoretic limits of supervised learning in committee machines, where a network is trained on $P$ independent $N$-dimensional examples, with the \unsure{sample-to-parameter} ratio $\alpha:=P/N =O_N(1)$ and $N\to\infty$ . The general picture emerging is that generalization performance is controlled by the ratio of the dataset size and the number of network parameters. In particular, \emph{specialization} of hidden neurons in the direction of the ones in the teacher can only occur when the number of samples is of the order of the number of network parameters, i.e. $P \propto N K$.

Classical results exist for the simpler case of \emph{tree} architectures (with non-overlapping receptive fields of the hidden units) and for \emph{fully connected} (FC) committees with sign activation function~\cite{schwarze_statistical_1992, barkai_broken_1992, schwarze_learning_1993, schwarze_generalization_1992}. While recent studies \cite{oostwal_hidden_2021,citton_phase_2025} started addressing the case of arbitrary activations, a general theory of learning using empirical risk minimization (ERM) is not yet present in the statistical mechanics literature. This gap is particularly relevant for modeling more realistic machine-learning tasks, where the learner lacks access to the true data-generating distribution—unlike in Bayes-optimal settings. This framework allows for the exploration of mismatched priors and provides a pathway to understand how emergent neural representations adapt to the underlying structure of the data.
In this work, we study the learning problem in the limit where the input dimension $N$ diverge and the number of hidden units $K$ is large, yet much smaller than $N$. \unsure{More precisely, we consider the regime $1 \ll K \ll N$, taking $N \to \infty$ first and subsequently analyzing the large-$K$ limit. While this asymptotic setting formally allows for joint scalings such as $K=o(N)$, our analysis is controlled in the sequential limit described above, and the joint-scaling regime should be understood as a consistent asymptotic extension. This scaling enables a typical-case analysis via the replica method in terms of a finite number of \emph{order parameters} (i.e., sufficient statistics that characterize the generalization error).}

We therefore study the typical performance of ERM estimators in a teacher-student setup with generic activation function \unsure{and fixed (non-trainable) second-layer weights}, investigating the emergence of learning phase transitions.
We obtain closed form expressions for the weight statistics and the overlap between the teacher and student weight vectors, thus providing a theoretical prediction for the generalization error as a function of \unsure{sample-to-parameter ratio $\tilde{\alpha}:=\alpha/K$}, expressed as the ratio of training set size and the total number of adjustable weights in the network. We show that our framework accurately predicts the performance of Langevin dynamics (LD), including in the vanishing-noise limit, and also captures the test-time behavior of networks trained with pure gradient descent (GD). Our analysis identifies distinct learning regimes, notably a computationally hard phase (i.e., a statistical--computational gap) in which posterior sampling methods such as LD exhibit exponentially slow convergence. To probe this regime, we consider an informed initialization of LD, where the student weights are initialized in the vicinity of the teacher configuration, enabling the dynamics to reach specialized states that are otherwise algorithmically \unsure{hard to sample.}

\paragraph{Related works ---} Shallow neural networks have been extensively studied using a statistical physics framework~\cite{Cui_narrow_2025}, where a description of the typical learning and generalization behavior over a joint input-label ensemble can be provided in the limit of large input dimension $N \to \infty$. The specific phenomenology of such a problem is related to the number of hidden units.

In the $K = O_N(1)$ regime, a two-layer neural network falls into the category of \emph{multi-index functions}. Classical works studied online learning dynamics in such a model, deriving a set of ordinary differential equations for the sufficient statistics of the hidden representations~\cite{saad_solla_online,saad_solla_solution,Riegler_online,Biehl_online,goldt_dynamics}. In the \textit{teacher-student} setting, where all information about the teacher is available to the student except for the input-to-hidden-layer weights, a rigorous classification of the difficulty in recovering the target function via approximate message passing (AMP)/TAP-like algorithms has been established in \cite{troiani_fundamental_2024}. The work~\cite{abbe_sgd_2023} systematically classified SGD performance on two-layer neural networks for any class of multi-index functions based on the \textit{leap complexity} of the target function.

While many questions have been resolved for finite index, the large-index $K$ regime remains to be explored. The regime $1\ll K\ll N$ has been addressed both in the tree~\cite{monasson_learning_1995, sollich_learning_1996, urbanczik1994tree} and the FC case~\cite{schwarze_statistical_1992, schwarze_learning_1993, engel_storage_1992, schwarze_discontinuous_1993,schwarze_learning_1993-1,urbanczik1997full}. These classical studies focused on \textit{sign} (or \textit{erf} ~\cite{ahr1999statistical}) activation function -- with either binary or spherically constrained first-layer weights -- and i.i.d. inputs. In the FC case, \cite{Ahr_committee_1999} \unsure{studied the} effect of $L_2$ regularization using an annealed approximation, corresponding to the \textit{large-dataset limit}.
A FC model with a bounded $\mathcal{C}^2$ activation function with bounded first and second derivatives was studied from a Bayesian-optimal perspective in~\cite{aubin_committee_2018}, where the authors supported replica predictions with rigorous theorems, proposing an efficient inference algorithm based on AMP. In the ERM context, a simple kernel order parameter has been shown to control the storage capacity in the simplified case of a tree committee with generic activation functions~\cite{baldassi_properties_2019,zavathone_activation,annesi_fullrsb}. Using similar methods, a recent work identified analogous kernel order parameters
for generic activation function and a particular choice of second-layer weights~\cite{nishiyama_solution_2025}. We further note the works~\cite{oostwal_hidden_2021, citton_phase_2025}, which analyzed networks with square-summable activation functions using Hermite expansions and a simplifying approximation in the limit of a very large random dataset.
Finally, we also mention a recent burst of interest for the \emph{extensive-width regime} $K=O(N)$ \cite{maillard_quadratic,barbier2025statistical,montanari2025dynamical,ren2025emergence}. In particular, when comparing  our work with \cite{barbier2025statistical}, a key distinction in the extensive-$K$ regime is the emergence of kernel-like learning in the small-data regime (which does not occur in our narrow-but-large-$K$ setting).
Conversely, both scalings converge on \emph{specialization} in the high-$\alpha$ regime.
\unsure{We emphasize that, based on our theoretical observations, for large sample-to-parameter ratio, the Bayes-optimal generalization error (BOE) in our framework is consistent with the extensive width result of \cite{barbier2025statistical}. However, exact analytical agreement for all values of $\alpha$ requires 
$K/N\to 0$. In other cases, while the BOE may match, the underlying order parameters differ. Our approach remains analytically tractable by reducing the problem to a fixed set of scalar order parameters. This avoids the complex \emph{functional matrix order parameters}-- and the necessary integration of replica theory with random matrix models, specifically the  Harish Chandra–Itzykson–Zuber (HCIZ) “spherical” integral \cite{itzyksonPlanarApproximationII1980, guionnetLargeDeviationsAsymptotics2002} -- required in the $K=O(N)$ case.}

\textbf{Our contributions ---} In this work, we develop a complete theoretical characterization of the typical performance of a one-hidden-layer neural network in the regime $1\ll K\ll N$ --again formally interpreted as a sequential limit $N$ large first then $K$ large-- in the presence of i.i.d. inputs and responses generated from a teacher network with the same structure.

Our contributions are the following:
\begin{itemize}
    \item We compute the generalization error and the corresponding learning curves as a function of the number of samples, for \unsure{narrow and shallow networks} with generic activation functions and a generic loss, in the classic teacher-student setting. Unlike previous works, which used an \emph{annealed} calculation~\cite{oostwal_hidden_2021, citton_phase_2025} for regression, our approach relies on a \emph{quenched} computation of the free entropy using the replica method, whose predictions have been validated through Langevin-based simulations. While our results agree with the annealed computation for very large dataset regime, finite datasets can display a continuous-to-discontinous behavior, as we report for the ReLU activation function.
    This computation is of independent interest to the machine learning community, and we anticipate that it will have further applications in learning theory. 
    \item  \unsure{We study the limit in which the Bayesian posterior concentrates on minima of the empirical loss, and experimentally compare its predictions to solutions found by ERM using GD  and LD.} While LD can sample from the posterior when run long enough, GD is a deterministic optimizer and does not explore the full posterior. Remarkably, we find that GD converges to solutions whose generalization performance is equivalent to those obtained by LD, suggesting that optimization and sampling dynamics can yield statistically similar learned representations.

    \item \unsure{ We provide a phase diagram highlighting the role of the data-to-parameters ratio $\tilde{\alpha}$ and the $L_2$ regularization strength $\lambda$ in shaping the properties of the equilibrium solution. This characterization identifies the optimal regularization required to approach the Bayes-optimal error.}

    \item \unsure{In the Bayes-optimal setting, taking the $K\to\infty$ limit after $N\to\infty$ recovers the BOE reported in ~\cite{barbier2025statistical} for activation functions without second Hermite polynomial in their Hermite decomposition. This suggests that for such activations, the large-width ($K$) and large-dimension ($N$) limits commute. For a general activation function, while both theories predict asymptotically (as $\tilde \alpha \to \infty$) the same generalization error, our results exactly match those of ~\cite{barbier2025statistical} at all $\Tilde{\alpha}$ only when $\gamma := K/N\to 0$.} 
    
\end{itemize}

\section{Setting and main results}
\subsection{Empirical risk minimization, and statistical physics formulation}\label{sec. model}

Throughout the paper, we consider the standard supervised learning setup with a synthetic dataset $\mathcal{D} := \{(\mathbf{x}^\mu, y^\mu_\star)\}_{\mu=1}^P$, with responses generated by a two-layer teacher neural network. The samples are thus generated as follows: (i) Construct the target function by sampling entry-wise i.i.d. the first-layer teacher weights $\bW^\star=(\mathbf{w}_k^\star\in \mathbb{R}^N)_{k=1}^K\in \mathbb{R}^{K\times N}$ from  $P_{W^\star}$, and the second-layer weights $\boldsymbol{A}^\star = (A_k^\star)_{k=1}^K$ from $P_{A^\star}$, with finite first and second moments. (ii) The inputs $\mathbf{x}^\mu$ are i.i.d. standard Gaussian vectors: $\mathbf{x}^\mu \sim \mathcal{N}(0,\mathbf{I}_N)$. (iii) The responses/labels $(y^\mu_\star)_{\mu=1}^P$ are generated as
\begin{equation}
      y^\mu_\star =   \varphi_{\mathbf{A}^\star}(\bW^\star \mathbf{x}^\mu,\,z^\mu\sqrt{\Delta^\star}) := f\Big(\frac{1}{\sqrt{K}}\sum_{k\le K}A_k^\star\sigma\Big(\frac{\mathbf{w}_k^\star\cdot \mathbf{x}^\mu}{\sqrt{N}}\Big) -  B^\star\sqrt{K}   + z^\mu\sqrt{\Delta^\star} \Big) .
      \label{Eq. Teacher labels - sm}
\end{equation}
The activation function $\sigma:\mathbb{R} \to \mathbb{R}$ acts entry-wise and is differentiable almost everywhere, and the readout function $f:\mathbb{R} \to \mathbb{R}$ defines the task;  e.g. in the case of regression one has $f(x) = x$ whereas in the case of binary classification $f(\cdot) = \operatorname{sign}(\cdot)$. The i.i.d. $z^\mu\sim \mathcal{N}(0,1)$ are label noise whose standard deviation is controlled by $\Delta^\star$. For analytical convenience, we introduce the term $B^\star$ in the second layer, fixed to remove the mean of the second-layer pre-activation.

Given the dataset $\mathcal{D}$, we study the problem of learning the target in a teacher-student (realizable) setting under the empirical risk minimization (ERM) framework, using a student (trainable) model with same parametric form as the target. We focus on the case where the trainable model is already set with the correct readout weights (which are few compared to the inner ones) and we thus skip the dependency on those weights to lighten notations, i.e. $\mathbf{A}^\star=\mathbf{A}$ and $\varphi_{\bf A^\star}(\cdot,\cdot) = \varphi_{\bf A}(\cdot,\cdot)$; our main running example will be all-ones readouts, as is usually the case in the literature on committee machines. The student's bias $B$, like $B^\star$, is adjusted to remove the mean of the readout pre-activation. In practice, $B$ is updated separately during learning once per epoch, based on the current state of the weights. The ERM with weight decay we consider reads 
\begin{equation}
     \bW_{\rm erm} \in {\rm argmin} \,\mathcal{L}_P(\bW), \quad \text{with}\quad \mathcal{L}_P(\bW) :=  \sum_{\mu\le P} \ell\left(y^\mu_\star, \varphi_{\bf A}\left(\bW \mathbf{x}^\mu,0\right)\right) + \frac{\lambda}{2} \|\mathbf{W}\|_{\rm F}^2.\label{Eq. risk}
\end{equation}
We consider the standard notion of mean-square generalization error $\epsilon_g$: for a test sample $(\mathbf{x}^{\mathrm{new}}, y^{\mathrm{new}}_\star)$ with same law as the samples in the training data and a given $\bW$, we define
\begin{equation}
    \epsilon_g(\bW) := \frac{1}{4^{l}}\mathbb{E}_{\mathbf{x}^{\mathrm{new}},y^{\mathrm{new}}_\star}\left[ (\hat{y}_{\bW}(\mathbf{x}^{\mathrm{new}})- y^{\mathrm{new}}_\star)^2\right]\label{Eq. gen error}
\end{equation}
with $l=0$ for regression and $l=1$ for classification, $\hat{y}_\bW (\mathbf{x}) := \varphi(\mathbf{W}\mathbf{x},0)$ is the student prediction. 

\paragraph{Statistical mechanics formulation ---}
We analyze this setup in the high-dimensional regime where the network architecture (matching the target) has a large width, yet vanishingly small compared to the input data dimension, and the number of training samples $P=P_N$ scales linearly with the number $NK$ of trainable parameters:
 \begin{align}
     N\to\infty, \ \ \text{with} \ \ K\gg 1 \ \ \text{(so that $N\gg K$)}, \ \ \text{and finite} \ \ \Tilde{\alpha}:=\lim_{N\to\infty} \frac{P_N}{NK} \in (0,\infty).\label{thermoLimi}
 \end{align}In order to analyze the problem in this asymptotic limit using statistical mechanics, it is useful to introduce a temperature (later taken small) and the associated Gibbs-Boltzmann measure:
\begin{equation}
    P_\beta(\bW\mid\mathcal{D}) = \frac{1}{\mathcal{Z}_\beta}\exp\big(-\beta \mathcal{L}_P(\bW)\big) \label{Eq. Gibbs-measure},
\end{equation}
where $\mathcal{Z}_\beta$ is the partition function (normalization factor), and  $\beta>0$ is an inverse temperature that controls the measure's concentration around solutions with small loss, going from Bayesian learning at finite $\beta$ to ERM when it diverges. 

The starting point of the statistical physics approach to learning is to compute the log-partition function, the so-called free entropy: $\ln \mathcal{Z}_\beta/(NK)$. We expect it to be self-averaging, i.e. its variance w.r.t. the realization of the problem (teacher and data) vanishes as $N\to\infty$ (this is proven using standard concentration techniques, see e.g. \cite{barbier2019optimal}). This allows us to focus on its expectation value 
\begin{equation}
    \Phi_{\beta,K} := \lim_{N\to \infty} \frac{1}{NK }\mathbb{E}_{\,\bW^\star,\,\mathcal{D},\,\mathbf{z}}\ln \mathcal{Z}_\beta,\label{Eq. Free entropy definition}
\end{equation}
where $\mathbb{E}_\mathbf{z}$ denotes the expectation w.r.t. to the label noise. Computing $\Phi_{\beta,K}$ yields the order parameters of the problem --sufficient statistics of interest--   particularly those related to the generalization error, whose characterization is our main goal. The finite temperature equivalent of the generalization error is the average Gibbs error, defined as
\begin{equation}
    \epsilon_{\rm Gibbs}(\beta) :=\frac{1}{4^l} \mathbb{E}_{\,\bW^\star,\,\mathcal{D},\,\mathbf{z},\,\mathbf{x}^{\rm new}_\star,\,y^{\rm new}_\star} \big[\big\langle (\hat{y}_{\bW}(\mathbf{x}^{\mathrm{new}})- y^{\mathrm{new}}_\star)^2  \big\rangle\big],\label{Eq. Gibbs error}
\end{equation}
where $\langle \,\cdot\, \rangle$ denotes the
average w.r.t. the Gibbs-Boltzmann measure \eqref{Eq. Gibbs-measure}. Again by the self-averaging property, in the aforementioned high-dimensional limit, the Gibbs error approaches $\epsilon_g(\bW_{\rm typ})$ where $\bW_{\rm typ}$ is a typical sample from the Gibbs measure. This latter error converges to $\epsilon_g( \bW_{\rm erm})$ as $\beta\to\infty$. This makes the finite (but low) temperature analysis suitable to study the performance of ERM. By the same argument we can approximate the training error of ERM using 
\begin{align}
 \epsilon_{\rm tr}(\beta) &:= \mathbb{E}_{\,\bW^\star,\,\mathcal{D},\,\mathbf{z}}\Big[\Big\langle \frac{1}{P}\sum_{\mu\le P} \ell(y^\mu_\star, \varphi_{\mathbf{A}}(\bW\mathbf{x}^\mu,0))\Big\rangle\Big] .   
\end{align}

\unsure{The asymptotic regime in Eq.~\eqref{thermoLimi} formally allows for joint scaling of the network width with the input dimension, such as $K\propto N^\delta$ with $\delta <1$. However, the replica computation presented below is controlled in the sequential limit, where $N,P \to \infty$ a fix (but arbitrary) $K$. In this limit, the free entropy is obtained from an extremization over $K\times K$ overlap matrices. In a second step, we consider the regime of large but finite $K$. Here, we assume the overlap matrices are parameterized by a uniform diagonal and off-diagonal structure, as suggested in~\cite{schwarze_generalization_1992} and detailed in the following section. Upon rescaling the free entropy by $K$, the stationary equations admit a well-defined large-$K$ limit, yielding asymptotically equivalent results. Consequently, the joint-scaling regime ($K\propto N^{\delta}$ with $\delta<1$) should be interpreted a heuristic extension consistent with these asymptotic equations, rather than a rigorously derived limit within the current framework.}

\subsection{Main results: closed form formulas for the free entropy, test and training errors}
\paragraph{Replica symmetric free entropy, and order parameters ---}
In order to state our main results we need to introduce some definitions. Define the so-called \emph{replica symmetric (RS) potential}
\begin{equation}
      K \Phi^{\rm RS}_{\beta, K}(m,q,v)  := \mathcal{G}_{SI}(m,q,v)  + \tilde{\alpha} K \mathcal{G}_E(m,q,v)  ,\label{Eq. Variational free entropy}
\end{equation}
where $m$ denotes a $K\times K$ matrix whose elements can be expressed in terms of two parameters $m_d$ and $m_a$ as $m=m_d\mathbf{I}_K +(m_a /K)\mathbf{1}_K\mathbf{1}_K^\intercal$ and similarly for $v,q$. 
\unsure{This parametrization of the overlap matrices can be justified in the regime $K\ll N$. By invoking statistical permutation symmetry -- naturally suggested by the indistinguishable nature of the hidden units -- we expect overlaps between distinct units to concentrate. This allows us to describe the system using a reduced set of scalar order parameters rather than tracking every individual pair. 
This choice is further constrained by asymptotic self-consistency: to ensure the network's output variance remain $\mathcal{O}(1)$ in the large $K$ limit, these off-diagonal overlaps must scales as $\mathcal{O}(1/K)$. If they were $\mathcal{O}(1)$, the $K(K-1)$ correlations between different units sum would cause a divergence. Thus, the $1/K$ scaling represents a structurally stable state where units are collectively balanced, and where individual fluctuations around their mean become negligible in the thermodynamic limit. 
}

Then, the ``entropic potential'' is

\begin{equation}
    \begin{split}
        \mathcal{G}_{SI}(m, q, v) &:= \frac{K}{2}\Big[1 + \frac{q_a - m_d^2}{v_d} - (q_d+v_d)\beta\lambda + \ln(2\pi v_d)\Big] - \frac{1}{2}\log\Big(\frac{v_d}{v_d+v_a}\Big)\\
        &+ \frac{1}{2(v_d + v_a)}\Big[ q_a - 2m_dm_a - m_a^2 - \frac{v_a}{v_d}(q_d - m_d^2) - (v_d + v_a)(q_a + v_a)\beta\lambda    \Big],
    \end{split}
\end{equation}
whereas the ``energetic potential'' reads
\begin{equation}
\begin{split}
       &\mathcal{G}_E(m,q,v) := \EE_{\boldsymbol{\xi} \sim\mathcal{N}(0,\mathbf{I}_K)}\int\mathrm{d}y\,Z_{\rm T}(m,q,v;y,\boldsymbol{\xi},\mathbf{A})\ln Z_{\rm S}(m,q,v;y,\boldsymbol{\xi},\mathbf{A}), \\
       &Z_{\rm T} := \EE_{z\sim\mathcal{N}(0,1)}\int_{\mathbb{R}^K}\mathrm{d}\mathbf{T} \, \mathcal{N}\big(\mathbf{T};\,mq^{-1/2}\boldsymbol{\xi},\,\mathbf{I}_K - mq^{-1}m\big) \delta(y - \varphi_\mathbf{A}(\mathbf{T},z\sqrt{\Delta^\star})),\\
       &Z_{\rm S} := \int_{\mathbb{R}^K}\mathrm{d}\mathbf{Z} \, \mathcal{N}(\mathbf{Z};\,q^{1/2}\boldsymbol{\xi},\,v)\exp\big[-\beta\ell\left(y,\varphi_\mathbf{A}(\mathbf{Z},0)\right)\big] ,
\end{split}\label{Eq. Energetic potential}
\end{equation}
where $\mathcal{N}(\,\cdot\,;\boldsymbol{\mu},\Omega)$, denotes the normal p.d.f. with mean vector $\boldsymbol{\mu}\in \mathbb{R}^K$ and covariance matrix $\Omega$. The RS potential allows us to approximate the free entropy through its extremization. Most importantly, the stationary equations constraining its extremizers give us directly access to the sufficient statistics and, consequently, the training and test errors. The RS approximation to the free entropy thus reads
\begin{equation}
     \Phi_{\beta,K} = {\rm extr}_{m,q,v} \Phi_{\beta,K}^{\rm RS}(m,q,v).\label{Eq. Variational free entropy problem}
\end{equation}
The extremization $\rm{ extr}$ selects a solution ($m^*, q^*,v^*)$ of the stationary equations (called ``RS equations''), obtained from $\nabla \Phi_{\beta,K}^{\rm RS}=\mathbf{0}$, maximizing the RS potential. The replica method predicts that with high probability, in the high-dimensional limit the order parameters (called overlaps) have the following limits:
\begin{equation}
    \frac{\bW^\star\bW_a^\intercal }{N} \to m^*,\;\; \frac{\bW_a \bW_b^\intercal}{N}\to q^* ,\;\; \frac{\bW_a \bW_a^\intercal}{N}\to q^*_{\mathrm{self}}:= q^*+ v^*  ,  \label{Eq. Overlap concentration}
\end{equation}
where ${\bW}_a$ and ${\bW}_b$ are conditionally (on $\mathcal{D}$) i.i.d. samples from the Gibbs measure. The sufficient statistics are thus three $K\times K$ matrices whose physical meaning are: $q^*$, the student-student overlap, quantifies the average correlation between two different weight samples from the Gibbs measure \eqref{Eq. Gibbs-measure}; $q_{\text{self}}^*$, the self-student overlap, measures the correlation matrix between learned weight vectors of a single configuration; and finally $m^*$, the teacher-student overlap, describes the alignment between the learned weights and the teacher's.
In the high-dimensional limit, our theory assumes that these matrices have constant diagonal and off-diagonal entries (up to irrelevant fluctuations), with the latter scaling differently with $K$. This assumption, along with the application of the central limit theorem following the approach in tree-like committee machines~\cite{baldassi_properties_2019,zavathone_activation}, leads to a substantial simplification of the RS entropy and enables the treatment of general activation functions for large $K$, see the supplementary material (SM) for the complete derivation.

\unsure{We note that the overlap structure assumed here is similar to that used in the capacity calculations for this architecture~\cite{seung1992statistical, engel_storage_1992} and in the teacher-student setting with spherical constraint~\cite{schwarze_generalization_1992} in the large-$K$ limit. Nevertheless, the absence of the spherical constraint, together  with the introduction of the bias term in a non-Bayes optimal setting, distinguishes the present derivation. These additions are crucial to obtain well-defined RS equations for general activation functions and to explicitly analyze the specialization transition. Moreover, this setting is arguably relevant from a machine-learning perspective, as it corresponds to ERM rather than Bayesian learning.}

\paragraph{Generalization error ---\label{sec: gen_error}}
Let $\mu_A$, $\nu_A$ be the first and second moments of $P_A$, respectively.
Let also $(t,s)$ be a centered Gaussian vector with covariance elements $\mathbb{E}[t^2] = \mathcal{E}(1,1,1,0) + \Delta^\star$, $ \mathbb{E}[s^2] =  \mathcal{E}(q_{{\rm self},d}^*,q_{{\rm self},d}^*,q_{{\rm self},d}^*,q_{{\rm self},a}^*)$ and $\mathbb{E}[t s]=\mathcal{E}(1,q_{{\rm self},d}^*,m_d^*,m_a^*)$ where, in the large width limit $K\to \infty$ (taken \emph{after} $N\to\infty$), the function $\mathcal{E}(x_d,y_d,z_d,z_a)$ is 
\begin{equation}
    \begin{split}
    &\mathcal{E}(x_d,y_d,z_d,z_a) = 
    \nu_A\mathcal{K}_0(x_d,y_d,z_d) - 
    \mu_A^2\mathcal{K}_0(x_d,y_d,0) - \frac{z_a\mu_A^2}{x_d y_d}
    \mathcal{K}_1(x_d,y_d,0),\\
    &\mathcal{K}_p(d_1,d_2,a) := \mathbb{E}_{(x_1\; x_2)\sim \mathcal{N}(0,\Omega)} [(x_1x_2)^p\sigma(x_1)\sigma(x_2)] \ \ \text{with} \ \ \Omega = \begin{pmatrix}
            d_1 &   a   \\
            a   &   d_2
        \end{pmatrix}.
    \end{split}
\end{equation}
Given $(m^*,q^*,v^*)$, a solution of the RS equations maximizing the RS potential \eqref{Eq. Variational free entropy}, the RS approximation to the (Gibbs) generalisation error $\lim_{N\to\infty} \epsilon_{\rm Gibbs}$ is given by 
$\mathbb{E}_{(t,s)}[(f(t) - f(s))^2] $. Notice that $\mathcal{K}_0(d_1,d_2,a)$ coincides with the NNGP kernel~\cite{Neal1996,Williams1996} of large-width neural networks at initialization. Similarly, the RS approximation to the training error $\lim_{N\to\infty}\epsilon_{\rm tr}$ reads $\partial_\beta \mathcal{G}_E|_{m^*,q^*,v^*}$.

\begin{figure}
    \centering
    \includegraphics[width=1.0\linewidth]{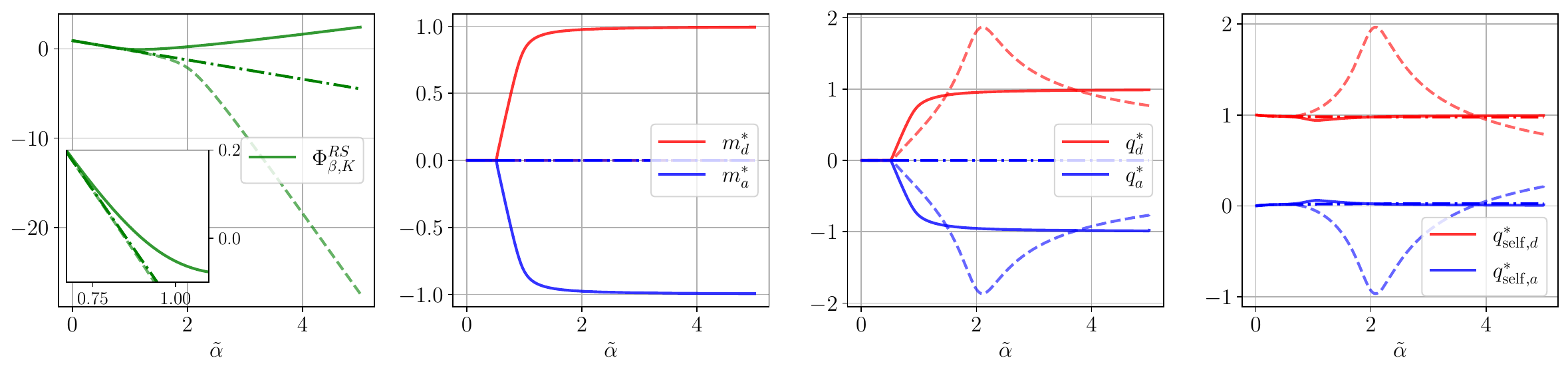}
    \caption{
    \justifying  
    Free entropy $ \Phi_{\beta,K}^{\rm RS}$ (first panel) and order parameters associated with the stationary solutions of the RS equations, as a function of the \unsure{sample-to-parameter ratio} $\tilde{\alpha}$, for quadratic activation function $\sigma(x) = x^2$, MSE loss, weight decay $\lambda = 0.1$, number of hidden units $K = 10$ and $\beta=10$. Blue curves represent off-diagonal components of the teacher-student overlap $m^*$, the student-student overlap $q^*$, and the self-student overlap $q_{\rm self}^*$.
    Red curves represent the dominant diagonal components of these quantities. The dotted-dashed lines correspond to the \emph{permutation-symmetric branch} (where $m_d^* = q_d^* = 0$), the solid lines to the \emph{specialization branch} ($m_d^*>0, q_d^*>0$) and the dashed one to the \emph{memorization branch} ($m_d^*=0, q_d^*>0$).}
    \label{fig:quadratic_activation}
\end{figure}
\section{Numerical experiments and comparison with theoretical predictions}
\label{algorithm_motivation}

As our analysis describes a Bayesian posterior over the first-layer weights with a $\beta$-dependent likelihood, our primary algorithm for testing is Langevin dynamics (LD), due to its explicit temperature dependence and theoretical guarantee of sampling the Gibbs measure when it reaches equilibrium. We consider two distinct initializations: \textit{LD Planted Init}, where the student parameters are initialized near the ground truth, and \textit{LD Random Init}, with i.i.d. $\mathcal{N}(0,1)$ initial weights. Across data regimes, at least one of these LD variants consistently displays a test error matching well the theoretical prediction extracted from a solution of the RS equations that \emph{locally} maximizes $\Phi_{\beta,K}^{\rm RS}$. 

\begin{figure}[!]
    \centering
    {\includegraphics[scale=0.755]{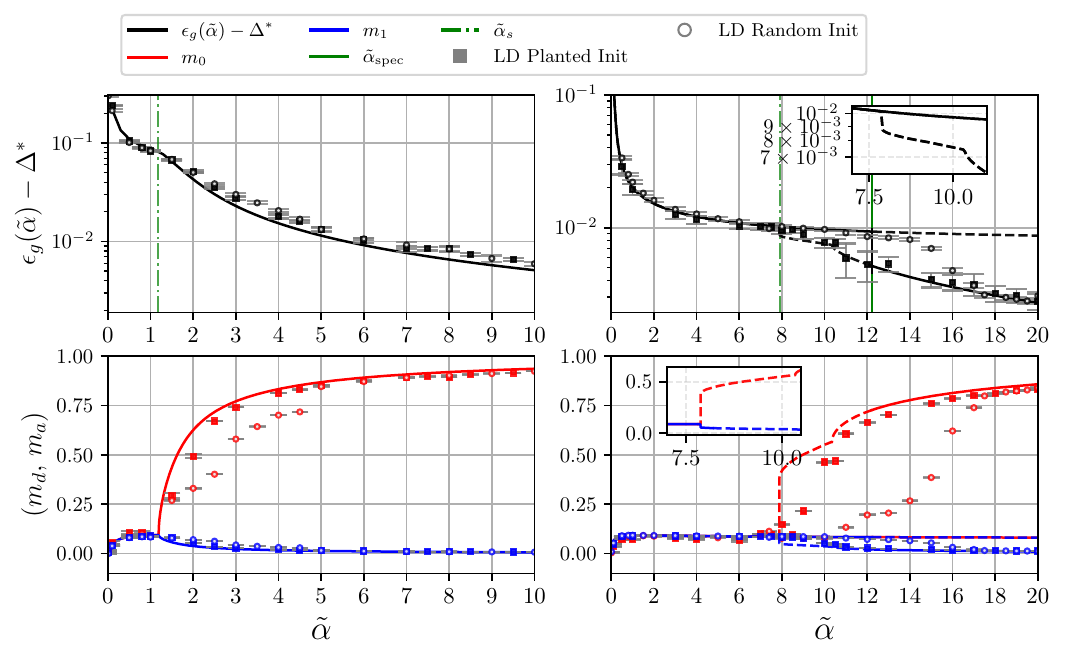}}
    \vspace{-1.0em}
    \caption{
    \justifying  
    Generalization error $\epsilon_g$ and teacher-student overlap  for noise-free regression task ($\Delta^\star = 0$) as functions of the sample-to-parameter ration $\Tilde{\alpha}$. Results are shown for inverse temperature $\beta=10$, weight decay $\lambda=0.1$, and activation functions   $\sigma(x) ={\rm ReLU}(x)$ (\textbf{Left}), \unsure{which exhibit a continuous specialization transition}, and  $\sigma(x) ={\rm Erf}(x/\sqrt{2})$ (\textbf{Right}) \unsure{which exhibit a discontinuous specialization transition}. Theoretical curves are derived from the RS equations. The teacher-student overlaps are parametrized as $m_0^*= m_d^*+m_a^*/K$ (diagonal) and $m_1^*= m_a^*/K$ (off-diagonal). The solid lines represent the equilibrium RS solutions (globally maximizing free entropy), while the dashed lines for the Erf case denote metastable solutions. The specialized branch becomes the equilibrium at $\tilde{\alpha}\geq \Tilde{\alpha}_{\rm spec}$; for Erf, this branch emerges at a spinodal point $\Tilde{\alpha}_s<\Tilde{\alpha}_{\rm spec}$. Markers represent averages over 500 LD samples with $K=10$ and $N=500$, for a single dataset. Generalization error is computed on 1000 test samples; error bars show standard deviations. Notably, for $\Tilde{\alpha} > \Tilde{\alpha}_{\rm spec}$, LD with random initialization fails to follow the RS equilibrium branch, suggesting metastability effects or dynamical barriers not fully captured by the RS equilibrium analysis.}
    \label{fig: Finite beta - Gen Error and T-S Overlap}
\end{figure}

\unsure{We perform numerical simulations at $K=10$ and $N=500$. While the theoretical equations predict a slight shift in the specialization threshold as $K$ increases towards the asymptotic limit, $K=10$ and $N=500$ is sufficient to capture the relevant physical features—such as the specialization transition and the emergence of a hard phase—while allowing for extensive LD sampling within reasonable computational limits.}

The \emph{global} maximizer corresponds to the ``equilibrium branch'' of the theory, describing typical student weights dominating the Gibbs measure, which may coexist with other solutions associated with metastable states, as described below. We also compare our $\beta\to\infty$ theory to gradient descent (GD), identifying regimes where GD performance aligns well with the theory, despite lacking guarantees for solving the non-convex ERM problem. Although neither GD nor LD consistently reach the minimal generalization error of ERM across all $\Tilde{\alpha}$, our theoretical framework reliably describes their performance.  A single GD run took at most $16$\,h, while LD required up to $58$\,h, including GPU and CPU time. Total compute time for all figures is $42$ effective GPU hours (see SM).

\begin{figure}[!]
    \centering
   \includegraphics[scale=0.45]{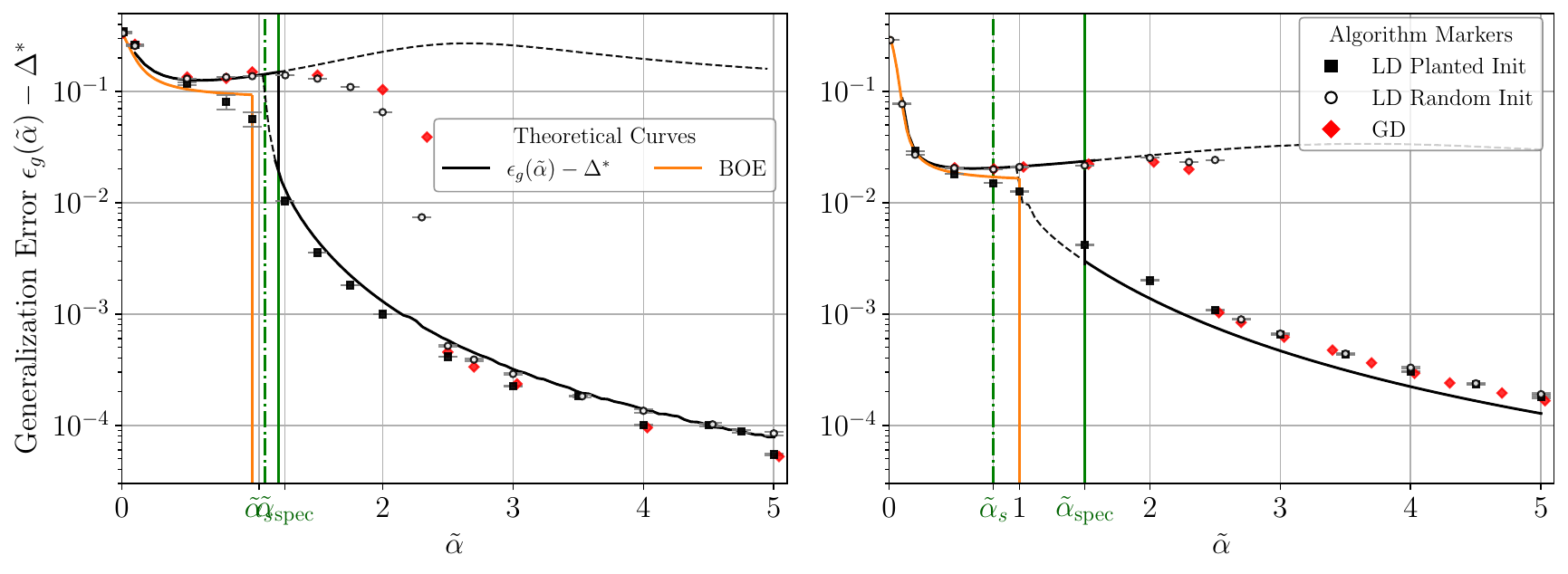}
   \vspace{-10pt}
    \caption{
    \justifying  
    Generalization error $\epsilon_g$  for noise-free regression ($\Delta^\star = 0$) as a function of the \unsure{sample-to-parameter ratio} $\Tilde{\alpha}$. (\textbf{Left}) $\sigma(x) = \mathrm{ReLU}(x)$;  (\textbf{Right}) $\sigma(x) = \mathrm{Erf}(x/\sqrt{2})$. \unsure{At large $\beta$ both activations exhibit a discontinuous specialization transition}. Results corresponds to ERM ($\beta\to\infty$ limit) with weight decay $\lambda=0.01$, where typical configurations are sampled from the memorization state for $\Tilde{\alpha} < \Tilde{\alpha}_{\rm spec}$. The equilibrium and metastable RS branches are denoted by solid and dashed black lines, respectively, while the orange line represents the Bayes-optimal error (BOE), which vanishes for $\tilde{\alpha}\geq 1$. Numerical results for a single dataset with $K = 10$ and $N = 500$, are obtained via LD at $\beta = 10^4$ (averaging over $500$ samples) and a single GD estimator with random initialization. Generalization error is computed on 1000 test samples; error bars indicate standard deviation. \unsure{Notably, for $\Tilde{\alpha}<\Tilde{\alpha}_{\rm spec}$, LD results whith planted initialized can fall below the Bayes-optimal prediction. This apparent violation is likely due to initialization bias and slow relaxation dynamics, which prevent the system from reaching the true equilibrium within the simulation time.}
    } 
    \label{fig: T_zero}
\end{figure}

\begin{figure}[!]
    \centering
    \includegraphics[scale=0.45]{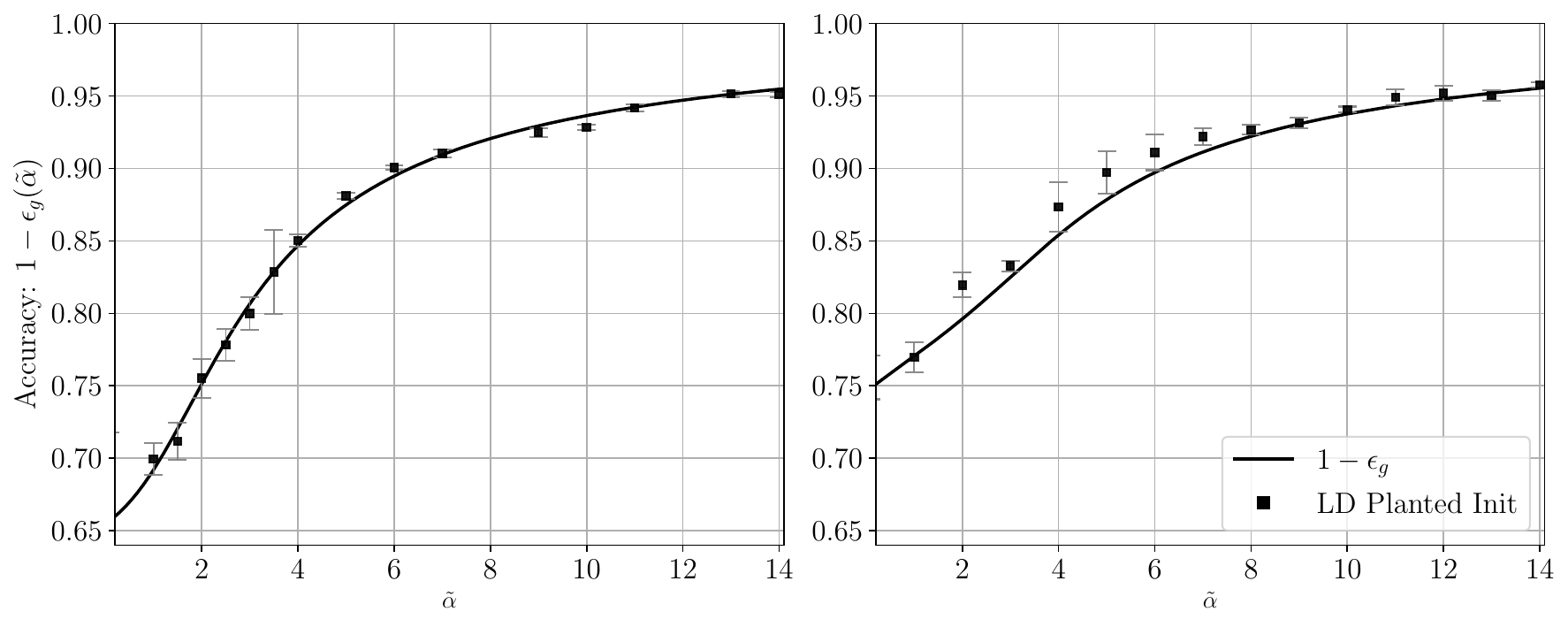}
    \caption{\justifying  
    Accuracy as a function of  $\tilde{\alpha}$ for classification with $f(x) = \mathrm{sign}(x)$ and no label noise ($\Delta^\star = 0$). Activations: (\textbf{Left}) $\sigma(x) = \mathrm{ReLU}(x)$ and (\textbf{Right}) $\sigma(x) = \mathrm{Erf}(x/\sqrt{2})$. Solid lines: ERM theory with Hinge loss in the Bayes-optinal setting (i.e. $\beta \lambda = 1$ and $\beta \to \infty$); markers: LD Planted Init. Accuracy is the average fraction of correctly classified samples by the student. Each point averages $500$ LD-sampled equilibria at $\beta=10^4$, $K=10$, $N=500$ for one dataset. Student weights are projected onto the unit sphere at each update, and teacher weights have unit norm.  Accuracy is computed on 1000 test samples; error bars show standard deviations.}
    \label{fig:Classification}
\end{figure}

While our theory extends to broader settings, for the sake of illustration we focus on regression with a mean-square (MSE) loss and classification with Hinge loss. Given the non-convex nature of the problem, our theory yields multiple stationary solutions of the RS free entropy potential, characterized by the order parameters $(m^*,q^*,v^*)$. We describe them in the next paragraph and Fig.~\ref{fig:quadratic_activation}. Once this picture is further detailed, we then provide numerical validations for all predicted observables for regression at finite $\beta$ (Fig.~\ref{fig: Finite beta - Gen Error and T-S Overlap}) and ERM (Fig.~\ref{fig: T_zero}). Finally, we numerically validate our theory adapted to study the Bayes-optimal (BO) setting for classification (Fig.~\ref{fig:Classification}).

\paragraph{Solutions of the replica symmetric equations and the specialization transition ---} 
To begin illustrating our theory, we present a complete description of all stationary solutions of the RS equations in Fig.~\ref{fig:quadratic_activation}, for regression with MSE loss  $\ell(y,x) =(y-x)^2/2$, and  quadratic activation $\sigma(x) = x^2$, for which the solutions are well separated. The phenomenology discussed here remains broadly general for different activation functions. 
The first panel in Fig.~\ref{fig:quadratic_activation} shows the value of the free entropy, while the order parameters corresponding to each stationary solution are shown in the remaining panels (with matching style of curve). In the particular setting shown here, a unique stationary solution exists up to $\tilde{\alpha} \approx 0.7$, which we refer to as \textit{permutation-symmetric (PS)}. There, the diagonal overlap $q_d^* = 0$, implying that the solution exhibits global permutation symmetry: permuting the hidden units in a typical weight configuration sampled from the Gibbs measure yields an equivalent configuration. At the same time, as we show in the second panel, in this phase each hidden unit of the student is equally correlated with every hidden unit of the teacher, resulting in $m_d^* = 0$. We refer to this as the \emph{PS branch} (dotted-dashed lines). Notice that this PS branch persists for all values of $\tilde{\alpha}$. For $\tilde{\alpha} \gtrsim 0.7$, two new solutions of the saddle point equations appear. The first one, that we name \textit{specialized branch} (solid lines), corresponds to a phase where the hidden units in the student begin to align with specific teacher units as $\tilde{\alpha}$ increases ($m_d^*>0$, second panel), leading to PS breaking ($q_d^* >0$, third panel). The second type of solution (dashed lines) is what we call the \emph{memorization branch}, in analogy to the storage capacity problem: despite exhibiting PS breaking ($q_d^*>0$, third panel), in this phase the student does not align with the teacher, resulting in poor generalization abilities. As we show in the first panel, for $\tilde{\alpha}>0.7$, the solution with the largest free entropy corresponds to the \emph{specialized branch}, meaning that the student finally starts learning the teacher rule.

\paragraph{Comparison with experiments ---} 

In Fig.~\ref{fig: Finite beta - Gen Error and T-S Overlap}, we illustrate the remarkable agreement between the analytical results for the overlaps, Eq.~\eqref{Eq. Overlap concentration}, and the generalization error described in Sec.~\ref{sec: gen_error}, with numerical results obtained from LD. Notably, systems with as few as $K=10$ hidden units are already well described by our large-width theory. Previous work using a simplified \emph{annealed} calculation~\cite{oostwal_hidden_2021} showed that the choice of activation function influences the nature of the learning phase transition. In a similar regression setting, the authors found that networks with ReLU activation undergo a continuous phase transition at a critical value of $\Tilde{\alpha}$. We recover this result, as shown in the upper-left panel in Fig.~\ref{fig: Finite beta - Gen Error and T-S Overlap}, where we observe a \emph{specialization transition} at $\Tilde{\alpha}_{\rm spec}$. \unsure{In this large-dataset regime, the quenched computation effectively reproduces the annealed prediction, which corresponds to the small $\beta$ limit}. However, our more accurate \emph{quenched} computation predicts that for sufficiently large $\beta$, i.e. close to the ERM setting, ReLU networks exhibit a discontinuous phase transition (see left panel in Fig.~\ref{fig: T_zero}).
In contrast, for networks with an Erf activation, we observe a discontinuous transition, independently of $\beta$ (upper-right panel of Fig.~\ref{fig: Finite beta - Gen Error and T-S Overlap} and right panel in Fig.~\ref{fig: T_zero}). Discontinuous transitions are associated with the presence of metastable thermodynamic states. Depending on the initialization, these states can trap LD, preventing it from sampling the Gibbs measure. We identify the presence of metastability by running LD with both random and teacher-close initializations of the weights. In the latter case, LD follows the specialized branch for $\Tilde{\alpha}\geq \Tilde{\alpha}_s$, where the teacher-student overlap increases with the training set size. Conversely, with random initialization, LD approaches the specialization branch at a value $\Tilde{\alpha}>\Tilde{\alpha}_{\rm spec}$ and not at the theoretical transition. This signals the possible presence of a \textit{hard phase} for LD, where it fails to equilibrate when the weights are initialized randomly. A theoretical explanation for this may require a finer analysis taking into account the effects of ``replica symmetry breaking'' in physics jargon  \cite{mezard1987spin,antenucci2019glassy}.

The specific branches that becomes stable around the specialization transition $\Tilde{\alpha}_{\rm spec}$ depends on $\beta$. At intermediate temperatures (e.g. $\beta=10$), the transition occurs between the permutation-symmetric and specialized branches (see right-upper panel of Fig.~\ref{fig: Finite beta - Gen Error and T-S Overlap}). In contrast, in the $\beta\to\infty$ limit, the transition shifts to one between the memorization and specialized branches (see right panel of Fig.~\ref{fig: T_zero}). This indicates that at low temperatures the PS branch loses stability to the memorization branch. 

The results obtained using LD with random initializations support this; in the $\beta\to\infty$ limit, both the generalization error and order parameters converge to the corresponding memorization branch for $\tilde{\alpha}<\Tilde{\alpha}_{\rm spec}$   (Fig.~\ref{fig: T_zero}). Notably, we observe in the same figure that GD with random initialization recovers the statistics of typical ERM minimizers.
Such observation aligns with recent work suggesting that (S)GD concentrates on solutions with probabilities close to the Bayesian posterior \cite{hennick2025almost,mingard2021sgd, smith2018stochastic,JMLR:v18:17-214}, providing a possible explanation for why GD reliably finds typical ERM minimizers despite non-convex landscape. \unsure{ However, at finite system size, initializing at the planted solution alongside incomplete equilibration may trap the dynamics in atypical low-error states. Consequently, the observed generalization error can dip below the asymptotic Bayes-optimal bound, reflecting the lack of full posterior sampling.}

A similar behavior was observed for two-layer neural networks in the extensive-width regime in the Bayes-optimal setting  with $\sigma(x)=x^2$ \cite{maillard_quadratic} \footnote{Maillard et al. \cite{maillard_quadratic} observed a gap between the performance of GD with small regularization and the Bayes-optimal one in the noisy teacher-student regression setting with $\sigma(x)=x^2$. Interestingly, in their paper GD follows a smooth trajectory, which may reflect the memorization branch we described here, a branch that could still exist in the extensive-width regime $K=O(N)$ they consider.}. In our case, we observed that GD reaches the specialization branch at values significantly larger than $\Tilde{\alpha}_{\rm spec}$-- specifically, at $\tilde{\alpha}^{\text{ReLU}}_{\rm GD} \simeq 2.3$ and $\tilde{\alpha}^{\text{Erf}}_{\rm GD} \simeq 2.5$. This discrepancy likely stems from subdominant, atypical configurations that attract GD but are not captured by our theory. 
\unsure{While the BOE \cite{aubin_committee_2018}  remains a strict lower bound in the thermodynamic limit, finite-size effects and biased (planted) initialization can lead to apparent violations in LD simulations at small $\Tilde{\alpha}$ due to incomplete equilibration. Nevertheless, as discussed in the next section, ERM can closely approach the BOE when the regularization $\lambda$ is properly tuned.}

\paragraph{Classification ---}
Analogous comparisons between theory and simulations can be done in the classification case (where $f=\mathrm{sign}$). In Fig.~\ref{fig:Classification} we show the accuracy ($1-\epsilon_g$) as a function of $\tilde{\alpha}$ for both ReLU and Erf activations in the noiseless case ($\Delta^\star=0$) found by doing ERM on the hinge loss, in the BO setting (obtained by setting $\lambda \beta = 1$ and $\beta \to \infty$). 
In both cases the theory predicts that the equilibrium is given by the specialized branch. This is in contrast with the $\mathrm{sign}$ activation, which was investigated in~\cite{schwarze_learning_1993,schwarze_discontinuous_1993}, for which one has a discontinuous transition from the permutation-symmetric to the specialized branch. A study of the small data regime where the size of the dataset is proportional to the size of the input (i.e. $\alpha = P/N = O(1)$), shows that in both those cases the transition from the PS solution is continuous (see SM for additional plots and details). 

\section{Phase diagram and extensive-width limit}

\unsure{Understanding how the regularization strength and the data-to-parameter ratio shape the equilibrium solution is a central question in learning theory. 
In this section, we characterize the resulting phase diagram and investigate how the limit $K \to \infty$ in our setting relates to the $N \to \infty$ limit, clarifying the interplay between width and input dimension.}

\begin{figure}[!]
    \centering
    \includegraphics[scale=0.45]{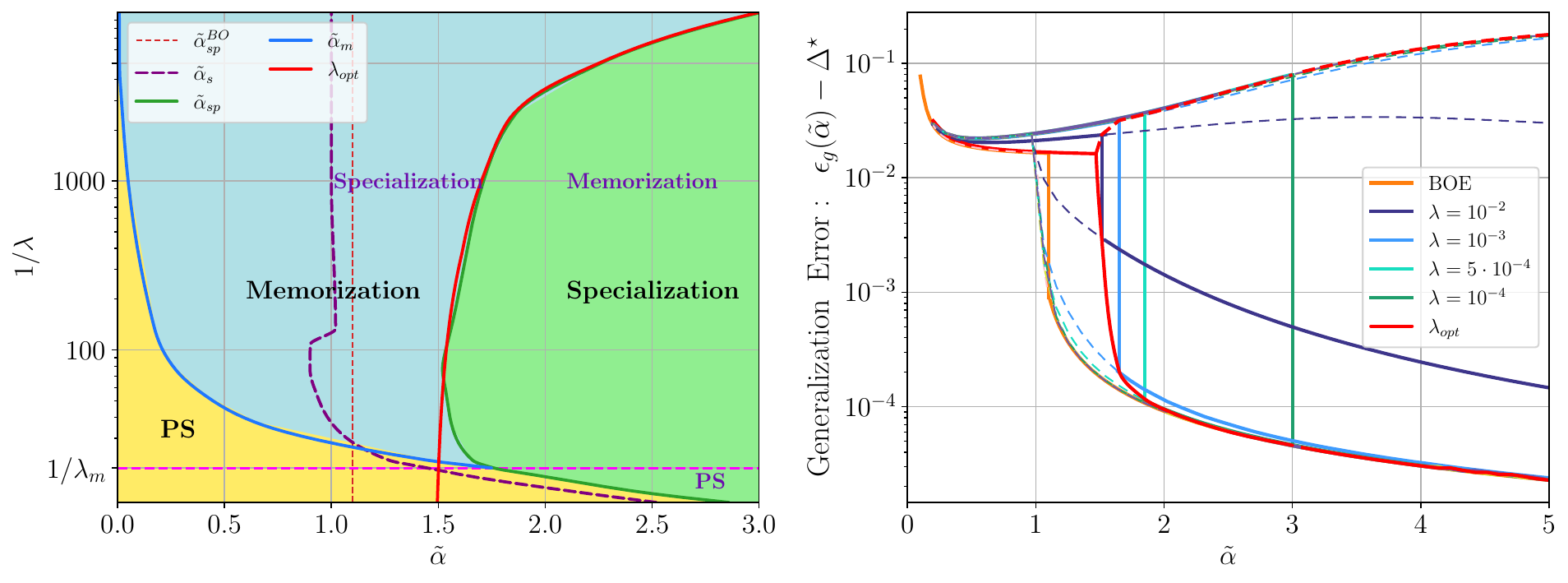}
    \vspace{-10pt}
    \caption{\justifying  Phase diagram (\textbf{Left}) and generalization error (\textbf{Right}) of a two-layer student neural network with activation function $\mathrm{Erf}(x/\sqrt{2})$, trained on a regression task via ERM ($\beta \to \infty$) with MSE loss $\ell(x,y) = (x-y)^2/2$, as a function of the inverse regularization $\lambda^{-1}$ and the sample-to-parameter ration $\tilde{\alpha}$. The training data are generated by a teacher network with matching architecture, label noise $\Delta^{*} = 10^{-4}$, and $K=10$ hidden units. \textbf{(Left)} Phase diagram in the $(\tilde{\alpha},\lambda^{-1})$ plane showing three learning regimes. In the \textbf{permutation-symmetric (PS)} phase, hidden units across different student networks—initialized differently but trained on data generated by the same teacher—remain uncorrelated ($q_d^{*}\simeq 0$); this occurs for small $\tilde{\alpha}$ and $\lambda > \lambda_m$. In the \textbf{memorization} phase, hidden units across different student networks—initialized differently but trained on data generated by the same teacher—become mutually correlated, yet remain uncorrelated with the teacher’s hidden units ($q_d^{*}>0$, $m_d^{*}\simeq 0$); this corresponds to intermediate $\tilde{\alpha}$ and $\lambda < \lambda_m$. The bold blue line at $\tilde{\alpha}_m$ marks the onset of stability of this solution. In the \textbf{specialization} phase, student units align with teacher units ($m_d^{*}>0$). The bold green line $\tilde{\alpha}_{sp}$ marks the stability threshold, while the dashed purple line $\tilde{\alpha}_s$ indicates where the specialized solution first appears (metastable until crossing the green boundary). The dashed vertical red line $\tilde{\alpha}^{BO}_{sp}$ is the Bayes-optimal specialization threshold, and the horizontal dashed pink line $\lambda_m$ marks where the regularization becomes beneficial as a network can escape to go from the \textbf{PS} to \textbf{memorization} region. Bold purple scripts indicates that these solutions are present only as metastable states. \textbf{(Right)} Bayesian optimal error (BOE, orange) and Gibbs generalization error as functions of $\tilde{\alpha}$. Solid curves denote equilibrium solutions, while dashed curves represent metastable ones. Across different $\tilde{\alpha}$ (highlighted in red in both panels), $\lambda_{\mathrm{opt}}$ denotes the value of the regularization strength for which the student's generalization error best approaches the BOE.}
    \label{fig:Phase_diagram_optimal_generalization}
\end{figure}

\paragraph{Phase diagram---}

This diagram is a key outcome of the replica analysis: it identifies which of the PS, Memorization, and Specialization learning phases is the equilibrium state, and which are metastable, as a function of training set size and the inverse of the regularization strength. Specifically, it shows when the system admits multiple coexisting phases --some close to yielding Bayes-optimal (BO) performance, and others only locally stable--, and moreover allows us to determine the optimal regularization strength in terms of generalization performance as a function of $\tilde{\alpha}$.\\

The optimal regularization strength depends on the sample-to-parameter ration $\tilde{\alpha}$, reducing overfitting in low-data regimes while enabling alignment with the teacher in high-data regimes. Contrary to convex optimization problems \cite{krogh1991simple}, the replica analysis reveals a discontinuous behaviour of the optimal regularization as a function of the training set size, a behaviour generally difficult to grasp using statistical learning bounds. In particular, for $\tilde{\alpha} < \tilde{\alpha}^{\mathrm{BO}}_{\mathrm{sp}}$ there coexist two learning phases: Memorization and PS. Although neither describes a phase where the student aligns with the teacher, the latter is still preferable in terms of generalization performance since it matches the BO performance for small $\tilde{\alpha}$. In this regime, a large regularization strength is optimal: it keeps the system in the PS phase, avoiding overfitting and preventing harmful memorization.

For $\tilde{\alpha} > \tilde{\alpha}^{\mathrm{BO}}_{\mathrm{sp}}$, the student starts having enough samples to align with the teacher, causing the specialization phase to emerge and gradually become stable. To enter this phase, $\lambda$ must be drastically reduced: small enough to allow alignment with the teacher, yet large enough to avoid entering the Memorization phase, which can coexist with specialization in the large-data regime and give rises to metastable configurations where algorithms like GD or LD can easily get trapped. In this regime, the replica analysis shows that the best regularization is found near the boundary in the $(\lambda^{-1},\tilde{\alpha})$ plane where Memorization loses global stability and Specialization becomes the only stable learning phase. This can be better observed in the right panel of Fig.~\ref{fig:Phase_diagram_optimal_generalization}, where no single value of $\lambda$ saturates the BOE across all $\tilde{\alpha}$. Based on these fixed-$\lambda$ curves, we reconstruct in red the effective learning curve that a student trained with the $\lambda_{opt}$ across $\tilde{\alpha}$ would follow.

The replica analysis thus predicts non-trivial, discontinuous behavior of the optimal regularization as a function of training set size, due to the presence of distinct learning phases --PS, Memorization, and Specialization--across data regimes. This phase diagram provides key theoretical insights into the algorithmic limitations of standard training procedures such as GD and LD, illustrating when the specialized phase is metastable, i.e., exists but hard to reach via common optimization methods.

\begin{figure}[!]
    \centering
    \includegraphics[scale=0.4]{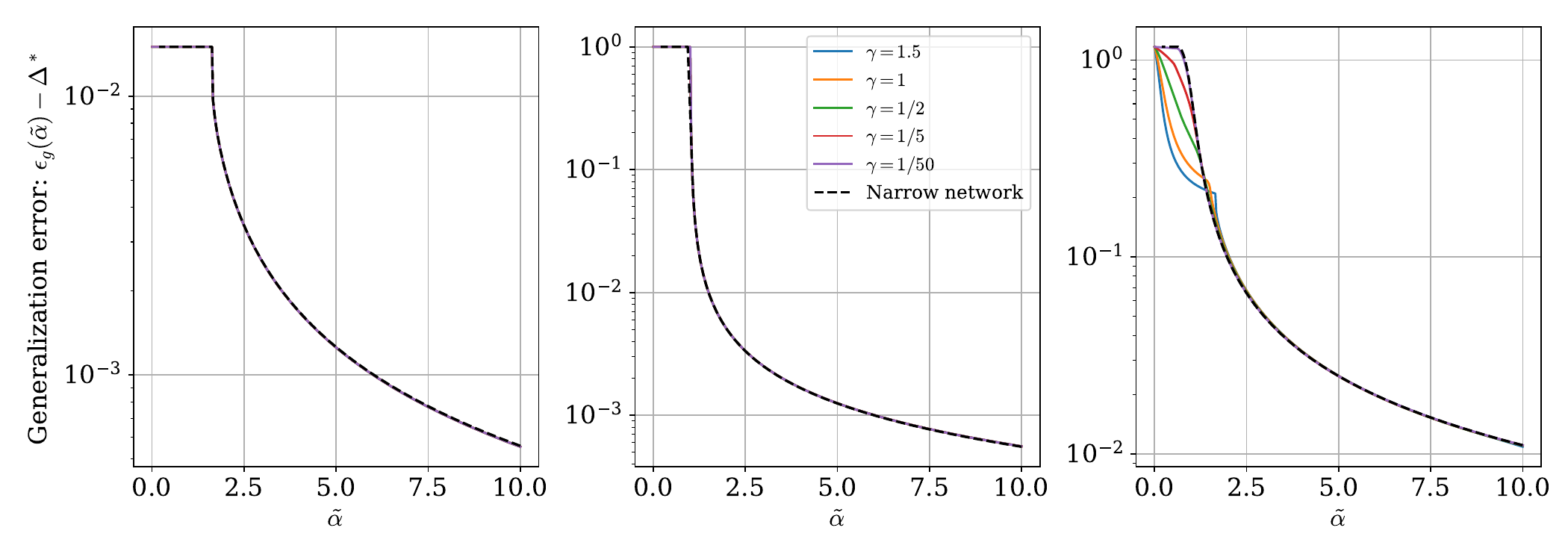}
    \vspace{-15pt}
    \caption{Comparison of the generalization error $\epsilon_g-\Delta^*$ as a function of the sample-to-parameter ratio $\tilde{\alpha} = P/(NK)$ in the Bayes-optimal setting, between our theory for narrow-networks (dashed black line, corresponding to the limit $K \to \infty$ taken after $N \to \infty$) and the extensive-width theory of~\cite{barbier2025statistical} (colored lines, corresponding to the limit $K,N \to \infty$ with $K=\gamma N$). 
(\textbf{Left}) $\sigma(x) = \mathrm{Erf}(x/\sqrt{2})$ and $\Delta^* = 0.005$, 
(\textbf{Middle}) $\sigma(x) = \He_{3}(x)$ and $\Delta^* = 0.005$, 
(\textbf{Right}) $\sigma(x) = \He_{2}(x) + \frac{1}{\sqrt{6}} \He_{3}(x)$ and $\Delta^* = 0.1$.}
    \label{fig:extensive_width_limit}
\end{figure}

\paragraph{The extensive width limit---} \unsure{It is natural to ask how the following two limits are related: the one treated in the present paper, namely $K \to \infty$ \emph{after} $N \to \infty$, and the small-but-extensive width limit considered in~\cite{barbier2025statistical}, where $N, K \to \infty$ jointly with $K = \gamma N$ and $\gamma \ll 1$ but independent of $N$. In Fig.~\ref{fig:extensive_width_limit}, we address this question in the Bayes-optimal setting by plotting the Bayes-optimal generalization error for three activation functions, from left to right:
$$
\sigma(x) = \mathrm{Erf}\!\left(\frac{x}{\sqrt{2}}\right),
\qquad 
\sigma(x) = \He_{3}(x), 
\qquad 
\sigma(x) = \He_{2}(x) + \frac{1}{\sqrt{6}} \He_{3}(x),
$$
where we use the normalized probabilist's Hermite polynomials
$\He_n(x) := \frac{(-1)^n}{\sqrt{n!}}\, e^{x^2/2}\, \frac{d^n}{dx^n} e^{-x^2/2}.$ 
Across the panels, the solid lines for different values of $\gamma$ correspond to the extensive-width prediction of~\cite{barbier2025statistical}, while the dashed lines represent the results obtained from our theory of narrow-network. In the left and middle panels of Fig.~\ref{fig:extensive_width_limit}, the two theories coincide for all values of the sample-to-parameter ratio $\tilde{\alpha}$. In these cases, the collapse of the predictions for different $\gamma$ in~\cite{barbier2025statistical} is a consequence of the absence of the second Hermite coefficient in the activation functions. In particular, the specialization transition matches that of the extensive-width case. In the right panel, we observe that as $\gamma \to 0$ the two theories agree across all data-to-parameter ratios (here, taking $\gamma = 1/50$ already yields a quasi-perfect match), while for non-vanishing $\gamma$ they (quickly) converge to the same limit as $\tilde{\alpha}\to \infty$. These observations suggest that the large-$N$ and large-$K$ limits commute for any $\tilde{\alpha}$ when the activation function has zero second Hermite coefficient, or ``quasi commute'' when $\Tilde{\alpha}$ diverges.}

\section{Discussion and conclusion}
We developed a theory for the typical generalization performance of a narrow one-hidden-layer network in the teacher–student setting, under both finite-temperature Bayesian learning and empirical risk minimization. Our theory highlights the presence of distinct phases and thermodynamic states, including metastable ones that can trap the dynamics of full-batch GD, both with and without explicit noise. We further showed that the activation function is responsible for the appearance of different types of transitions in a Bayesian setting, in a manner that depends on the the temperature parameter. Interestingly, we identified a memorization phase in which the student network breaks permutation symmetry among its hidden units without aligning to the teacher's weights. This study differs from the recent work of \cite{oostwal_hidden_2021, citton_phase_2025}, which focuses on the large-temperature, infinite-data regime, and from \cite{nishiyama_solution_2025}, which investigates the storage capacity problem. It also goes beyond  classical committee-machine analysis \cite{schwarze_statistical_1992, schwarze_learning_1993, engel_storage_1992, schwarze_discontinuous_1993, schwarze_learning_1993-1, urbanczik1997full, ahr1999statistical} by offering a theoretical framework valid for arbitrary inner and outer activation functions with regularized weights.

Our work paves the way for additional investigations of more complex network architectures and input distributions. A key limitation of present analysis is the focus on i.i.d. inputs, which considerably simplifies analytical derivations. Extending the framework to structured inputs is crucial to model the learning performance on real datasets and is left for future work. For instance, we expect our analytical approach to be generalizable to input distributions modeled as mixtures of Gaussian with generic mean vectors and covariance matrices.

Learning in two-layer networks in the $K\ll N$ regime has been recently analyzed using dynamical mean field theory for gradient flow~\cite{montanari_dfmt_2025}. The authors consider both first- and second-layer learnable weights and identify different dynamical regimes using a separation of time-scales argument. It would be interesting to characterize the steady states of such learning dynamics, in relation to the equilibrium distribution induced by Langevin-based sampling of the weight posterior. Doing so would require extending our framework to the case of plastic second-layer weights, which we treated as known (quenched) for simplicity. This extension  also help in fully clarifying the difference between the $K\ll N$ case and the so-called extensive-width regime, i.e. with $K=O\left(N\right)$ and $P=O\left( N \right)$, see \cite{LiSompolinsky2021, pacelli2023statistical} and references therein.

\acksection
F.G. is supported by project SERICS (PE00000014) under the MUR National Recovery and Resilience Plan funded by the European Union- NextGenerationEU. FG also acknowledges GNFM-Indam. E.M.M. acknowledges the MUR-Prin 2022 funding Prot. 20229T9EAT, financed by the European Union (Next Generation EU). J.B. and R.P. were funded by the European Union (ERC, CHORAL, project number 101039794). Views and opinions expressed are however those of the authors only and do not necessarily reflect those of the European Union or the European Research Council. Neither the European Union nor the granting authority can be held responsible for them. C.L. is supported by DARPA Award DIAL-FP-038, and
The William F. Milton Fund from Harvard University.

\newpage
\footnotesize{
\bibliography{references}} 


\newpage 

\clearpage
\appendix
\newgeometry{
	top=1.0in,
	bottom=1.0in,
	left=0.75in,
	right=0.75in,
	headheight=14pt,   
}

\let\addcontentsline\saveaddcontentsline  
\clearpage
\appendix

\vspace*{2em}
\begin{center}
	{\LARGE \bfseries Generalization performance of narrow one-hidden layer networks in the teacher-student setting \par}
	\vspace{0.5em}
	{\large Supplementary Material \par}
\end{center}
\vspace{2em}

\addcontentsline{toc}{section}{Supplementary Material}

\tableofcontents{}

\section{Replica Analysis}

In this section, we provide the full derivation of the free-energy density in eq.~\eqref{Eq. Variational free entropy} in the main. The starting point of the derivation is the Gibbs formulation of the optimization problem in eq.~\eqref{Eq. Free entropy definition} in the main. The resulting calculation is performed using the replica theory toolbox from the statistical physics of disordered systems.

\subsection{Recap of the learning setting} \label{sec:recap_learning_setting}

\paragraph{\textbf{Data Model.}} We consider a dataset $ \mathcal{D} = \{\bx^{\mu}, y^{\mu}\}_{\mu=1}^P$, consisting of $P$ examples. Each data point $\bx^{\mu} \in \mathbb{R}^N$ is i.i.d. random Gaussian $x_{i}^{\mu}\sim \mathcal{N}(0,1)$ and the labels $y^\mu$ are generated by a two-layer \emph{teacher} neural network:
\begin{equation}
	y^{\mu} = \varphi_{\mathbf{A}^\star}^\star(\bW ^\star\bx^\mu, z^\mu\sqrt{\Delta^\star}):=f^{\star} \left[ \frac{1}{\sqrt{K}}\sum_{k=1}^K \, A^{\star}_{k} \sigma \left( \frac{1}{\sqrt{N}} \bw^{\star}_{k} \cdot \bx^{\mu} \right) - \sqrt{K} B^\star + z^\mu\sqrt{\Delta^\star} \right] 
	\label{eq: teacher_label}
\end{equation}
where the first and second-layer teacher weights are denoted by $\bW^\star=(\bw_k^\star\in \mathbb{R}^N)_{k=1}^K\in \mathbb{R}^{K\times N}$ and $\mathbf{A}^{\star} \in \mathbb{R}^K$ respectively, while $\sigma(\cdot)$ and $f^{\star}(\cdot)$ are generic point-wise activation functions and $B^{\star}$ is a bias term.  The i.i.d. $z^\mu\sim \mathcal{N}(0,1)$ are label noise whose standard deviation is controlled by $\Delta^\star$. We will assume the first-layer weights of the teacher to be extracted from a generic prior $P_{\bW^\star}(\bW)$ which is factorized over the $N$ inputs
\begin{equation}
	P_{\bW^\star}(\bW) = \prod_{i = 1}^N P_{\bw^\star}(\bw_i).
\end{equation}
The second layer weights are extracted from a generic prior 
\begin{equation}
	P_{\mathbf{A}}(\mathbf{A}^\star) = \prod_{i=1}^K P_{A}(A^\star_i)    .
\end{equation}

The bias $B^\star$ is fixed to remove the mean of the second layer pre-activation. Therefore, given the input data and weight's distribution:
\begin{equation}
	B^\star = \frac{1}{K}\sum_{k=1}^K\mathbb{E}_{\bx^\mu,A_k^\star,\bw^\star}\left[a_k^\star\sigma\left(\frac{\bw ^\star_k \cdot \bx^\mu}{\sqrt{N}}\right)\right] =  \frac{\mu_A}{K}\sum_{k=1}^K\mathbb{E}_{x,\bw^\star}\left[\sigma\left(x\sqrt{\frac{\bw^\star_k \cdot \bw^\star_k}{
			N}}\right)\right] \overset{N\to\infty}{=} \mu_A \mathbb{E}_x\left[\sigma\left(x\sqrt{\frac{\mathbb{E}[\bw_k^2]}{N}}\right)\right]\label{eq:bias}.
\end{equation}
In the last equality, we use the assumption that each weight vector $\bw_k^\star$ is drawn independently and identically distributed (i.i.d.), and that the $\bw_{ki}$ are independently sampled. We denote by $\mu_{A}$ the mean of $P_A$.\\

\paragraph{\textbf{Learning Model.}} We want to fit the dataset $\mathcal{D}$ with a two-layer \emph{student} network:
\begin{equation}
	\hat{y}^{\mu} = \varphi_{\mathbf{A}}(\bW\bx^\mu,0) = f \left[ \frac{1}{\sqrt{K}}\sum_{k=1}^K \, A_{k} \sigma_k \left(\frac{1}{\sqrt{N}} \bw_{k} \cdot \bx^{\mu} \right) - \sqrt{K}B \right]\label{eq:predictions}
\end{equation}
whose second-layer weights $\mathbf{A} \in \mathbb{R}^K$ are generic but fixed and only the first layer $\bW \in \mathbb{R}^{K \times N}$ is learned through the training set. This learning model is also known as committee machine. The activation functions $f(\cdot)$ is not necessarily the same as the ones of the teacher network; and the bias term $B$ depends on the weights $\bW$. In practice, $B$ is updated separately during learning once per epoch, based on the current state of the weights (sec. \ref{sec:appendix_numerics}).\\ 

\paragraph{\textbf{The task.}} We are interested in analytically characterize the generalization performances of the Empirical Risk Minimization estimator:
\begin{equation}
	\hat{\bW} = \underset{\bW}{\mbox{argmin}} \left[ \sum_{\mu = 1}^P \ell \left( y^{\mu}, \hat{y}^{\mu} \left(\bx^{\mu}; \bW, \boldsymbol{A}  \right) \right) + \lambda r( \bW )\right] \label{eq: non-convex_opt_problem}
\end{equation}
for generic loss functions $\ell \left( \cdot, \cdot \right)$ and regularization $r(\cdot)$, with weight decay $\lambda\geq0$.  We focus on the case where the trainable model is already
set with the correct readout weights (which are few compared to the inner ones) and we thus 
the dependency on those weights to lighten notations, i.e. $\boldsymbol{A}^\star=\boldsymbol{A}$. 

\subsection{Gibbs formulation}

Given the learning setting defined in sec. \ref{sec:recap_learning_setting}, we define the following Gibbs measure over the student first-layer weights $\bW \in \mathbb{R}^{N \times K}$:
\begin{equation}
	\pi_{\beta} \left( \bW |  \mathcal{D}; \mathbf{A} \right) = \frac{1}{Z_{\beta}} e^{-\beta \sum_{\mu = 1}^P \ell \left( y^{\mu}, \ \hat{y}^{\mu}\left( \bx^{\mu}; \bW, \mathbf{A}\right) \right) - r\left( \bW \right)} 
	= \frac{1}{Z_{\beta}} P_{\bW}(\bW) \prod_{\mu = 1}^P P_y \left( y^{\mu} \Big\vert  \frac{\bW\bx^{\mu}}{\sqrt{N}}; \boldsymbol{A} \right) \label{eq:posterior}  
\end{equation}
with $\beta >0$ being the inverse temperature. In the low-temperature limit (e.g. $\beta \rightarrow \infty$), $\pi_{\beta}$ peaks in the solutions of the non-convex optimization problem in eq.~\eqref{eq: non-convex_opt_problem}. Note that, the second equality suggests that $\pi_{\beta}$ can be interpreted as a posterior distribution over the first layer $\bW$, with $P_\bW$ and $P_y$ being the prior and the likelihood respectively. We take $r$ to be a component-wise regularizer, which implies that the prior factorizes as:
\begin{equation}
	P_{\bW}(\bW) = \prod_{i=1}^N P_{\bw}(\bw_i) .
\end{equation}

Obtaining an exact analytical expression for $\pi_{\beta}$ when $N$, $P$ and $K$ are large is unpracticable. However, in this high-dimensional limit, it can be shown that the free-entropy density averaged over the training set distribution:
\begin{equation}
	\langle \Phi_{\beta} \rangle  =  \underset{N \rightarrow \infty}{\lim} \frac{1}{N K} \mathbb{E}_{\mathcal{D}}\left[\mbox{log}Z_{\beta}\right]  
\end{equation}
can be explicitly computed by means of the replica method. Here $\mathbb{E}_\mathcal{D}$ denotes the expectation over the data $\mathcal{D}$, which implicitly includes averaging over the weights $\bw^\star$ and the label noise $\mathbf{z}=(z^1,\hdots,z^P)$. This quantity is central because key observables characterizing the learning problem in sec. \ref{sec:recap_learning_setting} -- such as training loss or generalization error-- can be obtained from partial derivatives of the free-energy density. We will perform this computation in the thermodynamic limit $N, P, K \to \infty$, but in a regime where the number of hidden neurons is small compared to the input dimension and sample dimensions: $\frac{K}{N} \to 0$ and $\frac{K}{P} \to 0$. Moreover, we  distinguish between two scaling data regimes:\begin{itemize}
	\item A \textit{small-sample} regime, where the number of samples scales linearly with the input dimension: $P = \alpha N$, with $\alpha$ being the constraint density also known as the sample complexity.
	\item A \textit{large-sample} regime, where the sample complexity is controlled by $\Tilde{\alpha} \equiv \frac{P}{NK } = O (1) $
\end{itemize}

\subsection{Replica computation of the free-energy density} 

The free-entropy density $\Phi_{\beta}(\mathcal{D}) =  \mbox{log}\left( Z_{\beta}(\mathcal{D}) \right)$ is a random variable over different realizations of the training set $ \mathcal{D} = \{\bx^{\mu}, y^{\mu}\}_{\mu=1}^P$. We assume that its associated distribution $P(\Phi)$ satisfies a large deviation principle:
\begin{equation}
	P(\Phi) \simeq e^{-NK\Phi}.
\end{equation}
In the high-dimensional limit $N \rightarrow \infty$, this hypothesis implies that $P(\Phi)$ is peaked on $\langle \Phi_{\beta}\rangle$ and the fluctuations around this mean go to zero, a property known as \emph{self-averaging}. 

To compute $\langle \Phi_{\beta} \rangle$ we need to take the expectation of a logarithm. To overcome this difficulty, we can use replica theory and therefore write the mean free-entropy density in terms of $n>0$ different copies of the same learning system:
\begin{equation}
	K \langle \Phi_{\beta} \rangle =  \underset{N \rightarrow \infty}{\lim} \frac{1}{N} \underset{n \rightarrow 0^{+}}{\lim} \frac{ \mathbb{E}_{\mathcal{D}}\left[Z^n_{\beta}\right]-1}{n}
\end{equation}
where the replicated partition function  is:
\begin{equation}
	Z^n_{\beta} = \prod_{a=1}^n \int_{\mathbb{R}^{K\times N}} d\bw^a P_w\left( \bw^a \right) \prod_{\mu = 1}^P P_y \left( y^{\mu} \vert \left\{ \frac{\bw_k^a \cdot \bx^{\mu}}{\sqrt{N}} \right\}_{k=1}^K, \mathbf{A} \right).
\end{equation}

\paragraph{Average over the training set.} To average the replicated partition function over the training set, we can first write the expectation over the labels explicitly, using the generative model in eq.\eqref{eq: teacher_label}, and then introduce the hidden layer pre-activations for both the teacher and student network:
\begin{equation}
	\nu_{\mu k} = \frac{\bw_k^{\star} \cdot \bx^{\mu}}{\sqrt{N}} ,\hspace{10mm} \lambda^a_{\mu k} = \frac{\bw_k^{a} \cdot \bx^{\mu}}{\sqrt{N}} 
\end{equation}
by means of Dirac $\delta$-functions. This allows to simplify the expectation over the input data $\bx^{\mu}$:
\begin{equation}
	\begin{split}
		\mathbb{E}_{\mathcal{D}} \left[Z_{\beta}^n \right] &= \int \prod_{\mu = 1}^P dy^{\mu} \int d\bW^\star P_{w^\star}(\bW^\star) \prod_{\mu = 1}^P \int \prod_{k=1}^K \frac{d\nu_{\mu k}}{\sqrt{2\pi}} P^{\star}_y \left( y^{\mu} \vert \{ \nu_{\mu k}\}_{k=1}^K ,\,\mathbf{A} \right)\\ 
		&\times\int \prod_{a=1}^n d\bW^a P_w(\bW^a) \prod_{a=1}^n\prod_{\mu = 1}^P \prod_{k=1}^K \frac{d\lambda^a_{\mu k}}{\sqrt{2\pi}} P_y \left( y^{\mu} \vert \{ \lambda^a_{\mu k}\}_{k=1}^K,\,\mathbf{A}  \right) \\
		&\times \prod_{\mu = 1}^n \mathbb{E}_{\bx^{\mu}} \left[ \prod_{k=1}^K \delta\left( \nu_{\mu k} - \frac{\bw_k^{\star} \cdot \bx^{\mu}}{\sqrt{N}} \right) \prod_{a=1}^n \delta \left( \lambda^a_{\mu k} - \frac{\bw_k^{a} \cdot \bx^{\mu}}{\sqrt{N}} \right)\right].
	\end{split}
\end{equation}
By introducing the integral representation of the Dirac $\delta$-functions:
\begin{equation}
	\begin{split}
		\delta\left( \nu_{\mu k} - \frac{\bw_k^{\star} \cdot \bx^{\mu}}{\sqrt{N}} \right) &= \int \frac{d\hat{\nu}_{\mu k}}{\sqrt{2\pi}} \mbox{exp}\left( -i\hat{\nu}_{\mu k} \left( \nu_{\mu k} - \frac{\bw_k^{\star} \cdot \bx^{\mu}}{\sqrt{N}} \right) \right),\\
		\delta \left( \lambda^a_{\mu k} - \frac{\bw_k^{a} \cdot \bx^{\mu}}{\sqrt{N}} \right) &= \int \frac{d\hat{\lambda}^a_{\mu k}}{\sqrt{2\pi}} \mbox{exp}\left( -i\hat{\lambda}_{\mu k} \left( \lambda^a_{\mu k} - \frac{\bw_k^{a} \cdot \bx^{\mu}}{\sqrt{N}} \right) \right), 
	\end{split}
\end{equation}
we can finally perform the expectation over $\bx^{\mu}$, which is now reduced to a simple Gaussian integral:
\begin{equation}
	\begin{split}
		\mathbb{E}_{\mathcal{D}} \left[Z_{\beta}^n \right] &= \int d\bW^\star P_{w^\star}(\bW^\star) \prod_{\mu = 1}^P \int dy^{\mu} \int \prod_{k=1}^K \frac{d\nu_{\mu k} d\hat{\nu}_{\mu k}}{2\pi} e^{i \sum_{k=1}^K \hat{\nu}_{\mu k} \nu_{\mu k}} P^{\star}_y \left( y^{\mu} \vert \{ \nu_{\mu k}\}_{k=1}^K,\,\mathbf{A}  \right) \\ 
		&\times \int \prod_{a=1}^n d\bW^a P_w(\bW^a) \prod_{a=1}^n\prod_{\mu = 1}^P \int \prod_{k=1}^K \frac{d\lambda^a_{\mu k} d\hat{\lambda}^a_{\mu k}}{2\pi} e^{i \sum_{k=1}^K \hat{\lambda}^a_{\mu k} \lambda^a_{\mu k}}  P_y \left( y^{\mu} \vert \{ \lambda^a_{\mu k}\}_{k=1}^K,\,\mathbf{A}  \right) \\
		& \times \mbox{exp}\left(-\frac{1}{2} \sum_{k, k^\prime=1}^K \hat{\nu}_{\mu k} \hat{\nu}_{\mu k^\prime} \frac{\bw^\star_{k} \cdot \bw^\star_{k^{\prime}}}{N} - \frac{1}{2} \sum_{a,b = 1}^n \sum_{k, k^\prime=1}^K \hat{\lambda}^a_{\mu k} \hat{\lambda}^b_{\mu k^\prime} \frac{\bw^a_{k} \cdot \bw^b_{k^{\prime}}}{N} - \sum_{a=1}^n \sum_{k, k^\prime=1}^K \hat{\lambda}^a_{\mu k} \hat{\nu}_{\mu k^\prime} \frac{\bw^a_{k} \cdot \bw^\star_{k^\prime}}{N} \right).
	\end{split} \label{eq: replicated_pf_after_data_average}
\end{equation}

\paragraph{Order parameters.} After averaging over the training set, we note that the integrals in eq.~\eqref{eq: replicated_pf_after_data_average} get coupled via the following set of overlap parameters:
\begin{equation}
	\rho_{k k^\prime} \defeq \frac{\bw^\star_{k} \cdot \bw^\star_{k^{\prime}}}{N} ,\hspace{10mm} q^{ab}_{k k^\prime} \defeq \frac{\bw^a_{k} \cdot \bw^b_{k^{\prime}}}{N} ,\hspace{10mm} m^a_{k k^\prime} \defeq \frac{\bw^a_{k} \cdot \bw^\star_{k^\prime}}{N}.
\end{equation}
To decouple the integrals, it is useful to introduce the overlap definition by means of Dirac $\delta$-functions and their integral representations:
\begin{equation}
	\begin{split}
		1 &\propto \int \prod_{kk^\prime = 1}^K \frac{d\rho_{kk^\prime} d\hat{\rho}_{kk^\prime}}{2\pi} \mbox{exp}\left( -i \sum_{k k^\prime = 1}^K\hat{\rho}_{kk^\prime}\left(\rho_{kk^\prime} - \frac{\bw^\star_{k} \cdot \bw^\star_{k^{\prime}}}{N} \right) \right) \\
		& \times\int \prod_{kk^\prime = 1}^K  \prod_{a=1}^n \frac{dm^a_{kk^\prime} d\hat{m}^a_{kk^\prime}}{2\pi} \mbox{exp}\left( -i \sum_{k k^\prime = 1}^K\sum_{a=1}^n\hat{m}^a_{kk^\prime}\left(m^a_{kk^\prime} - \frac{\bw^a_{k} \cdot \bw^\star_{k^{\prime}}}{N} \right) \right)\\
		&\times \int \prod_{kk^\prime = 1}^K  \prod_{a \leq b=1}^n \frac{dq^{ab}_{kk^\prime} d\hat{q}^{ab}_{kk^\prime}}{2\pi} \mbox{exp}\left( -i \sum_{k k^\prime = 1}^K\sum_{a\leq b = 1}^n\hat{q}^{ab}_{kk^\prime}\left(q^{ab}_{kk^\prime} - \frac{\bw^a_{k} \cdot \bw^b_{k^{\prime}}}{N} \right) \right). 
	\end{split}
\end{equation}
Introducing the above in eq.~\eqref{eq: replicated_pf_after_data_average}, we notice that the integrals factorize over the index $i = 1,..., N$ and $\mu = 1, ..., P = \alpha N$. With the change of variables $i\hat{\rho}_{kk^\prime} \rightarrow  \hat{\rho}_{kk^\prime} $, $i\hat{m}^a_{kk^\prime} \rightarrow  \hat{m}^a_{kk^\prime}$ and $i\hat{q}^{ab}_{kk^\prime} \rightarrow  \hat{q}^{ab}_{kk^\prime}$, we can then express the replicated partition function in terms of saddle-point integrals:
\begin{equation}
	\mathbb{E}_{\mathcal{D}} \left[Z_{\beta}^n \right] = \int \prod_{kk^\prime = 1}^K \frac{d\rho_{kk^\prime} d\hat{\rho}_{kk^\prime}}{2\pi} \int \prod_{kk^\prime = 1}^K  \prod_{a=1}^n \frac{dm^a_{kk^\prime} d\hat{m}^a_{kk^\prime}}{2\pi} \int \prod_{kk^\prime = 1}^K  \prod_{a,b=1}^n \frac{dq^{ab}_{kk^\prime} d\hat{q}^{ab}_{kk^\prime}}{2\pi} \mbox{exp}\left( N \phi \right)
\end{equation}
where the potential $\phi$ is given by the sum of three distinct terms:
\begin{equation}
	\phi = G_I \left( \rho_{kk^\prime}, m^a_{kk^\prime}, q^{ab}_{kk^\prime}, \hat{\rho}_{kk^\prime}, \hat{m}^a_{kk^\prime}, \hat{q}^{ab}_{kk^\prime} \right) + G_S \left( \hat{\rho}_{kk^\prime}, \hat{m}^a_{kk^\prime}, \hat{q}^{ab}_{kk^\prime} \right) + \alpha G_E \left( \rho_{kk^\prime}, m^a_{kk^\prime}, q^{ab}_{kk^\prime} \right),
	\label{eq:action}
\end{equation}
with the ``interaction'', ``entropic'' and ``energetic'' potentials being respectively:
\begin{equation}
	\begin{split}
		&G_I = - \sum_{k k^\prime} \hat{\rho}_{kk^\prime} \rho_{kk^\prime} - \sum_{k k^\prime} \sum_{a=1}^n \hat{m}^a_{k k^\prime} m^a_{k k^\prime} - \sum_{k k^\prime}\sum_{a \leq b=1}^n \hat{q}^{ab}_{k k^\prime}q^{ab}_{k k^\prime},\\
		&G_S = \mbox{log} \int_{\mathbb{R}^K} d\bw^\star P_{w^\star}(\bw^\star) \prod_{a=1}^n \int_{\mathbb{R}^K} d\bw^a P_w(\bw^a) \ \mbox{exp}\left( \sum_{kk^\prime} \hat{\rho}_{kk^\prime} w_k^\star w^\star_{k^{\prime}} + \sum_{a=1}^n \sum_{k k^\prime} \hat{m}_{kk^\prime}^a w^a_k w^\star_{k^\prime} + \sum_{a \leq b = 1}^n \hat{q}^{ab}_{kk^\prime} w_k^a w^b_{k^\prime} \right),\\
		&G_E = \mbox{log}\int dy \int \prod_{k=1}^K \frac{d\nu_k d\hat{\nu}_k}{2\pi} P^{\star}_y \left( y \vert \{ \nu_{k}\}_{k=1}^K,\,\mathbf{A}  \right) \ \mbox{exp}\left(-\frac{1}{2} \sum_{k k^\prime = 1}^K \rho_{kk^\prime} \hat{\nu}_k \hat{\nu}_k^\prime + i \sum_{k=1}^K \hat{\nu}_k \nu_k \right) \times\\
		&\times \int \prod_{k=1}^K \prod_{a=1}^n \frac{d\lambda_k^a d\hat{\lambda}^a_k}{2\pi}  \, P_y \left( y \vert \{ \lambda^a_{ k}\}_{k=1}^K,\,\mathbf{A}  \right) \ \mbox{exp}\left( -\frac{1}{2} \sum_{ab=1}^n \sum_{kk^\prime = 1}^K q_{kk^\prime}^{ab} \hat{\lambda}_k^a \hat{\lambda}^b_{k^\prime} - \sum_{a=1}^n \sum_{kk^\prime = 1}^K m_{k k^\prime}^a \hat{\lambda}_k^a \hat{\nu}_{k^\prime} + i \sum_{a=1}^n \sum_{k=1}^K \hat{\lambda}^a_k \lambda_k^a\right).
	\end{split}
\end{equation}
In the high-dimensional limit, the integrals over the overlap parameters and their conjugates can be solved by saddle-point, so that the mean free-entropy density can be determined through the following optimization problem:
\begin{equation}
	\langle \Phi_{\beta} \rangle = \underset{N \rightarrow \infty}{\lim} \frac{1}{N} \underset{n \rightarrow 0^{+}}{\lim} \frac{\mathbb{E}_{\mathcal{D}}\left[Z^n_{\beta}\right] - 1}{n} = \underset{\rho_{kk^\prime}, m^a_{kk^\prime}, q^{ab}_{kk^\prime}, \hat{\rho}_{kk^\prime}, \hat{m}^a_{kk^\prime}, \hat{q}^{ab}_{kk^\prime}}{\mbox{extr}} \left[\underset{n \rightarrow 0^{+}}{\lim} \frac{1}{n} \phi \left( \rho_{kk^\prime}, m^a_{kk^\prime}, q^{ab}_{kk^\prime}, \hat{\rho}_{kk^\prime}, \hat{m}^a_{kk^\prime}, \hat{q}^{ab}_{kk^\prime} \right)\right].
\end{equation}
This leads to a system of coupled saddle-point equations, whose unknowns are precisely the overlap parameters and their conjugates.\\

\paragraph{Replica-Symmetry assumption.} To proceed further in the calculation, we can assume that all replicas are symmetric with respect to permutations:
\begin{equation}
	\begin{split}
		q_{kk^\prime}^{ab} &= r_{kk^\prime} \delta_{ab} + q_{kk^\prime} (1 - \delta_{ab}) ,\hspace{10mm} m_{kk^\prime}^{a} = m_{kk^\prime}, \\
		\hat{q}_{kk^\prime}^{ab} &= -\frac{1}{2}\hat{r}_{kk^\prime} \delta_{ab} + \hat{q}_{kk^\prime} (1 - \delta_{ab}), \hspace{4.3mm} \hat{m}_{kk^\prime}^{a} = \hat{m}_{kk^\prime}  .     
	\end{split}
\end{equation}
Indeed, since replicas have been introduced as different copies of the same system, we can think that they are all equivalent. Applying this ansatz in eq.\eqref{eq:action}, we can write the replica-symmetric interaction, entropic and energetic potential as:
\begin{equation}
	\begin{split}
		&G_I = - \sum_{k k^\prime} \hat{\rho}_{kk^\prime} \rho_{kk^\prime} - n\sum_{k k^\prime} \hat{m}_{k k^\prime} m_{k k^\prime} + \frac{n}{2} \sum_{k k^\prime} \hat{r}_{k k^\prime}r_{k k^\prime} - \frac{n(n-1)}{2} \sum_{kk^\prime} \hat{q}_{kk^\prime} q_{kk^\prime},\\
		&G_S = \mbox{log} \int_{\mathbb{R}^K} d\bw^\star P_{w^\star}(\bw^\star) e^{(\bw^\star)^t \hat{\rho} \bw^\star} \prod_{a=1}^n \int_{\mathbb{R}^K} d\bw^a P_w(\bw^a) \ \mbox{exp}\left( -\frac{1}{2}\sum_{a=1}^n (\bw^a)^t \hat{v} \bw^a + \frac{1}{2} \sum_{a b = 1}^n (\bw^a)^t \hat{q} \bw^b  + \sum_{a=1}^n (\bw^\star)^t \hat{m} \bw^a \right),\\
		&G_E = \mbox{log} \int dy \int \prod_{k=1}^K \frac{d\nu_k d\hat{\nu}_k}{2\pi} P^{\star}_y \left( y \vert \{ \nu_{k}\}_{k=1}^K,\,\mathbf{A}  \right) \mbox{exp}\left(-\frac{1}{2} \hat{\boldsymbol{\nu}}^t \rho \hat{\boldsymbol{\nu}} + i \hat{\boldsymbol{\nu}}^t \boldsymbol{\nu} \right) \\
		&\times \int \prod_{k=1}^K \prod_{a=1}^n \frac{d\lambda_k^a d\hat{\lambda}^a_k}{2\pi}  P_y \left( y \vert \{ \lambda^a_{ k}\}_{k=1}^K,\,\mathbf{A}  \right) \mbox{exp}\left(- \frac{1}{2} \sum_{a=1}^n (\hat{\boldsymbol{\lambda}}^a)^t v \hat{\boldsymbol{\lambda}}^a -\frac{1}{2} \sum_{ab=1}^n (\hat{\boldsymbol{\lambda}}^a)^t q \hat{\boldsymbol{\lambda}}^b -\sum_{a=1}^n \hat{\boldsymbol{\nu}}^t m \hat{\boldsymbol{\lambda}}^a + i \sum_{a=1}^n (\hat{\boldsymbol{\lambda}}^a)^t \boldsymbol{\lambda}^a  \right).
	\end{split}
\end{equation}
with $\boldsymbol{\nu}, \hat{\boldsymbol{\nu}}, \boldsymbol{\lambda}^a, \hat{\boldsymbol{\lambda}}^a \in \mathbb{R}^K$ and $\rho, \hat{\rho}, q, \hat{q}, r, \hat r, m, \hat{m} \in \mathbb{R}^{K\times K}$. The variables $v$ and $\hat v$ appearing in the energetic and entropic terms are also defined as
\begin{subequations}
	\label{eq::change_of_variables}
	\begin{align}
		v &\equiv r- q, \\
		\hat v &\equiv \hat r + \hat q \,.
	\end{align}
\end{subequations}
We can now notice that, the quadratic sums appearing in both $G_S$ and $G_E$ can be linearized using the following Hubbard-Strat\"{o}novich transformations:
\begin{equation}
	\begin{split}
		&\mbox{exp}\left( \frac{1}{2} \sum_{ab=1}^n (\bw^a)^t \hat{q} \bw^b \right) = \int_{\mathbb{R}^K} D\boldsymbol{\xi}\ \mbox{exp} \left( \boldsymbol{\xi}^t \hat{q}^{1/2} \sum_{a=1}^n \bw^a\right),\\
		& \mbox{exp}\left( -\frac{1}{2} \sum_{ab=1}^n (\hat{\boldsymbol{\lambda}}^a)^t q \hat{\boldsymbol{\lambda}}^b \right)=\int_{\mathbb{R}^K} D\boldsymbol{\xi} \ \mbox{exp}\left( i \boldsymbol{\xi}^t q^{1/2} \sum_{a=1}^n \boldsymbol{\lambda}^a \right),
	\end{split}
\end{equation}
with $\boldsymbol{\xi} \in \mathbb{R}^K$ and $D\boldsymbol{\xi} \defeq{\mathrm{d}\boldsymbol{\xi}\,e^{-\frac{1}{2}\boldsymbol{\xi}^2}/(2\pi)^{K/2}}$. This allows us to factorize both $G_S$ and $G_E$ over the replica index $a=1,...,n$:
\begin{equation}
	\begin{split}
		&G_S = \mbox{log} \int_{\mathbb{R}^K} D\boldsymbol{\xi} \int_{\mathbb{R}^K} d\bw^\star P_{w^\star}(\bw^\star) e^{(\bw^\star)^t \hat{\rho} \bw^\star} \left[ \int_{\mathbb{R}^K} d\bw \, P_w(\bw) \ \mbox{exp}\left( -\frac{1}{2}\bw^t \hat{v} \bw + \mathbf{\xi}^t \hat{q}^{1/2} \bw +( \bw^\star)^t \hat{m} \bw \right)\right]^n,\\
		&G_E = \mbox{log} \int D\boldsymbol{\xi} \int dy \int \prod_{k=1}^K \frac{d\nu_k d\hat{\nu}_k}{2\pi} P^{\star}_y \left( y \vert \{ \nu_{k}\}_{k=1}^K,\,\mathbf{A}  \right) \mbox{exp}\left(-\frac{1}{2} \hat{\boldsymbol{\nu}}^t \rho \hat{\boldsymbol{\nu}} + i \hat{\boldsymbol{\nu}}^t \boldsymbol{\nu} \right) \times\\
		&\times \left[\int \prod_{k=1}^K \frac{d\lambda_k d\hat{\lambda}_k}{2\pi}  P_y \left( y \vert \{ \lambda_{ k}\}_{k=1}^K,\,\mathbf{A}  \right) \mbox{exp}\left(- \frac{1}{2} \hat{\boldsymbol{\lambda}}^t v \hat{\boldsymbol{\lambda}} +i\boldsymbol{\xi}^t q^{1/2} \hat{\boldsymbol{\lambda}} - \hat{\boldsymbol{\nu}}^t m \hat{\boldsymbol{\lambda}} + i \hat{\boldsymbol{\lambda}}^t \boldsymbol{\lambda}  \right)\right]^n.
	\end{split}
\end{equation}
We note that the teacher and the student are coupled in $G_S$ by the term $(\bw^\star)^t \hat{m} \bw$ and in $G_E$ by the term $\hat{\boldsymbol{\nu}}^t m \hat{\boldsymbol{\lambda}}$. To uncouple them and simplify $G_S$ and $G_E$ further, we can perform the change of variables $\boldsymbol{\xi} \rightarrow \boldsymbol{\xi} -  \hat{q}^{-1/2}\hat{m}^t\hat{\bw}^\star$ in $G_S$ and $\boldsymbol{\xi} \rightarrow \boldsymbol{\xi} - i q^{-1/2}m^t\hat{\boldsymbol{\nu}}$ in $G_E$. In this way, we get the following:
\begin{equation}
	\begin{split}
		&G_S = \mbox{log} \int \frac{d\boldsymbol{\xi}}{(2\pi)^{K/2}}  \left[ \int_{\mathbb{R}^K} d\bw^\star P_{w^\star}(\bw^\star) e^{-\frac{1}{2} \left(\boldsymbol{\xi} - q^{-1/2}\hat m^t\hat{\bw}^\star \right)^2 + (\bw^\star)^t \hat{\rho} \bw^\star} \right] \left[ \int_{\mathbb{R}^K} d\bw \, P_w(\bw) \ \mbox{exp}\left( -\frac{1}{2}\bw^t \hat{v} \bw + \boldsymbol{\xi}^t \hat{q}^{1/2} \bw \right)\right]^n ,\\
		&G_E = \mbox{log} \int \frac{d\boldsymbol{\xi}}{(2\pi)^{K/2}} \int dy \int \prod_{k=1}^K \frac{d\nu_k d\hat{\nu}_k}{2\pi} P^{\star}_y \left( y \vert \{ \nu_{k}\}_{k=1}^K,\mathbf{A}  \right) \exp\left(-\frac{1}{2} \left(\boldsymbol{\xi} - iq^{-1/2}m^t\hat{\boldsymbol{\nu}} \right)^2 -\frac{1}{2} \hat{\boldsymbol{\nu}}^t \rho \hat{\boldsymbol{\nu}} + i \hat{\boldsymbol{\nu}}^t \boldsymbol{\nu} \right) \\
		&\hspace{9mm}\times \left[\int \prod_{k=1}^K \frac{d\lambda_k d\hat{\lambda}_k}{2\pi}  P_y \left( y \vert \{ \lambda_{ k}\}_{k=1}^K,\,\mathbf{A}  \right) \exp\left(- \frac{1}{2} \hat{\boldsymbol{\lambda}}^t v \hat{\boldsymbol{\lambda}} +i\boldsymbol{\xi}^t q^{1/2} \hat{\boldsymbol{\lambda}} + i \hat{\boldsymbol{\lambda}}^t \boldsymbol{\lambda}  \right)\right]^n.
	\end{split}
\end{equation}

\paragraph{Limit $n \rightarrow 0$.} Through a series of Taylor's expansions around $n = 0$, we get the following expressions for the interaction, entropic and energetic potentials:
\begin{equation}
	\begin{split}
		&\mathcal{G}_I = \underset{n \rightarrow 0}{\lim} \frac{G_I}{n} = \underset{n \rightarrow 0}{\lim}\left[-\frac{1}{n} \sum_{kk^\prime} \hat{\rho}_{kk^\prime} \rho_{kk^\prime} - \sum_{k k^\prime} \hat{m}_{kk^\prime} m_{kk^\prime} + \frac{1}{2}\sum_{kk^\prime} (\hat{v} - \hat{q})_{kk^\prime} (v + q)_{kk^\prime} + \frac{1}{2} \sum_{k k'} \hat{q}_{kk^\prime} q_{kk^\prime}\right],\\
		&\mathcal{G}_S = \underset{n \rightarrow 0}{\lim} \frac{G_S}{n} = \underset{n \rightarrow 0}{\lim}\left[\frac{1}{n} \log \int_{\mathbb{R}^K} d\bw^\star P_{w^\star}(\bw^\star) e^{(\bw^\star)^t \hat{\rho} \bw^\star} + \frac{1}{n}\mbox{log}\left( 1 + n \mathcal{I}(\hat{\rho}, \hat{v}, \hat{q}, \hat{m})\right)\right],\\
		& \mathcal{G}_E = \underset{n \rightarrow 0}{\lim} \frac{G_E}{n} = \underset{n \rightarrow 0}{\lim}\left[ \int \frac{d\boldsymbol{\xi}}{\sqrt{(2\pi)^K}} \int dy \int \prod_{k=1}^K \frac{d\nu_k d\hat{\nu}_k}{2\pi} P^{\star}_y \left( y \vert \{ \nu_{k}\}_{k=1}^K,\,\mathbf{A}  \right) e^{-\frac{1}{2} \left(\boldsymbol{\xi} - iq^{-1/2}m^t\hat{\boldsymbol{\nu}} \right)^2 -\frac{1}{2} \hat{\boldsymbol{\nu}}^t \rho \hat{\boldsymbol{\nu}} + i \hat{\boldsymbol{\nu}}^t \boldsymbol{\nu}} \right.\\
		&\hspace{21.5mm}\times \left.\left( 1 + n \mbox{log} \left( \int \prod_{k=1}^K \frac{d \lambda_k d\hat{\lambda}_k}{2\pi} P_y \left( y \vert \{ \lambda_{k}\}_{k=1}^K,\mathbf{A}  \right) e^{-\frac{1}{2} \hat{\boldsymbol{\lambda}}^t v \hat{\boldsymbol{\lambda}} + i \boldsymbol{\xi}^t q^{1/2} \hat{\boldsymbol{\lambda}} + i \hat{\boldsymbol{\lambda}}^t \boldsymbol{\lambda} }  \right) \right) \right]
	\end{split}\label{eq:Taylor_potentials}
\end{equation}
where the function $\mathcal{I}(\hat{\rho}, \hat{v}, \hat{q}, \hat{m})$ is given by:
\begin{equation}
	\mathcal{I} = \frac{\int D \boldsymbol{\xi} \int d\bw^\star P_{w^\star}(\bw^\star) e^{(\bw^\star)^t \hat{\rho} \bw^\star} \mbox{log}\left[ \int d\bw P_w(\bw ) \mbox{exp}\left( -\frac{1}{2}\bw^t \hat{v} \bw + \boldsymbol{\xi}^t \hat{q}^{1/2} \bw +( \bw^\star)^t \hat{m} \bw \right)\right]}{\int d\bw^\star P_{w^\star}(\bw^\star) e^{(\bw^\star)^t \hat{\rho} \bw^\star}}.
\end{equation}
We note that there are two terms $\sim \mathcal{O}(n^{-1})$, one in the interaction and another in the entropic potential. To avoid divergent potentials in the limit $n\rightarrow 0$, we need to require both terms to vanish:
\begin{equation}
	-\frac{1}{n} \sum_{kk^\prime} \hat{\rho}_{kk^\prime} \rho_{kk^\prime} + \frac{1}{n} \mbox{log} \int_{\mathbb{R}^K} d\bw^\star P_{w^\star}(\bw^\star) e^{(\bw^\star)^t \hat{\rho} \bw^\star} = 0
\end{equation}
which is true only if $\hat{\rho} = 0$. This then determines the equation for $\rho$ at the saddle-point for $n \rightarrow 0$:
\begin{equation}
	\label{eq::teacher_teacher_overlap}
	\rho_{kk^\prime} = \frac{\partial \mathcal{G}_S}{\partial \hat{\rho}_{kk^\prime}}\biggm\vert_{\hat{\rho} = 0} = \mathbb{E}_{\bw^\star} \left[w_k^\star w^\star_{k^\prime} \right] .
\end{equation}
Thanks to that, we can simplify the expression for the entropic potential in the zero-replica limit:
\begin{equation}
	\mathcal{G}_S =  \int_{\mathbb{R}^K} D\boldsymbol{\xi} \int_{\mathbb{R}^K} d\bw^\star P_{w^\star}(\bw^\star) \, \mbox{log}\int_{\mathbb{R}^K} d\bw \, P_w(\bw) \, e^{-\frac{1}{2} \bw^t \hat{v} \bw + (\mathbf{w^\star})^t \hat{m} \bw + \boldsymbol{\xi}^t \hat{q}^{1/2} \bw}.
\end{equation}
The expression of the energetic potential in eq.~\eqref{eq:Taylor_potentials} can also be further simplified, by computing the $\hat{\boldsymbol{\nu}}, \hat{\boldsymbol{\lambda}}$ - integrals, which, at this stage, are just $K$-dimensional Gaussian integrals. This leads to the following energetic potential:
\begin{equation}
	\begin{split}
		\mathcal{G}_E = \int dy \int D\boldsymbol{\xi} \int D\boldsymbol{\nu} \ P^\star_{y}(y \vert v_\star^{1/2} \boldsymbol{\nu} + m (q^\star)^{-1/2} \boldsymbol{\xi}, \mathbf{A})\label{eq:replica_symmetric_potential} \ \mbox{log} \int D\boldsymbol{\lambda} \ P_{y}(y \vert v^{1/2} \boldsymbol{\lambda} + q^{1/2} \boldsymbol{\xi}, \mathbf{A}) 
	\end{split}
\end{equation}
where we have defined
\begin{equation}
	v^\star = \rho - q^\star \in \mathbb{R}^{K \times K}, \hspace{5em} q^\star = mq^{-1}m^t
	\label{eq:v_star}.
\end{equation}

\paragraph{Summary.} Combining all the expressions above for the interaction, entropic and energetic potential, we get the final form of the mean free-energy density in the replica-symmetry assumption:
\begin{equation}
	\langle f_{\beta} \rangle = \underset{q_{kk^\prime}, v_{kk^\prime}, m_{kk^\prime}, \hat{q}_{kk^\prime}, \hat{v}_{kk^\prime}, \hat{m}_{kk^\prime}}{\mbox{extr}} \left[ \phi^{(n=0)} \left( \rho_{kk^\prime}, q_{kk^\prime}, v_{kk^\prime}, m_{kk^\prime}, \hat{q}_{kk^\prime}, \hat{v}_{kk^\prime}, \hat{m}_{kk^\prime} \right)\right]\label{eq:final_mean_free_energy}
\end{equation}
with the potential $\phi^{(n=0)}$ being the sum of the following three contributions:
\begin{equation}
	\phi^{(n=0)} = \mathcal{G}_I(q_{kk^\prime}, v_{kk^\prime}, m_{kk^\prime}, \hat{q}_{kk^\prime}, \hat{v}_{kk^\prime}, \hat{m}_{kk^\prime}) + \mathcal{G}_S(\hat{q}_{kk^\prime}, \hat{v}_{kk^\prime}, \hat{m}_{kk^\prime}) + \alpha \mathcal{G}_E(\rho_{kk^\prime}, q_{kk^\prime}, v_{kk^\prime}, m_{kk^\prime})
\end{equation}
that is, the interaction, entropic and energetic contribution in the zero-replica limit:
\begin{equation}
	\begin{split}
		&\mathcal{G}_I = - \sum_{k k^\prime} \hat{m}_{kk^\prime} m_{kk^\prime} + \frac{1}{2}\sum_{kk^\prime} (\hat{v} - \hat{q})_{kk^\prime} (v + q)_{kk^\prime} + \frac{1}{2} \sum_{k k^\prime} \hat{q}_{kk^\prime} q_{kk^\prime},\\
		&\mathcal{G}_S = \int_{\mathbb{R}^K} D\boldsymbol{\xi} \int_{\mathbb{R}^K} d\bw^\star \, P_{w^\star}(\bw^\star) \, \mbox{log} \int_{\mathbb{R}^K} d\bw \, P_w(\bw) \, e^{-\frac{1}{2} \bw^t \hat{v} \bw + (\mathbf{w^\star})^t \hat{m} \bw + \boldsymbol{\xi}^t \hat{q}^{1/2} \bw},\\
		&\mathcal{G}_E = \int dy \int D\boldsymbol{\xi} \int D\boldsymbol{\nu} \ P^\star_{y} \left(y \vert v_\star^{1/2} \boldsymbol{\nu} + m q^{-1/2} \boldsymbol{\xi}, \mathbf{A}\right) \ \mbox{log} \int D\boldsymbol{\lambda} \ P_{y}\left(y \vert v^{1/2} \boldsymbol{\lambda} + q^{1/2} \boldsymbol{\xi}, \mathbf{A}\right) . 
	\end{split}    
	\label{eq:replica_symmetric_potential}  
\end{equation}

\subsection{Teacher and student prior}

The entropic potential in eq.~\eqref{eq:replica_symmetric_potential} holds for any prior of the teacher and the student. In the following, we will restrict the discussion to a Gaussian prior over the teacher 
\begin{equation}
	P_{w^\star}(\bw^\star) = \frac{1}{\sqrt{(2\pi)^K}} e^{-\frac{1}{2} (\bw^\star)^t \bw^\star}.
\end{equation}
This will enforce, because of equation~\eqref{eq::teacher_teacher_overlap}
\begin{equation}
	\label{eq::teacher_self_overlap}
	\rho_{k k'} = \delta_{k k'} .
\end{equation}
as expected for large $N$ by the central limit theorem.  In the following, we also consider the specific case of a 
\emph{Gaussian prior} for the student
\begin{equation}
	P_w(\bw) = e^{-\frac{\beta \lambda}{2} \bw^t \bw} .
\end{equation}
This corresponds to describing the empirical risk minimization in eq.~\eqref{eq: non-convex_opt_problem} with $L_2$-regularization $r(\bw) = \beta \lambda \vert \vert \bw\vert\vert_2^2$ of intensity controlled by the parameter $\beta \lambda$. 
Note that the prior of the teacher and the student are the same if $\beta \lambda = 1$. Equivalently, remind that in the large $N$ and $P$ limit, for each choice of $\lambda$ one selects a certain value of the (squared) norm of the $K$ set of student weights in the first layer
\begin{equation}
	r_{kk} = \frac{1}{N} \sum_{i=1}^N w_{ik}^2\,, \qquad \forall k = 1, \dots, K.
\end{equation}
Because of equation~\eqref{eq::teacher_self_overlap} tells us that the norm of the teacher weights is always 1, having matching teacher and student prior will also impose $r_{kk} = 1$. Notice that the matrix $r$ corresponds to the self-overlap written as $q_{{\rm self}}$ in the main. 

In the Gaussian prior case, the integrals over $\bw$, $\bw^\star$ and $\boldsymbol{\xi}$ in eq.~\eqref{eq:replica_symmetric_potential} are standard $K$-dimensional Gaussian integrals, that solved give the following expression for the entropic potential:

\begin{equation}
	\mathcal{G}_S = \frac{K}{2} \log(2\pi) - \frac{1}{2} \mbox{log}\left[ \mbox{det}\left( \beta \lambda \mathbb{I}_{K\times K} + \hat{v} \right) \right] + \frac{1}{2}\mbox{tr}\left[ \hat{q} \left(\beta \lambda \mathbb{I}_{K\times K} + \hat{v} \right)^{-1} \right] + \frac{1}{2} \mbox{tr} \left[ \hat{m} \left( \beta \lambda \mathbb{I}_{K\times K} + \hat{v} \right)^{-1} \hat{m}^t \right] \label{eq: replica_symmetric_committee_ansatz}.
\end{equation}

\section{The limit of large number of hidden units}
The replica analysis led to a major simplification: it allowed us to transform the original optimization problem in eq.\eqref{eq: non-convex_opt_problem} spanning over an $N\times K$-dimensional space, into the $K \times K$-dimensional extremization problem in eq.~\eqref{eq:final_mean_free_energy}. Solving this problem analytically for an arbitrary $K$ is generally intractable. In this work, we consider the limit where $K \rightarrow \infty$ but slower than $N$, meaning that $K/N \rightarrow 0$ as in classical and recent works on committee machines~\cite{barkai_broken_1992,engel_storage_1992,baldassi_properties_2019, schwarze_discontinuous_1993,schwarze_learning_1993-1,aubin_committee_2018, nishiyama_solution_2025}. In this limit, as suggested in~\cite{schwarze_learning_1993}, it is reasonable to look for solutions of eq.~\eqref{eq:final_mean_free_energy} of the form:
\begin{equation}
	\begin{split}
		o &= o_d \mathbb{I}_{K \times K} + \frac{o_a}{K} \textbf{1}_K \textbf{1}^t_K  \hspace{10mm} \forall o \in \{ q, v, m \},\\
		\hat{o} &= \hat{o}_d \mathbb{I}_{K \times K} + \frac{\hat{o}_a}{K} \textbf{1}_K \textbf{1}^t_K  \hspace{10mm} \forall \hat{o} \in \{ \hat{q}, \hat{v}, \hat{m} \},
	\end{split}   
	\label{eq:committe-symmetric_ansatz} 
\end{equation}
with the $K$-dimensional all-ones column vector $\textbf{1}_K$. Indeed, when the diagonal part $o_d$ is zero, the ansatz describes solutions that are symmetric under permutations of the hidden units. In contrast, when $o_d$ is nonzero, the ansatz captures solutions in which each student hidden unit specializes by correlating with a specific hidden unit of the teacher committee machine. Note that this symmetric-committee ansatz further reduces the dimensionality of the extremization problem in eq.~\eqref{eq:final_mean_free_energy} from $K \times K$ to $2$ for each overlap parameter.

In the following, we show in full detail how to apply this ansatz in the energetic potential. We instead just provide the final expression for the interaction and the entropic potential, since applying the symmetric-committee ansatz in this case is straightforward.

\subsection{Interaction and entropic potential}

The interaction potential is
\begin{equation}
	\begin{split}
		\mathcal{G}_I =& -K\Big(\hat{m}_d \, m_d - \frac{\hat{v}_d(q_d+v_d)}{2} + \frac{\hat{q}_d \, v_d}{2} \Big)\\
		&- \hat{m}_d \, m_a  + \frac{\hat{v}_d(q_a+v_a)}{2} - \frac{\hat{q}_d \, v_a}{2} - \hat{m}_a(m_d+m_a)  + \frac{\hat{v}_a(q_d+q_a+v_d+v_a)}{2} -\frac{\hat{q}_a(v_d+v_a)}{2}.
	\end{split}
\end{equation}
For the entropic potential, neglecting vanishing terms in $K$, one gets
\begin{equation}
	\begin{split}
		\mathcal{G}_S &= \frac{K}{2}\left[\frac{\hat{m}_d^2 + \hat{q}_d}{\beta\lambda + \hat{v}_d} + \log\left(\frac{2\pi}{\beta\lambda + \hat{v}_d}\right)\right] + \frac{1}{2} \log\left(\frac{\beta\lambda + \hat{v}_d}{\beta\lambda + \hat{v}_d+\hat{v}_a}\right) + \frac{\hat q_d + \hat q_a + (\hat m_d + \hat m_a)^2}{2(\beta \lambda + \hat v_d + \hat v_a )} - \frac{\hat q_d + \hat m_d^2}{2(\beta \lambda + \hat v_d)}. 
	\end{split}
\end{equation}

\subsubsection{Integrating the conjugated order parameters}

The interaction and entropic can be directly extremized with respect to the conjugated order parameters contained in the matrices $\hat{m}$, $\hat{q}$, and $\hat{v}$, as they do not appear in the energetic term that still needs to be analyzed. The corresponding extremization is sufficiently easy that can be solved analytically. One gets
\begin{equation}
	\begin{split}
		\mathcal{G}_{SI} &:= \max_{\hat{m},\hat{q},\hat{v}}\Big[\mathcal{G}_I + \mathcal{G}_S\Big]\\
		&= -\frac{K}{2} \left(q_d + v_d + \frac{q_a + v_a}{K}\right) \beta \lambda + \frac{K}{2}\Big[1 + \frac{q_d - m_d^2}{v_d} + \log(2\pi v_d)\Big] - \frac{1}{2} \log\Big(\frac{v_d}{v_d+v_a}\Big)\\
		& + \frac{1}{2(v_d + v_a)}\left[ q_a - 2m_d \, m_a - m_a^2 - \frac{v_a}{v_d}(q_d - m_d^2) \right]. \label{eq:entropic_channel}
	\end{split}
\end{equation}
With a slight abuse of language in the following we will call $\mathcal{G}_{SI}$ simply as the entropic potential. Note that, as expected, the term $q_d + v_d + \frac{q_a + v_a}{K}$ that multiplies the regularization strength $\beta \lambda$ is the squared norm $Q$. 

\subsection{Energetic potential}

We now analyze the large $K$ limit of the energetic potential in eq.\eqref{eq:replica_symmetric_potential}, which, making explicit the expression of the likelihood of the teacher and the student as in eq.\eqref{eq: teacher_label} and \eqref{eq:posterior} respectively, can be written as
\begin{equation}
	\begin{split}
		\mathcal{G}_E &= \int dy \int D\boldsymbol{\xi} \int D\boldsymbol{\nu} \ \int D z \, \delta \left(y - f_\star\left[\frac{1}{\sqrt{K}}\mathbf{A}^\top \sigma(v_\star^{1/2} \boldsymbol{\nu} + m q^{-1/2} \boldsymbol{\xi}) - B_\star\sqrt{K} + \sqrt{\Delta^\star} z \right] \right)  \\
		&\hspace{2em}\times \log \int D\boldsymbol{\lambda}\ \mbox{exp}\left(-\beta \ell\left(y,  f \left[ \frac{1}{\sqrt{K}} \mathbf{A}^\top  \sigma \left(v^{1/2} \boldsymbol{\lambda} + q^{1/2} \boldsymbol{\xi} \right) - B\sqrt{K} \right] \right) \right)  .
	\end{split}\label{eq: Ge_with_explicit_Pout}
\end{equation}
Here, $\sigma(\mathbf{u})$ acts element-wise on each component of $\mathbf{u}$. As a first step, we apply the symmetric-committee ansatz~\eqref{eq: replica_symmetric_committee_ansatz} in the energetic potential. We detail all the steps for the $\boldsymbol{\lambda}$-integral involving the student likelihood. The ansatz can be applied in the same way on the $\boldsymbol{\nu}$-integral involving the different teacher likelihood.   \\    

\paragraph{Student likelihood.} Using the ansatz~\eqref{eq:committe-symmetric_ansatz}, the matrix $v^{1/2}$ can be written as
\begin{equation}
	v^{1/2} = \sqrt{v_d} \mathbb{I}_K + \frac{\sqrt{v_a + v_d} - \sqrt{v_d}}{K} \boldsymbol{1}_K\boldsymbol{1}^t_K.
\end{equation}
The $\boldsymbol{\lambda}$-integral can then be expressed in terms of the diagonal and off-diagonal parts of $v$ as:
\begin{equation}
	I_{\boldsymbol{\lambda}} = \int D\boldsymbol{\lambda} \ \mbox{exp}\left(-\beta \ell\left(y,  f \left[ \frac{1}{\sqrt{K}}\sum_{k=1}^K \, A_{k} \sigma \left( \sqrt{v_d} \lambda_k + \left( \sqrt{v_d + v_a} - \sqrt{v_d} \right) \frac{\boldsymbol{1}^t_K \boldsymbol{\lambda}}{K} + \left(q^{1/2} \boldsymbol{\xi} \right)_k \right) - B\sqrt{K} \right] \right)\right).
\end{equation}
To factorize over the $k$-index, we can introduce the following definitions by means of Dirac $\delta$-functions and their integral representations:
\begin{equation}
	\begin{split}
		1 &\propto \int \frac{d\omega d\hat{\omega}}{2\pi} \mbox{exp}\left( i\hat{\omega} \left( \omega - \frac{\boldsymbol{1}^t_K \boldsymbol{\lambda}}{K} \right) \right) \\
		&\times \int \frac{du d\hat{u}}{2\pi} \mbox{exp}\left( i\hat{u} \left( u - \frac{1}{\sqrt{K}}\sum_{k=1}^K \, A_{k} \sigma \left( \sqrt{v_d} \lambda_k + \left( \sqrt{v_d + v_a} - \sqrt{v_d} \right) \frac{\omega}{\sqrt{K}} + \left(q^{1/2} \boldsymbol{\xi} \right)_k \right) + B\sqrt{K} \right) \right) 
	\end{split}
\end{equation}
leading to the following expression for the $\boldsymbol{\lambda}$-integral:
\begin{equation}
	\begin{split}
		I_{\boldsymbol{\lambda}} &= \int \frac{d\omega d\hat{\omega}}{2\pi} \int \frac{du d\hat{u}}{2\pi} \mbox{exp}\left(-\beta \ell\left(y,  f \left( u \right)\right)+ i\hat{\omega}\omega + i \hat{u}u \right) \\
		&\times\prod_{k=1}^K \int D\lambda_k \ \mbox{exp} \left( -i\hat{\omega} \frac{\lambda_k}{\sqrt{K}} - i \frac{\hat{u}}{\sqrt{K}} \left(A_k \sigma \left(\sqrt{v_d} \lambda_k + \left( \sqrt{v_d + v_a} - \sqrt{v_d} \right) \frac{\omega}{\sqrt{K}} + \left( q^{1/2} \boldsymbol{\xi} \right)_k \right) \right) + i\hat{u}B \sqrt{K} \right) .
	\end{split}\label{eq: lamb_int_v_ansatz}
\end{equation}
Using the ansatz~\eqref{eq: replica_symmetric_committee_ansatz}, the matrix $q^{1/2}$ can be written as
\begin{equation}
	q^{1/2} = \sqrt{q_d} \mathbb{I}_K + \frac{\sqrt{q_a + q_d} - \sqrt{q_d}}{K} \boldsymbol{1}_K\boldsymbol{1}^t_K
\end{equation}
so that each pre-activation $\lambda_k$ is a Gaussian variable of variance $\sqrt{v_d}$ and mean:
\begin{equation}
	\mu_k\left( \omega, \xi \right) = \sqrt{q_d} \xi_k + \frac{1}{\sqrt{K}}\left[ \left( \sqrt{v_d + v_a} - \sqrt{v_d} \right) \omega + \left( \sqrt{q_d + q_a} - \sqrt{q_d} \right) \omega_{\xi} \right] \coloneqq \mu_{k,0} \left( \xi_k\right) + \frac{1}{\sqrt{K}} \mu_1\left( \omega, \omega_{\xi} \right)
	\label{eq: mean_preact}
\end{equation}
where we defined $\omega_{\xi} = \frac{1}{\sqrt{K}} \sum_{k=1}^K \xi_k$. By plugging the definition of $\mu_k$ in eq.~\eqref{eq: lamb_int_v_ansatz} and performing the change of variable $\sqrt{v_d} \lambda_k + \mu_k \rightarrow \sqrt{v_d} \lambda_k$, we then get the following expression for $I_{\boldsymbol{\lambda}}$ under the committee-symmetric ansatz:  
\begin{equation}
	\begin{split}
		I_{\boldsymbol{\lambda}} &= \int \frac{d\omega d\hat{\omega}}{2\pi} \int \frac{du d\hat{u}}{2\pi} \mbox{exp}\left(-\beta \ell\left(y,  f \left( u \right)\right)+ i\hat{\omega}\omega + i \hat{u}u + i\hat{u} B \sqrt{K} \right) \\
		&\times\prod_{k=1}^K \mbox{exp}\left( -\frac{\mu^2_k}{2v_d} + i \frac{\hat{\omega}}{\sqrt{K}}\frac{\mu_k}{\sqrt{v_d}} \right)  \int D\lambda_k \ \mbox{exp} \left( \frac{\mu_k}{\sqrt{v_d}} \lambda_k - i\frac{\hat{\omega}}{\sqrt{K}}\lambda_k - i \frac{\hat{u}}{\sqrt{K}} A_k \sigma \left( \sqrt{v_d} \lambda_k \right) \right). 
	\end{split}\label{eq: lamb_int_q_ansatz}
\end{equation}

To expand $I_{\boldsymbol{\lambda}}$ in the limit of $K \rightarrow \infty$ we can re-write this integral as
\begin{equation}
	\begin{split}
		I_{\boldsymbol{\lambda}} &= \int \frac{d\omega d\hat{\omega}}{2\pi} \int \frac{du d\hat{u}}{2\pi} \mbox{exp}\left(-\beta \ell\left(y,  f \left( u \right)\right)+ i\hat{\omega}\omega + i \hat{u}u + i\hat{u} B \sqrt{K} +\sum_{k=1}^K \mbox{log} J_k\right)
	\end{split}\label{eq: int_lamb_explog_trick}    
\end{equation}
where we have used the standard log-exp trick so that
\begin{equation}
	\begin{split}
		\mbox{log}J_k &= -\frac{\mu_{k,0}^2}{2v_d} - \frac{\mu_{k,0} \mu_1}{v_d\sqrt{K}} + i\frac{\hat{\omega}\mu_{k,0}}{\sqrt{v_dK}} - \frac{\mu_1^2}{2v_dK} + i \frac{\hat{\omega} \mu_1}{\sqrt{v_d}K}+\\
		&+ \mbox{log} \int D\lambda_k \ \mbox{exp}\left( \frac{\mu_{k,0}}{\sqrt{v_d}} \lambda_k + \frac{1}{\sqrt{K}}\left[ \left( \frac{\mu_1}{\sqrt{v_d}} - i\hat{\omega} \right) \lambda_k - i\hat{u} A_k \sigma \left(\sqrt{v_d} \lambda_k \right) \right] \right)
	\end{split}
\end{equation}
and where we have further used the definition of $\mu_k$ in terms of $\mu_{k,0}$ and $\mu_1$ as in eq.~\eqref{eq: mean_preact}. At this point, we can expand $\mbox{log}J_k$ up to order $1/K$. Since the measure of $\lambda_k$ is normalized to $1$ up to order one in $K$, the expansion leads to the following expression:
\begin{equation}
	\mbox{log}J_k \simeq -\frac{\mu_{k,0} \mu_1}{v_d \sqrt{K}} + i \frac{\hat{\omega} \mu_{k,0}}{\sqrt{v_d K}} - \frac{\mu_1^2}{2v_d K} + i \frac{\hat{\omega} \mu_1}{\sqrt{v_d}K} + \frac{t_1}{\sqrt{K}} + \frac{t_2 - t^2_1}{2K}\label{eq: logJk_expansion}
\end{equation}
where we have defined the terms $t_1$ and $t_2$ as:
\begin{equation}
	\begin{split}
		t_1 &= \int D\lambda_k \left[ \left(\frac{\mu_1}{\sqrt{v_d}} - i \hat{\omega} \right)\lambda_k - i \hat{u} \left( A_k \sigma\left( \sqrt{v_d}\lambda_k\right)\right) \right] \mbox{exp} \left( -\frac{\mu_{k,0}^2}{2v_d} + \frac{\mu_{k,0}}{\sqrt{v_d}}\lambda_k\right),\\
		t_2 &= \int D\lambda_k \left[ \left(\frac{\mu_1}{\sqrt{v_d}} - i \hat{\omega} \right)\lambda_k - i \hat{u} \left( A_k \sigma\left( \sqrt{v_d}\lambda_k\right)\right) \right]^2 \mbox{exp} \left( -\frac{\mu_{k,0}^2}{2v_d} + \frac{\mu_{k,0}}{\sqrt{v_d}}\lambda_k\right).
	\end{split}
\end{equation}

We can now perform the integral over $\lambda_k$ in $t_1$ and $t_2$ whose result directly depends on the moments and their first derivative of the activation function with respect to the pre-activations distribution:

\begin{equation}
	\begin{split}
		t_1 &= \left( \frac{\mu_1}{\sqrt{v_d}} - i \hat{\omega} \right) \frac{\mu_{k,0}}{\sqrt{v_d}} - i \hat{u} A_k g_1 \left( \mu_{k,0}, v_d \right),\\
		t_2 &= \left( \frac{\mu_1}{\sqrt{v_d}} - i\hat{\omega} \right)^2 \left( 1 + \frac{\mu_{k,0}^2}{v_d}\right) - \hat{u}^2 A_k^2 g_2 \mu_{k,0}, v_d \\
		&- 2i\hat{u} \left( \frac{\mu_1}{\sqrt{v_d}} - i \hat{\omega}\right) \left( A_k \left( \sqrt{v_d} \Delta_1 \left(\mu_{k,0}, v_d \right) + \frac{\mu_{k,0}}{\sqrt{v_d}}  g_1 \left( \mu_{k,0}, v_d\right)\right) \right) 
	\end{split}
\end{equation}
where the functions $g_1$, $g_2$ and $\Delta_1$ are defined as it follows:
\begin{equation}
	\begin{split}
		g_1\left( \xi_k \sqrt{q_d}, v_d \right) &= \mathbb{E}_{\lambda_k\sim\mathcal{N}(0,1)} \left[\sigma(\xi_k \sqrt{q_d} + \sqrt{v_d}\lambda_k)\right],\\
		g_2\left( \xi_k \sqrt{q_d}, v_d \right) &=\mathbb{E}_{\lambda_k\sim\mathcal{N}(0,1)} \left[\sigma^2(\xi_k \sqrt{q_d} + \sqrt{v_d}\lambda_k)\right],\\
		\Delta_1\left( \xi_k \sqrt{q_d}, v_d \right) &= \mathbb{E}_{\lambda_k\sim\mathcal{N}(0,1)} \left[\sigma'(\xi_k \sqrt{q_d} + \sqrt{v_d}\lambda_k)\right].
	\end{split}
\end{equation}
We can replace these expressions for $t_1$ and $t_2$ back into eq.\eqref{eq: logJk_expansion} and massaging the final result with a bit of algebra and summing over $k=1,...,K$, we finally get: 
\begin{equation}
	\begin{split}
		\sum_{k=1}^K \mbox{log} J_k &\simeq -\frac{\hat{w}^2}{2} - i \hat{u}  \frac{1}{\sqrt{K}} \sum_{k=1}^K A_k g_1 \left( \mu_{k,0}, v_d \right) -\frac{\hat{u}^2}{2}  \frac{1}{K} \sum_{k=1}^K A_k^2 \left( g_2 \left( \mu_{k,0}, v_d\right) - g_1^2\left( \mu_{k,0}, v_d\right)\right)\\ 
		&- i\hat{u} \left( \mu_1 - i \hat{\omega}\sqrt{v_d} \right) \frac{1}{K} \sum_{k=1}^K A_k \Delta_1\left( \mu_{k,0}, v_d\right).
	\end{split}\label{eq: sum_logJk_expansion}
\end{equation}
If we now define the following auxiliary functions:
\begin{equation}
	\begin{split}
		G_1 &= \frac{1}{\sqrt{K}} \sum_{k=1}^K A_k g_1 \left( \mu_{k,0}, v_d \right),\hspace{10mm}
		G_2 = \frac{1}{K} \sum_{k=1}^K A_k^2 g_2 \left( \mu_{k,0}, v_d\right),\\ 
		G_{12} &= \frac{1}{K} \sum_{k=1}^K A_k^2 g_1^2\left( \mu_{k,0}, v_d\right),\hspace{10mm}
		D_1 = \frac{1}{K} \sum_{k=1}^K A_k \Delta_1\left( \mu_{k,0}, v_d\right),
	\end{split}
\end{equation}
we can then write $I_{\boldsymbol{\lambda}}$ as:
\begin{equation}
	\begin{split}
		I_{\boldsymbol{\lambda}} &= \int \frac{d\omega d\hat{\omega}}{2\pi}\int \frac{du d\hat{u}}{2\pi} \mbox{exp}\left(-\beta \ell \left(y, f(u) \right) \right)\\
		&\times \mbox{exp}\left( -\frac{\hat{\omega}^2}{2} + i \hat{\omega}\omega - \frac{\hat{u}^2}{2} \left( G_2 - G_{1,2}\right) + i \hat{u} \left(u - G_1 + \sqrt{K}B \right) - i\hat{u} \left( \mu_1 - i \hat{\omega}\sqrt{v_d} \right)D_1 \right).
	\end{split}
\end{equation}
By replacing the definition of $\mu_1$ as in eq.\eqref{eq: mean_preact} and then integrating over the Gaussian integrals in $\omega$ and $\hat{\omega}$, we get:
\begin{equation}
	I_{\boldsymbol{\lambda}} = \int \frac{du d\hat{u}}{2\pi} \mbox{exp}\left(-\beta \ell \left(y, f(u) \right) - \left( G_2 - G_{12} + v_a D^2_1\right) \frac{\hat{u}^2}{2} + i \left( u - \left( G_1 + \sqrt{K}B\right) - \left(\sqrt{q_a + q_d} - \sqrt{q_d} \right) \omega_{\xi} D_1\right) \hat{u}\right).
\end{equation}
At his point, we realize that the integral over $\hat{u}$ is also Gaussian and can be easily solved, leading to:
\begin{equation}
	I_{\boldsymbol{\lambda}} = \int \frac{du}{\sqrt{2\pi \sigma^2}} \mbox{exp}\left(-\frac{1}{2} \frac{\left(u - \mu\right)^2}{\sigma^2} -\beta \ell \left(y , f(u) \right)\right) \coloneqq \mathcal{G}\left(y, \omega_{\xi}, G_1, G_2, G_{12}, D_1 \right),
\end{equation}
where the mean and the variance of the second-layer pre-activation $u$ are given by:
\begin{equation}
	\begin{split}
		\mu &= G_1 + \sqrt{K}B + \left(\sqrt{q_a + q_d} - \sqrt{q_d} \right) \omega_{\xi} D_1,\\
		\sigma^2 &= G_2 - G_{12} + v_a D_1^2.
	\end{split}
\end{equation}
Depending on the choice of the loss function $\ell$, the integral over $u$ acquires different shapes. For instance, in the case of square loss, it turns into a simple Gaussian integral.\\

\paragraph{Teacher likelihood.} All the steps outlined in the paragraph above can be repeated in the same way for the $\boldsymbol{\nu}$-integral in eq.~\eqref{eq: Ge_with_explicit_Pout}, with the only expectation that the teacher likelihood is now a Dirac-$\delta$ centered on the teacher's labels. With this in mind, we then get:
\begin{equation}
	\begin{split}
		I_{\boldsymbol{\nu}} &= \int D \boldsymbol{\nu} \int D z \, \delta \left(y - f_\star\left[\frac{1}{\sqrt{K}}\sum_{k=1}^K \, A_{k} \sigma(v_\star^{1/2} \boldsymbol{\nu} +  (q^\star)^{-1/2} \boldsymbol{\xi}) - B_\star\sqrt{K} + \sqrt{\Delta^\star} z \right] \right)  \\    
		&= \int \frac{du_\star}{\sqrt{2\pi \sigma_\star}} \delta \left( y - f_\star \left( u_\star\right)\right) \mbox{exp}\left( -\frac{1}{2} \frac{\left( u_\star - \mu_\star\right)^2}{\sigma_\star^2}\right) \coloneqq \mathcal{G}_{\star} \left( y, \omega_\xi, G_1^\star, G_2^\star, G_{12}^\star, D_1^\star\right)
	\end{split}
\end{equation}
where, precisely as in the student case, we have that the mean and the variance of the pre-activations $u_\star$ of the second layer are:
\begin{equation}
	\begin{split}
		\mu_\star &= G^\star_1 + \sqrt{K}B^\star + \left(\sqrt{q^\star_a + q^\star_d} - \sqrt{q^\star_d} \right) \omega_{\xi} D^\star_1,\\
		\sigma^2_\star &= G^\star_2 - G^\star_{12} + v^\star_a (D_1^\star)^2 + \Delta^\star,
	\end{split}
\end{equation}
with the teacher auxiliary functions $G^\star_1$, $G^\star_2$, $G^\star_{12}$ and $\Delta^\star_1$ defined as:
\begin{equation}
	\begin{split}
		G^\star_1 &= \frac{1}{\sqrt{K}} \sum_{k=1}^K A_k g^\star_1 \left( \mu^\star_{k,0}, v^\star_d \right),\hspace{10mm}
		G^\star_2 = \frac{1}{K} \sum_{k=1}^K A^2_k g^\star_2 \left( \mu^\star_{k,0}, v^\star_d\right),\\ 
		G^\star_{12} &= \frac{1}{K} \sum_{k=1}^K A^2_k (g^\star_1\left( \mu^\star_{k,0}, v^\star_d\right))^2,\hspace{10mm}
		D^\star_1 = \frac{1}{K} \sum_{k=1}^K A_k \Delta^\star_1\left( \mu^\star_{k,0}, v^\star_d\right),
	\end{split}
\end{equation}
and the functions $g^\star_1$, $g^\star_2$ and $\Delta^\star_1$ given by: 
\begin{equation}
	\begin{split}
		g^\star_1\left( \sqrt{q^\star_d} \xi_k, v^\star_d \right) &= \mathbb{E}_{\nu_k\sim\mathcal{N}(0,1)} \left[\sigma(\xi_k \sqrt{q_d^\star} + \sqrt{v_d^\star}\nu_k)\right],\\
		g^\star_2\left(\sqrt{q^\star_d} \xi_k, v^\star_d \right) &=\mathbb{E}_{\nu_k\sim\mathcal{N}(0,1)} \left[\sigma^2(\xi_k \sqrt{q_d^\star} + \sqrt{v_d^\star}\nu_k)\right],\\
		\Delta^\star_1\left(\sqrt{q^\star_d} \xi_k, v^\star_d \right) &=\mathbb{E}_{\nu_k\sim\mathcal{N}(0,1)} \left[\sigma'(\xi_k \sqrt{q_d^\star} + \sqrt{v_d^\star}\nu_k)\right],
	\end{split}
\end{equation}
with $\mu^\star_{k,0} = \sqrt{q^\star_d} \xi_k$ and $v^\star_d$ and $q^\star_d$ the diagonal part of the matrices $q_\star^{1/2}$ and $v_\star^{1/2}$:
\begin{equation}
	\begin{split}
		q_\star^{1/2} &= \sqrt{q^\star_d} \mathbb{I}_K + \frac{\sqrt{q^\star_a + q^\star_d} - \sqrt{q^\star_d}}{K} \boldsymbol{1}_K\boldsymbol{1}^t_K,\\
		v_\star^{1/2} &= \sqrt{v^\star_d} \mathbb{I}_K + \frac{\sqrt{v^\star_a + v^\star_d} - \sqrt{v^\star_d}}{K} \boldsymbol{1}_K\boldsymbol{1}^t_K.
	\end{split}
\end{equation}

\paragraph{Integration over $\xi$.}
The integration over $\boldsymbol{\nu}$ and $\boldsymbol{\lambda}$, the pre-activations of the teacher and student network respectively, leads to expressing the energetic potential $\mathcal{G}_E$ in eq.~\eqref{eq: Ge_with_explicit_Pout} as:
\begin{equation}
	\mathcal{G}_E = \int dy \int D\boldsymbol{\xi} \ \mathcal{G}^\star(y, \omega_\xi, G_1^\star, G_2^\star, G^\star_{12}, D_1^\star) \mbox{log} \ \mathcal{G}(y, \omega_{\xi}, G_1, G_2, G_{12}, D_1).
\end{equation}
To integrate over $\xi$ we introduce the definition of $G_1, G_2, G_{12}, D_1$ and $G^\star_1, G^\star_2, G^\star_{12}, D^\star_1$ by means of Dirac-$\delta$s and their integral representations. In this way, the energetic potential $\mathcal{G}_E$ becomes:
\begin{equation}
	\begin{split}
		\mathcal{G}_E &= \int dy \int \frac{d\omega_\xi d\hat{\omega}_\xi}{\sqrt{2\pi}}\int \frac{dG^\star_1 d\hat{G}^\star_1 dG_1 d\hat{G}_1}{\sqrt{2 \pi}} \int \frac{dG^\star_2 d\hat{G}^\star_2 dG_2 d\hat{G}_2}{\sqrt{2 \pi}} \int \frac{dG^\star_{12} d\hat{G}^\star_{12} dG_{12} d\hat{G}_{12}}{\sqrt{2 \pi}} \int \frac{dD^\star_1 d\hat{D}^\star_1 dD_1 d\hat{D}_1}{\sqrt{2 \pi}} \\
		&\times \mathcal{J} \left(y, \omega_\xi, G_1, G_2, G_{12}, D_1, G^\star_1, G^\star_2, G^\star_{12}, D^\star_1 \right) I_{\boldsymbol{\xi}} \left(G_1, G_2, G_{12}, D_1, G^\star_1, G^\star_2, G^\star_{12}, D^\star_1 \right)\label{eq: Ge_before_int_xi}
	\end{split}
\end{equation}
where the functions $\mathcal{J}$ and $I_{\boldsymbol{\xi}}$ are:
\begin{equation}
	\begin{split}
		\mathcal{J} &= e^{i\hat{\omega}_{\xi}\omega_{\xi}+i \hat{G}^\star_1 G^\star_1 + i \hat{G}^\star_2 G^\star_2 + i \hat{G}^\star_{12} G^\star_{12} + i \hat{D}^\star_1 D^\star_1 +i \hat{G}_1 G_1 + i \hat{G}_2 G_2 + i \hat{G}_{12} G_{12} + i \hat{D}_1 D_1}\\
		&\hspace{60mm}\times \mathcal{G}^\star (y, \omega_\xi, G^\star_1, G^\star_2, G^\star_{12}, D^\star_1) \mbox{log} \mathcal{G} (y, \omega_\xi, G_1, G_2, G_{12}, D_1),\\
		I_{\boldsymbol{\xi}} &= \int D\boldsymbol{\xi} \ e^{-i\frac{\hat{\omega}_\xi}{\sqrt{K}}\sum_{k=1}^K \xi_k -i \frac{\hat{G}^\star_1}{\sqrt{K}} \sum_{k=1}^K A_k g^\star_1  - i \frac{\hat{G}^\star_2}{K} \sum_{k=1}^K A_k^2 g^\star_2 - i \frac{\hat{G}^\star_{12}}{K} \sum_{k=1}^K A_k^2 (g^\star_1)^2 - i \frac{\hat{D}^\star_1}{K} \sum_{k=1}^K A_k \Delta^\star_1  }\\
		& \hspace{60mm}\times e^{-i \frac{\hat{G}_1}{\sqrt{K}} \sum_{k=1}^K A_k g_1  - i \frac{\hat{G}_2}{K} \sum_{k=1}^K A_k^2 g_2 - i \frac{\hat{G}_{12}}{K} \sum_{k=1}^K A_k^2 g_1^2 - i \frac{\hat{D}_1}{K} \sum_{k=1}^K A_k \Delta_1  }.
	\end{split} 
\end{equation}

To solve $I_{\xi}$, we first notice that it factorizes over the index $k$. Because of that, we can use the exp-log trick and write it as:
\begin{equation}    I_{\boldsymbol{\xi}} = \mbox{exp} \left( \sum_{k=1}^K \mbox{log}\int D\xi_k \mbox{exp}\left( \frac{t_1}{\sqrt{K}} + \frac{t_2}{K}\right)\right) = \mbox{exp}\left(\sum_{k=1}^K \mbox{log}J_k \right)\label{eq:int_xi}
\end{equation}
where we have isolated the terms of order $K^{-1/2}$ from those of order $K^{-1}$ with:
\begin{equation}
	\begin{split}
		t_1 &= - i \hat{\omega}_\xi \xi_k - i \hat{G}_1^\star A_k g_1^\star - i \hat{G}_1 A_k g_1,\\
		t_2 &= -i\hat{G}_2^\star A_k^2 g_2^\star + i \hat{G}_{12}^\star A_k^2 (g_1^\star)^2 - i \hat{D}_1^\star A_k \Delta_1^\star - i \hat{G}_2 A_k^2g_2 + i \hat{G}_{12} A_k^2 g_1^2 - i \hat{D}_1 A_k \Delta_1.
	\end{split}
\end{equation}
Expanding the logarithm up to order $K^{-1}$ we get:
\begin{equation}
	\mbox{log}J_k \simeq \frac{\langle t_1 \rangle_\xi}{\sqrt{K}} + \frac{\langle t_2 \rangle_\xi}{K} + \frac{\langle t_1^2 \rangle_\xi - \langle t_1 \rangle^2_\xi}{2K} ,
\end{equation}
where $\langle \cdot \rangle_{\xi}$ is the expectation over the Gaussian variable $\xi$. We can then compute the expectation over $\xi$, thus getting:
\begin{equation}
	\begin{split}
		\langle t_1 \rangle_\xi &= -i\hat{G}_1^\star A_k \langle g_1^\star \rangle_\xi -i\hat{G}_1 A_k \langle g_1^\star \rangle_\xi, \\
		\langle t_2 \rangle_\xi &= -i\hat{G}_2^\star A_k^2 \langle g_2^\star \rangle_\xi +i\hat{G}_{12}^\star A_k^2 \langle (g_1^\star)^2  \rangle_{\xi} - i  \hat{D}_1^\star A_k \langle \Delta_1^\star \rangle_\xi - i\hat{G}_2 A_k^2 \langle g_2 \rangle_\xi +i\hat{G}_{12} A_k^2 \langle (g_1)^2  \rangle_\xi - i  \hat{D}_1 A_k \langle \Delta_1 \rangle_\xi,\\
		\langle t_1^2 \rangle_\xi &= - \hat{\omega}_\xi^2 - (\hat{G}_1^\star)^2 A_k^2 \langle (g_1^\star)^2\rangle_\xi - \hat{G}_1^2 A_k^2 \langle g_1^2\rangle_\xi - 2 \hat{G}_1^\star \hat{G}_1 A^2_k \langle g_1^\star g_1\rangle_\xi - 2\hat{\omega}_\xi \hat{G}_1^\star A_k \langle g_1^\star \xi_k\rangle_\xi - 2\hat{\omega}_\xi \hat{G}_1 A_k \langle g_1 \xi_k\rangle_\xi,\\
		\langle t_1\rangle^2_{\xi} &= (\hat{G}_1^\star)^2 A_k^2 \langle g_1^\star \rangle^2_\xi + \hat{G}_1^2 A_k^2 \langle g_1\rangle^2_\xi + 2 \hat{G}_1^\star \hat{G}_1 A_k^2 \langle g_1^\star\rangle_\xi \langle g_1\rangle_\xi.
	\end{split}
\end{equation}

We can now plug the result of this expansion back into eq.~\eqref{eq:int_xi} to get the final expression for $I_{\boldsymbol{\xi}}$ and then plug $I_{\boldsymbol{\xi}}$ back into eq.~\eqref{eq: Ge_before_int_xi}. This gives the final expression for the energetic potential after the integration over $\boldsymbol{\xi}$:
\begin{equation}
	\begin{split}
		\mathcal{G}_E &= \int dy \int Dz \int \frac{d\omega_\xi d\hat{\omega}_\xi}{\sqrt{2\pi}}\int \frac{dG^\star_1 d\hat{G}^\star_1 dG_1 d\hat{G}_1}{\sqrt{2 \pi}} \int \frac{dG^\star_2 d\hat{G}^\star_2 dG_2 d\hat{G}_2}{\sqrt{2 \pi}} \int \frac{dG^\star_{12} d\hat{G}^\star_{12} dG_{12} d\hat{G}_{12}}{\sqrt{2 \pi}} \int \frac{dD^\star_1 d\hat{D}^\star_1 dD_1 d\hat{D}_1}{\sqrt{2 \pi}} \\
		&\hspace{20mm}\times \mbox{exp}\left(\mathcal{W}(G^\star_1, G^\star_2, G^\star_{12}, D^\star_1, \hat{G}^\star_1, \hat{G}^\star_2, \hat{G}^\star_{12}, \hat{D}^\star_1, G_1, G_2, G_{12}, D_1, \hat{G}_1, \hat{G}_2, \hat{G}_{12}, \hat{D}_1)\right) \\
		&\hspace{50mm}\times\mathcal{G}^\star (y, \omega_\xi, G^\star_1, G^\star_2, G^\star_{12}, D^\star_1, z) \mbox{log} \mathcal{G} (y, \omega_\xi, G_1, G_2, G_{12}, D_1)\label{eq: Ge_after_int_xi}
	\end{split}
\end{equation}
where we defined the function $\mathcal{W}$ as:
\begin{equation}
	\begin{split}
		\mathcal{W} &= - \frac{\hat{\omega}^2_\xi}{2} - \frac{(\hat{G}_1^\star)^2}{2}\nu_A\left( \langle (g_1^\star)^2\rangle_\xi - \langle g_1^\star\rangle_\xi^2 \right) - \frac{\hat{G}_1^2}{2}\nu_A\left( \langle g_1^2\rangle_\xi - \langle g_1\rangle_\xi^2 \right) - \hat{G}_1^\star \hat{G}_1 \nu_A(\langle g_1^\star g_1\rangle_\xi - \langle g_1^\star \rangle_\xi \langle g_1\rangle_\xi ) \\
		& - \hat{G}_1^\star \mu_A (\hat{\omega}_\xi \sqrt{q_d^\star} \langle \Delta^\star_1 \rangle_\xi + i \sqrt{K} \langle g_1^\star\rangle_\xi)- \hat{G}_1 \mu_A(\hat{\omega}_\xi \sqrt{q_d} \langle \Delta_1 \rangle_\xi + i \sqrt{K} \langle g_1\rangle_\xi)\\
		& -i\hat{G}_2^\star \nu_A \langle g_2^\star\rangle +i\hat{G}_{12}^\star \nu_A\langle (g_1^\star)^2\rangle -i\hat{G}_2 \nu_A \langle g_2^\star\rangle +i\hat{G}_{12}\nu_A\langle g_1^2\rangle  -i\hat{D}_1^\star \mu_A\langle \Delta_1^\star\rangle_\xi  -i\hat{D}_1\mu_A \langle \Delta_1\rangle_\xi\\
		&+i\hat{\omega}_{\xi}\omega_{\xi}+i \hat{G}^\star_1 G^\star_1 + i \hat{G}^\star_2 G^\star_2 + i \hat{G}^\star_{12} G^\star_{12} + i \hat{D}^\star_1 D^\star_1 +i \hat{G}_1 G_1 + i \hat{G}_2 G_2 + i \hat{G}_{12} G_{12} + i \hat{D}_1 D_1
	\end{split}
\end{equation}
with $\mu_A = \langle A_k \rangle_{P_A(\mathbf{A})}$ and $\nu_A = \langle A^2_k \rangle_{P_A(\mathbf{A})}$, since we expect, in the large $K$ limit, that the empirical mean and variance of the second-layer weights, that is $ K^{-1}\sum_{k=1}^K A_k$ and $K^{-1}\sum_{k=1}^K A^2_k$, converge to the first and second moment of the second-layer weights distribution $P_A(\mathbf{A})$. In order to integrate over the hat-variables, we realize that some of these integrals are of the form:
\begin{equation}
	I_{\hat{x}} = \int \frac{d\hat{x}}{\sqrt{2\pi}} \mbox{exp}(i \hat{x} (x - \langle c \rangle_{\xi})) = \delta(x - \langle c \rangle_{\xi}).
\end{equation}
Because of that, these integrals can be directly solved by simply setting $x = \langle c \rangle_{\xi} = \bar{x}$. This is the case for the hat-variables $\hat{D}_1, \ \hat{D}_1^\star, \ \hat{G}_2, \ \hat{G}^\star_2, \ \hat{G}_{12}, \ \hat{G}^\star_{12}$ and, consequently, $D_1 = \mu_A\langle \Delta_1 \rangle_\xi = \mu_A\bar{D}_1, \ D^\star_1 = \mu_A\langle \Delta^\star_1 \rangle_\xi = \mu_A \bar{D}_1^\star, \ G_2 = \nu_A\langle g_2 \rangle_\xi =  \nu_A\bar{G}_2, \ G^\star_2 = \nu_A\langle g^\star_2 \rangle_\xi = \nu_A \bar{G}^\star_2,\ G_{12} =  \nu_A\langle g_1^2\rangle = \nu_A\bar{G}_{12}, \ \hat{G}^\star_{12} = \nu_A\langle (g_1^\star)^2\rangle = \nu_A \bar{G}^\star_{12}$. This means that the fluctuations of these random variables with respect to the randomness induced by $\xi$ can be neglected when we expand the energetic potential up to order $K^{-1}$. In this way, $\mathcal{G}_E$ in eq.~\eqref{eq: Ge_after_int_xi} consistently simplifies as:
\begin{equation}
	\begin{split}
		\mathcal{G}_E &= \int dy \int Dz \int \frac{d\omega_\xi d\hat{\omega}_\xi}{\sqrt{2\pi}}\int \frac{dG^\star_1 d\hat{G}^\star_1}{\sqrt{2 \pi}} \int \frac{dG_1 d\hat{G}_1}{\sqrt{2\pi}} \mbox{exp}\left(\mathcal{W}(G^\star_1, G_1; \bar{G}_{12}^\star, \bar{G}_{12}, \bar{G}_1^\star, \bar{G}_1, \bar{G}_{\star1}\right) \\ 
		&\hspace{70mm}\times \mathcal{G}^\star (y, \omega_\xi, G^\star_1; \bar{G}_2^\star, \bar{G}_{12}^\star, \bar{D}_1^\star) \mbox{log} \mathcal{G} (y, \omega_\xi, G_1; \bar{G}_2, \bar{G}_{12}, \bar{D}_1)   
	\end{split}    
\end{equation}
where the function $\mathcal{W}$ is given by:
\begin{equation}
	\begin{split}
		\mathcal{W} &= -\frac{\hat{\omega}_\xi^2}{2} - \frac{(\hat{G}_1^\star)^2}{2} \nu_A (\bar{G}_{12}^\star - (\bar{G}_1^\star)^2) - \frac{\hat{G}^2_1}{2} \nu_A(\bar{G}_{12} - \bar{G}_1^2) - \hat{G}_1^\star \hat{G}_1 \nu_A (\bar{G}_1^\star - \bar{G}_1^\star \bar{G}_1)-\hat{G}_1^\star \mu_A (\hat{\omega}_{\xi} \sqrt{q_d^\star} \bar{D}_1^\star + i\sqrt{K}\bar{G}_1^\star) \\
		&\hspace{70mm}-\hat{G}_1 \mu_A (\hat{\omega}_{\xi} \sqrt{q_d} \bar{D}_1 + i\sqrt{K}\bar{G}_1) + i\hat{\omega}_\xi \omega_\xi + i\hat{G}_1 G_1 + i\hat{G}_1^\star G_1^\star.
	\end{split}
\end{equation}
We have set $\bar{G}_1^\star =  \langle g_1^\star\rangle_\xi$, $\bar{G}_1 = \langle g_1\rangle_\xi$ and $\bar{G}_{\star1} = \langle g_1^\star g_1\rangle_\xi$. At this point, performing the change of variables $G_1^\star -\mu_A\sqrt{K}\bar{G}_1^\star \rightarrow G_1^\star $ and $G_1 -\mu_A\sqrt{K}\bar{G}_1 \rightarrow G_1$ and exploiting the identity:
\begin{equation}
	\int_{\mathbb{R}^n} \frac{d\bx d\boldsymbol{\hat{x}}}{(2\pi)^d} \mbox{exp}\left( -\frac{1}{2}\hat{\bx}^t \Sigma \hat{\bx} + i \bx^t \hat{\bx} \right) f(\bx) = \langle f(\bx)\rangle_{\bx}
\end{equation}
with $\bx = (\omega_\xi, G_1, G_1^\star)$ and $d=3$, we notice that $\mathcal{G}_E$ can be finally written as an expectation over the jointly Gaussian random variables $\omega_\xi, G_1$ and $G_1^\star$ as:
\begin{equation}
	\mathcal{G}_E = \biggl<  \int dy \int Dz \ \mathcal{G}^\star (y, \omega_\xi, z, G^\star_1 +\mu_A \sqrt{K} \bar{G}_1^\star; \bar{G}_2^\star, \bar{G}_{12}^\star, \bar{D}_1^\star) \ \mbox{log}\ \mathcal{G} (y, \omega_\xi, G_1 + \mu_A\sqrt{K} \bar{G}_1; \bar{G}_2, \bar{G}_{12}, \bar{D}_1) \biggl>_{\omega_\xi, G_1^\star, G_1}\label{eq:final_finally_Ge}
\end{equation}
with zero mean and covariance matrix:
\begin{equation}
	\Sigma := \begin{pmatrix}
		\sigma_{\omega_\xi}^2& \sigma^2_{\omega_\xi G_1} & \sigma^2_{\omega_\xi G^\star_1}\\
		\sigma^2_{ \omega_\xi G_1}  & \sigma_{G_1}^2 & \sigma^2_{G_1^\star G_1 }\\
		\sigma^2_{ \omega_\xi 
			G_1^\star} & \sigma^2_{G_1^\star G_1} &
		\sigma^2_{G_1^\star}
	\end{pmatrix}\label{eq: Sigma_cov}    
\end{equation}
whose elements are: 
\begin{equation}
	\begin{aligned}[c]
		\sigma_{\omega_\xi}^2 &= 1, \\
		\sigma_{\omega_\xi G_1}^2 &= \mu_A\sqrt{q_d}\bar{D}_1,\\
		\sigma_{\omega_\xi G_1^\star }^2 &= \mu_A\sqrt{q_d}\bar{D}^\star_1,\\
	\end{aligned}
	\hspace{5em}
	\begin{aligned}[c]
		\sigma_{G_1}^2 &= \nu_A(\bar{G}_{12} - \bar{G}_1^2),\\
		\sigma_{G_1^\star G_1}^2 &= \nu_A(\bar{G}_{\star1} - \bar{G}_1^\star \bar{G}_1) ,\\
		\sigma_{G_1^\star}^2 &= \nu_A(\bar{G}^\star_{12} - (\bar{G}_1^\star)^2). 
	\end{aligned}
\end{equation}  
Note that, after the change of variables in $G_1^\star$ and $G_1$, the mean and the variance of the teacher and student second-layer pre-activation $u^\star$ and $u$ are given by:
\begin{equation}
	\begin{aligned}
		\mu &= G_1 + \sqrt{K}\mu_A\bar{G}_1 + \sqrt{K}B + \left(\sqrt{q_a + q_d} - \sqrt{q_d} \right) \mu_A \bar{D}_1 \omega_{\xi},\\ 
		\mu_\star &= G^\star_1 + \sqrt{K} \mu_A\bar{G}^\star_1 + \sqrt{K}B^\star + \left(\sqrt{q^\star_a + q^\star_d} - \sqrt{q^\star_d} \right) \mu_A \bar{D}^\star_1 \omega_{\xi},
	\end{aligned}
	\hspace{4em}
	\begin{aligned}
		\sigma^2 &= \nu_A(\bar{G}_2 - \bar{G}_{12}) + v_a \mu_A^2 \bar{D}_1^2 ,\\
		\sigma^2_\star &= \nu_A(\bar{G}^\star_2 - \bar{G}^\star_{12}) + v^\star_a \mu_A^2 (\bar{D}_1^\star)^2 + \Delta^\star .
	\end{aligned}
\end{equation}
The bias terms $B$ and $B^\star$ must then be chosen to cancel the $K \rightarrow \infty$ divergence induced by $\sqrt{K}\bar{G}_1$ and $\sqrt{K}\bar{G}^\star_1$ terms. At the leading order $K$, we should then fix $B^\star \sim \bar{G}^\star_1$ and $B \sim \bar{G}_1$, which is consistent with the definition in \eqref{eq:bias}.\\

\paragraph{Kernel equivalence.} The energetic potential in eq.~\eqref{eq:final_finally_Ge} can be written in terms of the \emph{Kernel} function:
\begin{equation}
	\mathcal{K}(d_1,d_2,a): = \mathbb{E}_{(x_1\; x_2)\sim \mathcal{N}(0,\Omega)} [\sigma(x_1)\sigma(x_2)]  \ \ \text{with} \ \ \Omega = \begin{pmatrix}
		d_1 &   a   \\
		a   &   d_2
	\end{pmatrix},
\end{equation}
which, making the expectation over $x_1$ and $x_2$ explicit, acquires the shape:
\begin{equation*}
	\mathcal{K}(d_1,d_2,a) =\int Dx_1 Dx_2 \, \sigma\left(\sqrt{d_1}x_1\right)\sigma\left(\frac{a}{\sqrt{d_1}}x_1 + \sqrt{\frac{d_1d_2 - a^2}{d_1}}x_2 \right) .
\end{equation*}
This Kernel function is called Neural Network Gaussian Process (NNGP) and describes the covariance of the function implemented by a neural network at initialization (i.e., with random weights) in the infinite-width limit, evaluated at two different inputs~\cite{Neal1996, Williams1996}. 

Indeed, by applying a linear transformation to the Gaussian random variables $\nu_k$, $\lambda_k$ and $\xi_k$, we can rewrite the auxiliary functions as:
\begin{equation}
	\begin{split}
		\Bar{G}_1 &= \int Dx \, \sigma\left(x \sqrt{q_d + v_d } \right) = \sqrt{\mathcal{K}(q_d+v_d,q_d+v_d,0)} ,\\
		\overline{G}_{2} &= \int Dx \, \sigma^2\left( \sqrt{q_d + v_d} x\right) =\mathcal{K}(q_d+v_d,q_d+v_d,q_d+v_d),\\
		\overline{G}_{12} &= \int Dx \left[ \int Dy \, \sigma\left(\sqrt{q_d} x + \sqrt{v_d} y \right) \right]^2 =\mathcal{K}(q_d+v_d,q_d+v_d,q_d),\\
		\overline{D}_1 &= \int Dx \, \sigma' \left(x\sqrt{q_d + v_d}  \right)  = (q_d+v_d)\sqrt{\partial_a\mathcal{K}(q_d+v_d,q_d+v_d,a)\vert_{a=0}}   ,     
	\end{split}
\end{equation}
the same holds for the teacher auxiliary functions by mapping $q\mapsto q^\star$ and $v\mapsto v^\star$. The term corresponding to the correlation between teacher and student neurons is then:
\begin{equation}
	\overline{G}_{^\star1} = \int Dx \sigma(x\sqrt{q_d^\star+ v_d^\star}) \int Dy \, \sigma\left( \sqrt{\frac{q_dq_d^\star}{q_d^\star+v_d^\star}} x + \sqrt{q_d+v_d - \frac{(q_d^\star q_d)^2}{q_d^\star+v_d^\star}} y \right) =\mathcal{K}(q_d^\star+v_d^\star,q_d+v_d,\sqrt{q_dq_d^\star}).
\end{equation}
Taking into account the definition of $q^\star$ and $v^\star$ in eq.~\eqref{eq:v_star}, we have $v_d^\star + q_d^\star = \rho_d = 1$, and similarly $q_d+v_d =r_d$. We can then rewrite the elements of the covariance matrix $\Sigma$ in eq.~\eqref{eq: Sigma_cov} in terms of the Kernel function as:
\begin{equation}
	\begin{aligned}[c]
		\sigma_{\omega_\xi}^2 &= 1,\\
		\sigma_{\omega_\xi G_1}^2 &=  \sqrt{r_d^2\partial_a\mathcal{K}(r_d,r_d,a)\vert_{a=0}},\\
		\sigma_{\omega_\xi G_1^\star}^2 &= \sqrt{\partial_a\mathcal{K}(1,1,a)\vert_{a=0}},\\
	\end{aligned}
	\hspace{5em}
	\begin{aligned}[c]
		\sigma_{G_1}^2 &= \left(\mathcal{K}(r_d,r_d,q_d) -\mathcal{K}(r_d,r_d,0)\right),\\
		\sigma_{G_1^\star G_1}^2 &= \left(\mathcal{K}(1,r_d,m_d) - \mathcal{K}(1,q_d,0)\right),\\
		\sigma^2_{G_1^\star}&= \left(\mathcal{K}(1,1,1)-\mathcal{K}(1,1,0)\right).         
	\end{aligned}
\end{equation}

\subsubsection{Regression and $L_2$ loss}
The regression case can be obtained by setting $f^\star(\cdot) = f(\cdot) =\cdot$, and with the MSE loss $\ell(y,x) = \frac{1}{2}(y - x)^2$, we obtain that the energetic potential is reduced to:
\begin{equation}
	\mathcal{G}_E = -\frac{L_y + \sigma_\star^2 + \Delta^\star}{2(\beta^{-1}+\sigma^2)} - \frac{1}{2}\log(\beta^{-1} + \sigma^2) + \frac{1}{2}\log(2\pi) \ \ \text{with} \ \ L_y = \langle (\mu - \mu_\star)^2 \rangle_{\omega_\xi, G_1^\star, G_1}.
\end{equation}
Note that $(\mu-\mu_\star)^2$ is a degree-two polynomial in $\omega_\xi$, $G_1^\star$ and $G_1$, so its expectation is a linear combination of the elements of the covariance matrix $\Sigma$.

\color{black}
\subsubsection{Classification case}
The classification case can be obtained from the previous formulas by setting $f_\star(\cdot) = \text{sign}(\cdot) = f(\cdot)$. We also consider a loss that is only depending on the product of the label and the preactivation of the output i.e. $\ell(y, \hat y) = \ell(y \hat y)$.

After some simplifications the energetic term can be written in terms of two Gaussian integrals only as 
\begin{multline}
	\mathcal{G}_{E}=2\int Dx\,H\left(-\frac{\left(D + m_a V_\star^{2} \right) \sqrt{\eta}}{\sqrt{\gamma \left( \eta (\Sigma - V_\star^2) - (D - m_d V_\star^2)^2 \right) - \left( \eta V_\star - (D - m_d V V_\star) (\Delta V + m_a V) \right)^2 }} x \right) \\
	\times \ln\int Dh \, e^{-\beta\ell\left( \sqrt{\Delta_1 + v_a V_{\star}^{2}}h+\sqrt{ \Delta_0 +q_a V_{\star}^{2}}x \right)}
\end{multline}
where $H(x) \equiv \frac{1}{2} \mathrm{Erfc}\left( \frac{x}{\sqrt{2}} \right) = \int_x^\infty Dy$. We have defined the additional terms as
\begin{subequations}
	\begin{align}
		\eta &\equiv \mathcal{K}(r_d, r_d, q_d) - \mathcal{K}(r_d, r_d, 0)  + \left(q_a - (m_a + m_d)^2 \right) V^2 ,\\
		\gamma &\equiv \mathcal{K}(Q, Q, q_d) - \mathcal{K}(Q, Q, 0)  + q_a V^2 , \\
		\Sigma &=  \mathcal{K}_{1}(1) - \mathcal{K}_1(0) + \Delta^\star ,\\
		V_\star &= \sqrt{\left. \partial_a \mathcal{K}_1(1, 1, a) \right|_{a=0}}, \\
		V &= \sqrt{r_d^2\left. \partial_a \mathcal{K}(r_d, r_d, a) \right|_{a=0}} ,\\
		\Delta_0 &=  \mathcal{K}(Q, Q, q_d) -\mathcal{K}(Q, Q, 0), \\
		\Delta_1 &= \mathcal{K}(Q, Q, Q) -\mathcal{K}(Q, Q, q_d), \\
		D &= \mathcal{K}(1, r_d, m_d) - \mathcal{K}(1, r_d, 0).
	\end{align}
\end{subequations}
We have written the previous expressions in the case $a_k = a_k^\star = 1$ for simplicity. In the Bayes optimal case where the squared norm is $r_d=1$, the Nishimori conditions $q_d = m_d$, $q_a = m_a$, $r_a = 0$ hold, and we have $V=V_\star$. 
The energetic term can be then simplified as follows
\begin{equation}
	\mathcal{G}_{E}=2\int Dx\,H\left(-\frac{D+ q_a V^{2}}{\sqrt{\Sigma(\Delta_0 + q_a V^{2})-(D+q_a V^{2})^{2}}}x \right) \ln\int Dh \, e^{-\beta\ell\left( \sqrt{\Sigma - q_a V^{2}}h+\sqrt{\tilde \Delta+q_a V^{2}}x \right)}.
\end{equation}
Note that this expression reduces to the ones found by Schwarze~\cite{schwarze_learning_1993} in the case of sign activation function case, zero label noise $\Delta^\star$ and theta loss $\ell(x) = \Theta(-x)$ in the infinite $\beta$ limit. Similarly, this expression reduces to the one reported in a recent paper~\cite{nishiyama_solution_2025} studying the storage problem, where the labels are extracted completely random. This corresponds to the limit $\Delta^\star \to \infty$ of our expressions, where in addition has $m_a = m_d = 0$.

\section{The zero temperature limit}

The free-energy in the $K \gg 1$ limit is then given by:
\begin{equation}
	K \langle \Phi_{\beta} \rangle = \underset{q_d, q_a, v_d, v_a, m_d, m_a}{\mbox{extr}} \left[  \mathcal{G}_{SI} (q_d, q_a, v_d, v_a, m_d, m_a) + \mathcal{G}_E (q_d, q_a, v_d, v_a, m_d, m_a) \right]
\end{equation}
with $\mathcal{G}_{SI}$ and $\mathcal{G}_{E}$ defined, respectively, in eq.~\eqref{eq:entropic_channel} and eq.~\eqref{eq:final_finally_Ge}. In the zero-temperature limit, that is:
\begin{equation}
	K \langle \Phi_{\beta} \rangle = \underset{\beta \rightarrow \infty} {\mbox{lim}}\frac{1}{\beta}\underset{q_d, q_a, v_d, v_a, m_d, m_a}{\mbox{extr}} \left[  \mathcal{G}_{SI} (q_d, q_a, v_d, v_a, m_d, m_a) + \mathcal{G}_E (q_d, q_a, v_d, v_a, m_d, m_a) \right]\label{eq:free_energy_zero_t}
\end{equation}
both the entropic-interaction and the energetic potential need to scale at least proportionally with $\beta$ in order to avoid the free-energy diverging at zero temperature. This leads the order parameters to scale with $\beta$ as:
\begin{equation}
	q_{d/a} = O(1), \hspace{10mm} v_{d/a} = \beta \tilde{v}_{d/a},  \hspace{10mm} m_{d/a} = O(1). 
\end{equation}
By replacing this scaling in eq.~\eqref{eq:free_energy_zero_t}, we finally get the free energy in the zero temperature limit. From this quantity we can then compute all the observables of interest, such as the generalization error as in the next section. Indeed, the generalization error will depend on the fixed points of the corresponding saddle-point equations.

\section{Generalization error}
We detail here the computation of the Gibbs error, and how this is related to the order parameters in our system. Recall that we wish to estimate $\bw^\star$ from the observations $y^\mu = \varphi^\star_{\boldsymbol{a}^\star}(\bw^\star\bx^\mu,z^\mu\sqrt{\Delta^\star})$ with $\mu\in[P]$. Thus the goal is to characterize the mean-squared generalization error $\epsilon_g$, for a test sample $(\bx^{\rm new},\,y^{\rm new})$. As mentioned in the main, at finite temperature the Gibbs error is the equivalent to the generalization error and it's defined as 
\begin{equation}
	\epsilon_g = \frac{1}{4^\ell} \mathbb{E}\big[\big\langle \big(y^{\rm}  - \hat{y}_{\bw}(\bx^{\rm new}) \big)^2 \big\rangle\big]\label{equ:generalization_error}
\end{equation}
where $\hat{y}_{\bw}(\bx^{\rm new}) = \varphi_{\boldsymbol{a}}(\bw\bx, 0)$ is the prediction label,  with $\ell=0$ for regression and $\ell=1$ for classification.  $\mathbb{E}[\cdot]$ denotes the expectation value over the quenched variables: the outer weights $\boldsymbol{a}^\star$ and $\boldsymbol{a}$; $\bw^\star$ and  $z^{\rm new}$. Wherea $\langle \cdot\rangle$ denotes the Gibbs average. To light the notation we will  assume that there's no Bias term in $\varphi$  and we will refer to $z^{\rm new}$ as simply $z$, which is sampled from a standard gaussian and similar for $\bx^{\rm new}$. We additionally encoded the $1/\sqrt{N}$ of the first layer pre-activation in the gaussian distribution  of $\bx$
Then we have
\begin{align}
	\epsilon_g &\propto \mathbb{E}\big[\big\langle\big(y^{\rm new} - \hat{y}_{\bw }(\bx) \big)^2 \big\rangle\big]\\
	&= \mathbb{E}\Bigg[\Bigg\langle\Bigg(f^\star\Big(\frac{1}{\sqrt{K^\star}}\boldsymbol{a}^\star\cdot\sigma(\bw^\star\bx) + z\sqrt{\Delta^\star}\Big) - f\Big(\frac{1}{\sqrt{K}}\boldsymbol{a}\cdot\sigma(\bw\bx)\Big)\Bigg)^2 \Bigg\rangle\Bigg]\\
	&\overset{(a)}{=} \mathbb{E}\Bigg\langle\Bigg[ \int\frac{\mathrm{d}\mathbf{T}\mathrm{d}\hat{\mathbf{T}}}{(2\pi)^{K^\star}}\frac{\mathrm{d}\mathbf{Z}\mathrm{d}\hat{\mathbf{Z}}}{(2\pi)^K}\exp\left(i\hat{\mathbf{T}}(\mathbf{T}-\bw^\star \bx)^\top + i\hat{\mathbf{Z}}(\mathbf{Z}-\bw \bx)^\top\right)\notag\\
	&\hspace{15em} \cdot\Bigg(f^\star\Big( \frac{1}{\sqrt{K^\star}}\boldsymbol{a}^\star\cdot\sigma(\mathbf{T}) + z\sqrt{\Delta^\star} \Big)  -f\Big(\frac{1}{\sqrt{K}}\boldsymbol{a}\cdot\sigma(\mathbf{Z})\Big) \Bigg)^2\Bigg] \Bigg\rangle\\
	&=  \int\frac{\mathrm{d}\mathbf{T}\mathrm{d}\hat{\mathbf{T}}}{(2\pi)^{K^\star}}\frac{\mathrm{d}\mathbf{Z}\mathrm{d}\hat{\mathbf{Z}}}{(2\pi)^K} \exp\left(i\hat{\bf T}\mathbf{T}^\top + i \hat{\mathbf{Z}}\mathbf{Z}^\top\right)
	\mathbb{E}_{\bw^\star,\,\bx}\Bigg\langle\Bigg[\exp\left(-i\hat{\mathbf{T}}(\bw^\star \bx)^\top - i\hat{\mathbf{Z}}(\bw \bx)^\top\right)\Bigg]\Bigg\rangle\notag\\
	&\hspace{15em}\cdot\mathbb{E}_{\boldsymbol{a}^\star,\boldsymbol{a},z}\Bigg(f^\star\Big( \frac{1}{\sqrt{K^\star}}\boldsymbol{a}^\star\cdot\sigma(\mathbf{T}) + z\sqrt{\Delta^\star}
	\Big)  -f\Big(\frac{1}{\sqrt{K}}\boldsymbol{a}\cdot\sigma(\mathbf{Z})\Big) \Bigg)^2\Bigg] \\
	&\overset{(b)}{=}  \int\frac{\mathrm{d}\mathbf{T}\mathrm{d}\hat{\mathbf{T}}}{(2\pi)^{K^\star}}\frac{\mathrm{d}\mathbf{Z}\mathrm{d}\hat{\mathbf{Z}}}{(2\pi)^K} \exp\left(i\hat{\bf T}\mathbf{T}^\top + i \hat{\mathbf{Z}}\mathbf{Z}^\top\right)
	\mathbb{E}_{\bw^\star}\Bigg\langle\Bigg[\exp\left(-\frac{1}{2}\hat{\mathbf{T}}\frac{\bw^\star(\bw^\star)^\top}{N}\mathbf{T}^\top - \frac{1}{2}\hat{\mathbf{Z}}\frac{\bw\bw^\top}{N}\mathbf{Z}^\top - \hat{\mathbf{T}}\frac{\bw^\star\bw^\top}{N}\hat{\mathbf{Z}} \right)\Bigg]\Bigg\rangle\notag\\
	&\hspace{15em}\cdot\mathbb{E}_{\boldsymbol{a}^\star,\boldsymbol{a},z}\Bigg(f^\star\Big( \frac{1}{\sqrt{K^\star}}\boldsymbol{a}^\star\cdot\sigma(\mathbf{T}) + z\sqrt{\Delta^\star}
	\Big)  -f\Big(\frac{1}{\sqrt{K}}\boldsymbol{a}\cdot\sigma(\mathbf{Z})\Big) \Bigg)^2\Bigg] \\
	&\overset{(c)}{=} \int_{\mathbb{R}^{K^\star}\times \mathbb{R}^K}\mathrm{d}\mathbf{T}\mathrm{d}\mathbf{Z}\,\mathcal{N}\left(\mathbf{T},\,\mathbf{Z}|\mathbf{0}, Q\right)\,\mathbb{E}_{\boldsymbol{a}^\star,\boldsymbol{a},z}\Bigg(f^\star\Big( \frac{1}{\sqrt{K^\star}}\boldsymbol{a}^\star\cdot\sigma(\mathbf{T}) + z\sqrt{\Delta^\star}
	\Big)  -f\Big(\frac{1}{\sqrt{K}}\boldsymbol{a}\cdot\sigma(\mathbf{Z})\Big) \Bigg)^2\Bigg] . 
\end{align}
In (a), we introduce Dirac delta functions to facilitate the computation of the expectation value over the input $\bx$. In (b), by taking the expectation with respect to the Gibbs measure and the teacher weight priors, the overlap matrices concentrate around the stationary points of the replica equations. As previously discussed, the equilibrium solution dominates the Gibbs measure, describing the most probable configurations. In what follows, we keep the overlap matrices generics, but we should evaluate them at their stationary values to obtain the corresponding generalization error. In (c), after integrating with respect to $\hat{\mathbf{T}}$ and $\hat{\mathbf{Z}}$, we obtain the the gaussian p.d.f. $\mathcal{N}(\mathbf{T},\mathbf{Z}|\mathbf{0},Q)$ for the $K^\star-$dimensional vector $\mathbf{T}$ and the $K-$dimensional vector $\mathbf{Z}$, with covariance matrix
\begin{equation}
	Q := \begin{pmatrix}
		\rho & m\\
		m & r
	\end{pmatrix}.
\end{equation}
Here $\rho$ is the $K^\star\times K^\star$ covariance matrix of the teacher weights, $m$ the $K^\star\times K$ matrix teacher-student overlap matrix and $r$ $K\times K$ the self-student overlap. For simplicity, in our setting we set $K^\star=K$, $f^\star=f$ and $\rho=\mathbf{I}_K$; then given the RS ansatz $r$ and $m$ are parametrized as in \eqref{eq: replica_symmetric_committee_ansatz}. 

Now including the Bias terms in our equations we obtain:
\begin{equation}
	\epsilon_g = \frac{1}{4^\ell} \int_{\mathbb{R}^{K}\times \mathbb{R}^K}\mathrm{d}\mathbf{T}\mathrm{d}\mathbf{Z}\,\mathcal{N}\left(\mathbf{0}, Q\right)\,\mathbb{E}_{\boldsymbol{a}^\star,\boldsymbol{a},z}\Bigg[\Bigg(f\Big( \frac{1}{\sqrt{K}}\boldsymbol{a}^\star\cdot\sigma(\mathbf{T})  - \sqrt{K^\star}B^\star+ z\sqrt{\Delta^\star}
	\Big)  - f\Big(\frac{1}{\sqrt{K}}\boldsymbol{a}\cdot\sigma(\mathbf{Z})- \sqrt{K}B \Big)  \Bigg)^2\Bigg].
\end{equation}
We aim to compute expectation value over the preactivatios of the second layer, and again by using Dirac-deltas we obtain:
\begin{equation}
	\begin{split}
		\epsilon_g&\propto\int\frac{\mathrm{d}\lambda\mathrm{d}\hat{\lambda}}{2\pi}\frac{\mathrm{d}\nu\mathrm{d}\hat{\nu}}{2\pi}\exp\left(i\hat{\nu}\nu + i\hat{\lambda}\lambda\right)\left(f^\star(\nu)-f(\lambda)\right)^2\\
		&\hspace{1em}\cdot \mathbb{E}_{\mathbf{T},\mathbf{Z},\boldsymbol{a}^\star,\boldsymbol{a},z}\,\left[\exp\left(-i\hat{\nu}\sum_{k^\star}\frac{(a_{k^\star}^\star\sigma(T_{k^\star})  - B^\star + z\sqrt{\Delta^\star}/\sqrt{K^\star})}{\sqrt{K^\star}} \right)\exp\left( -i\hat{\lambda} \sum_k\frac{(a_k\sigma(Z_k) -  B)}{\sqrt{K}}\right)\right].
	\end{split}
\end{equation}
To compute the expectation of the previous equation and for the last time, we make use of dirac-deltas for each site of the student and teacher neurons, by introducing:
\begin{equation}
	\begin{split}
		\chi_k&= \frac{a_k\sigma(Z_k) - B}{\sqrt{K}},\\
		\chi_{k}^\star&= \frac{a_{k}^\star \sigma(T_{k}) - B^\star + z\sqrt{\Delta^\star}/\sqrt{K}}{\sqrt{K}}.
	\end{split}
\end{equation}
After a lengthy computation involving (i) an expansion in powers of $1/\sqrt{K}$ of the exponential (ii) taking the expectation over the random variables, (iii) performing a ``Taylor contraction", and (iv) integrating over $\hat{\lambda}$ and $\hat{\nu}$, we obtain:
\begin{equation}
	\epsilon_g = \frac{1}{4^\ell} \int\mathrm{d}\lambda\mathrm{d}\nu\mathcal{N} \left(\lambda,\nu|\boldsymbol{\mu},\Omega_{\nu,\lambda}\right)\left(f^\star(\nu )-f(\lambda)\right)^2 \ \ \text{where} \ \ \Omega_{\nu,\lambda} := \begin{pmatrix}
		\varepsilon_T + \Delta^\star& \varepsilon_C\\
		\varepsilon_C &\varepsilon_S
	\end{pmatrix},
\end{equation}
\begin{equation}
	\label{eq:testError_defs}
	\begin{split} 
		\varepsilon_T &:=\nu_A g_1(1) + (K-1)\mu_A^2 \mathcal{K}(1, 1, 0) - K(B^\star)^2  , \\ 
		\varepsilon_S &:= \nu_A g_1(r_0 ) + (K-1)\mu_A^2 \mathcal{K}(r_0,r_0 ,r_1) - KB^2 , \\
		\varepsilon_C &:= \nu_A g_2(1,r_0, m_0) + (K-1)\mu_A^2 \mathcal{K}(1,r_0 ,m_1)  - K B^\star B ,
	\end{split}
\end{equation}
where $\mu_a$ and $\nu_a$ are the first and second moments of $P_a$; and  $m_0 = m_d + m_a/K$, $m_1 = m_a/K$, and the same for $r$ with the respective parameters. Additionally,  we defined the auxiliary Gaussian integrals: 
\begin{equation*}
	g_1(d) := \mathbb{E}_{x\sim \mathcal{N}(0,d)}[\sigma(x)^2]=\mathcal{K}(d,d,d), \ \ g_2(d_1,d_2,a) =: \mathbb{E}_{(x_1\; x_2)\sim \mathcal{N}(0,\Omega)} [\sigma(x_1)\sigma(x_2)] = \mathcal{K}(d_1,d_2,a) \ \ \text{with} \ \ \Omega = \begin{pmatrix}
		d_1 &   a   \\
		a   &   d_2
	\end{pmatrix}.
\end{equation*}
We can further verify that all elements of the covariance matrix $\Omega_{\nu,\lambda}$ are of order $O_K(1)$ by performing a Taylor expansion of the Gaussian probability density function as follows:
\begin{equation*}
	\mathcal{N}\left(x_1,\,x_2 \Big| 0 , \begin{pmatrix}
		d_1 & a/K\\
		a/K & d_2
	\end{pmatrix}\right) =  \mathcal{N}\left(x_1,\,x_2 \Big| 0 , \begin{pmatrix}
		d_1 & 0\\
		0 & d_2
	\end{pmatrix}\right) \left(1 - \frac{1}{K}\frac{ax_1x_2}{d_1d_2}\right) + \mathcal{O}(K^{-2})
\end{equation*}
and by using the definition of the bias:
\begin{equation*}
	\begin{split}
		B &= \mu_A \sqrt{\mathcal{K}\left(r_d + \nicefrac{r_a}{K},r_d + \nicefrac{r_a}{K},  0\right)} = \mu_A \sqrt{\mathcal{K}(r_d,r_d,0)} + O(K^{-1}),\\
		B^\star &= \mu_A \sqrt{\mathcal{K}(1,1,0)}.
	\end{split}
\end{equation*}
We obtain that the elements of the covariance matrix can be all writen in terms of the NNGP kernel $\mathcal{K}$ as:
\begin{equation}
	\label{eq:testError_defs2}
	\begin{split} 
		\varepsilon_T &:=\nu_A\mathcal{K}(1,1,1)  - \mu_A^2 \mathcal{K}(1, 1, 0)  , \\ 
		\varepsilon_S &:= \nu_A\mathcal{K}(r_d, r_d, r_d) - \mu_A^2 \mathcal{K}(r_d, r_d, 0) - r_a\mu_A^2  \partial_a\mathcal{K}(r_d,r_d,a)|_{a=0} , \\
		\varepsilon_C &:= \nu_A\mathcal{K}(1,r_d, m_d) - \mu_A^2 \mathcal{K}(1,r_d ,m_d)  - m_a \mu_A^2 \partial_a\mathcal{K}(1,r_d,a)|_{a=0}.
	\end{split}
\end{equation}

We now make a change of variables that allows us to replace the $2\times2$ covariance matrix $\Omega_{\nu,\lambda}$ with an identity covariance, i.e. we send $(\lambda, \nu) \to \sqrt{\Omega} \, (\lambda, \nu)$. After some simplifications one gets
\begin{equation}
	\begin{split}
		\epsilon_g &= \frac{1}{4^l} \int D\nu D \lambda \, \left[ f^\star\left(\sqrt{\varepsilon_T+ \Delta^\star} \nu \right) - f\left( \frac{\varepsilon_C}{\sqrt{\varepsilon_T+ \Delta^\star}} \nu + \sqrt{\frac{(\varepsilon_T+ \Delta^\star) \, \varepsilon_S - \varepsilon_C^2}{\varepsilon_T+ \Delta^\star}} \lambda \right)\right]^2 .
	\end{split}
\end{equation}
In regression and classification the generalization error can be explicitly computed as a function of the entries of $\Omega$.\\

\paragraph{In regression.} We use $l=0$ and $f(x)=x$, so we obtain
\begin{equation}
	\epsilon_g = \left(\sqrt{\varepsilon_T+ \Delta^\star} - \frac{\varepsilon_C}{\sqrt{\varepsilon_T+ \Delta^\star}}\right)^2 + \varepsilon_S - \frac{\varepsilon_C^2}{\varepsilon_T+ \Delta^\star} = \varepsilon_T + \varepsilon_S - 2 \varepsilon_C+ \Delta^\star .
\end{equation}
Since the generalization error includes an additive contribution from the label noise variance $\Delta^\star$, as shown above, we report in Figure 5 on the main, the generalization error minus $\Delta^\star$. This allows us to isolate the excess error due to learning and is standard practice in Bayesian regression and teacher-student models with known additive noise. We did not update the axis label to reflect this subtraction, but the plotted quantity corresponds to $\epsilon_g - \Delta^\star$.\\

\paragraph{In classification.} We use $l=1$ and $f(x)=\text{sign}(x)$, so we obtain
\begin{equation}
	\epsilon_g = \frac{1}{\pi} \arccos\left( \frac{\varepsilon_C}{\sqrt{(\varepsilon_T+\Delta^\star) \varepsilon_S }} \right)
\end{equation}
having used the identity $2\int_0^\infty Dx \, H\left( \frac{R x}{\sqrt{1-R^2}} \right) = \frac{1}{\pi} \arccos R$. 
\color{black}

\section{Saddle point equations}

Having written compactly the free entropy in terms of entropic $\mathcal{G}_{SI}$ and energetic $\mathcal{G}_E$ in the large $K$ limit, the next step is to solve the corresponding saddle point equations for the order parameter $o_p\in\{ m_d,m_a,q_d,q_a,v_d,v_a\}$. They are of the form
\begin{equation}
	\partial_{o_p}\Big[ \mathcal{G}_{SI}(m,q,v) + \alpha \mathcal{G}_E(m,q,v)  \Big] \overset{!}{=} 0.
\end{equation}
Those can be easily solved numerically. However the equations and their solutions crucially depends on how $\alpha = P/N$ scales with $K$. For each regime one needs to match the leading order of the derivatives of $\mathcal{G}_P$ with those of $\alpha \mathcal{G}_E$.

\subsection{Small data regime $\alpha = P/N = O(1)$}

\begin{figure}
	\centering
	\includegraphics[width=0.7\linewidth]{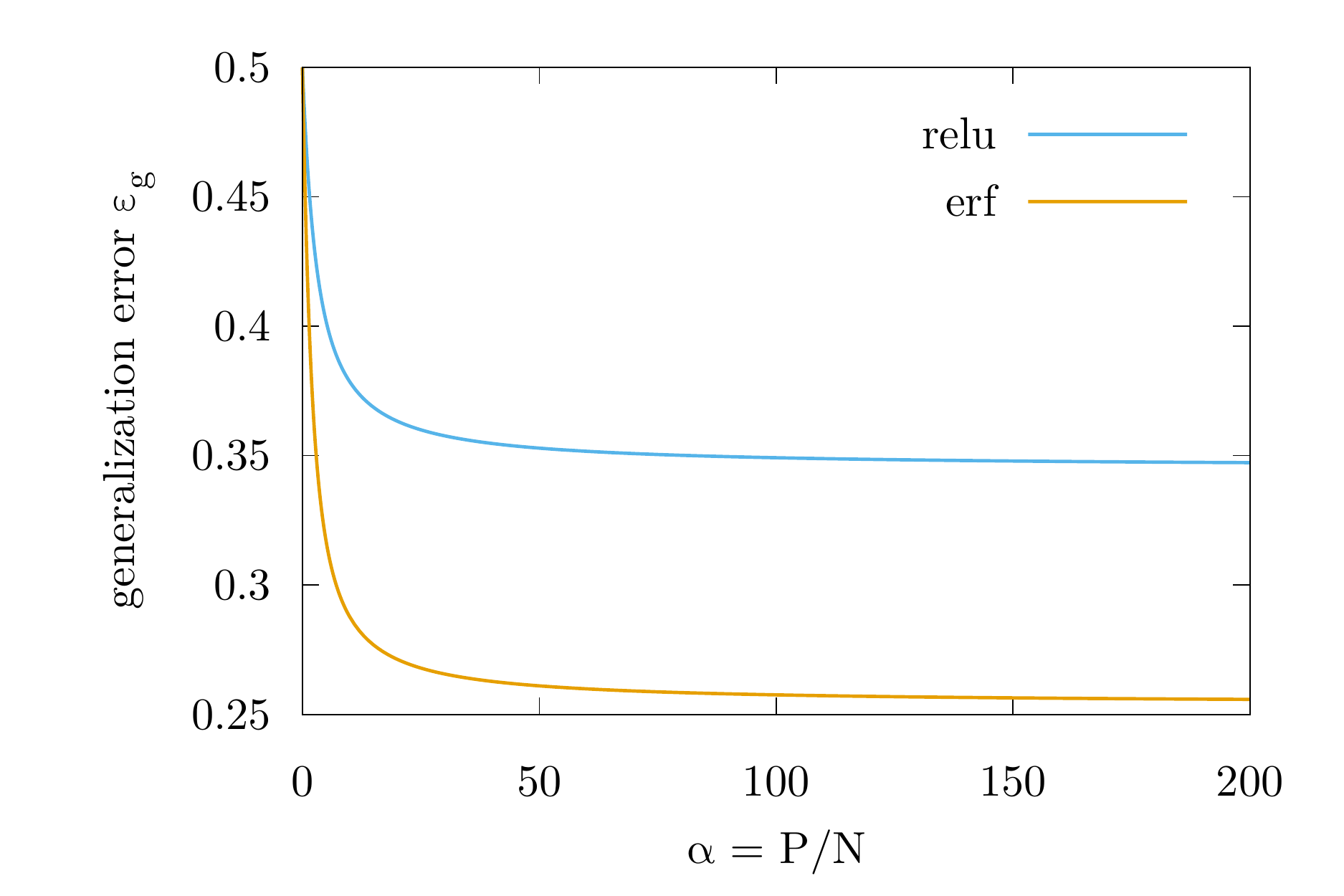}
	\caption{Generalization error in the small data regime (i.e. $\alpha = P/N = O(1)$) found by solving equations~\eqref{eq::BO_low_alpha_regime} for the ReLU $\sigma(x) = \max(0, x)$ and Erf $\sigma(x) = \text{Erf}(x)$ activation functions  and for the number of error loss function $\ell(x) = \Theta(-x)$ in the large $\beta$ limit. In this regime the unique saddle point is given by the \emph{non-specialized branch}. Here the two lines refer to the Bayes Optimal case case where the squared norm of the student $Q=1$ (or, equivalently the $L_2$ regularization is fixed as $\beta \lambda = 1$). For large $\alpha$ the generalization error tends to the generalization error found in the large data regime $\tilde \alpha = \frac{P}{NK}$ when $\tilde \alpha \to 0$. This means that in the Bayes Optimal case the transition to the specialized solution is continuous and it happens as soon as size of the training set $P$ becomes of the order of the number of parameter $NK$ learned.}
	\label{fig:classification_small_alpha}
\end{figure}

In the small data regime, where $\alpha = P /N = O(1)$ the leading order of $\partial_{o_p}[\alpha\mathcal{G}_E]$ is of order one. From the saddle point equations of the diagonal parameters, one finds
\begin{subequations}
	\begin{align}
		q_d &= m_d = 0 ,\\
		v_d &= \frac{1}{\beta \lambda},
	\end{align}
\end{subequations}
meaning that in this regime the only possibility is to be in the \emph{PS branch}, i.e. as defined in the main text, the student exhibits global permutation symmetry (PS) of the hidden units and at the same time all student's hidden units are correlated with the teacher's ones in the same way (i.e. there is no \emph{specialization}). The other saddle-point equations for $m_a$, $q_a$, and $v_a$ can be numerically solved. We show in Fig.~\ref{fig:classification_small_alpha} the plot of the generalization error for the ReLU and Erf activation functions in this regime.

\subsection{Large data regime $\tilde \alpha = \frac{P}{NK} = O(1)$}





In the regime where $\alpha = O(K)$, we naturally define $\Tilde{\alpha} = \alpha/K$, such that $\Tilde{\alpha}$ remains of order one. The equation \eqref{eq:entropic_channel} for $\mathcal{G}_P$ and its derivatives with respect to the order parameters imply that $v_d + v_a \sim \chi / K$, where $\chi$ remains finite as $K \to \infty$. This ensures that the derivatives of $\mathcal{G}_P$ and $\alpha \mathcal{G}_E$ with respect to the off-diagonal parameters match the leading order in $K$, yielding a closed form for the corresponding saddle point equations. Heuristically, we found that the parametrization in terms of $\chi$ is convenient to solve the saddle point numerically for large $K$, increasing the stability to the initialization for iterative methods e.g. Newton's method.

\subsection{Bayes optimal case}

In the Bayes optimal case the saddle point equations are particularly simply to write in both the small and large data regimes described above. Indeed, due to matching of the teacher and student prior and likelihood, one has the following \emph{Nishimori conditions}~\cite{Nishimori_1980}
\begin{subequations}
	\begin{align}
		m_d &= q_d, \\
		m_a &= q_a, \\
		r_a &= v_a + q_a = 0, \\
		r_d &= v_d + q_d ,=\rho_d = 1.
	\end{align}
\end{subequations}
Using those relations, the entropic factor greatly simplifies
\begin{equation}
	\begin{split}
		\mathcal{G}_{SI}
		&= \frac{K}{2}\Big[1 + \log(2\pi) + q_d + \log(1-q_d)\Big] - \frac{1}{2} \log\Big(\frac{1 - q_d}{1 - q_d - q_a}\Big) + \frac{q_a}{2} .\label{eq:entropic_channel_BO}
	\end{split}
\end{equation}
The saddle point equation read
\begin{subequations}
	\begin{align}
		\label{eq::BO_firstSP}
		\frac{q_d + q_a}{2 (1- q_d - q_a)} &= K \tilde \alpha \frac{\partial \mathcal{G}_E}{\partial q_a} ,\\
		\frac{1}{2} \left( \frac{K q_d}{1-q_d} + \frac{q_a}{(1-q_d)(1-q_d-q_a)} \right) &= K \tilde \alpha \frac{\partial \mathcal{G}_E}{\partial q_d}.
		\label{eq::BO_secondSP}
	\end{align}
\end{subequations}

In the small data regime $\tilde \alpha = \frac{\alpha}{K}$ with $\alpha = O(1)$, so one obtains that the only way to not have a divergent term in the second saddle point equation~\eqref{eq::BO_secondSP}, $q_d = 0$. The saddle points are therefore given by the two equations
\begin{subequations}
	\label{eq::BO_low_alpha_regime}
	\begin{align}
		\frac{q_a}{2 (1 - q_a)} &= \alpha \frac{\partial \mathcal{G}_E}{\partial q_a} ,\\
		q_d = 0.
	\end{align}
\end{subequations}

In the large data regime instead one has to impose, in order to have matching scalings in $K$ that $q_a + q_d = 1 - \frac{\chi}{K}$. The two saddle point equation therefore read
\begin{subequations}
	\begin{align}
		\frac{q_d + q_a}{2 \chi } &\simeq \frac{1}{2 \chi} =  \tilde \alpha \frac{\partial \mathcal{G}_E}{\partial q_a}, \\
		\frac{1}{2} \left( \frac{q_d}{1-q_d} + \frac{q_a}{(1-q_d) \chi} \right) &\simeq \frac{1}{2} \left( \frac{q_d}{1-q_d} + \frac{1}{\chi} \right) = \tilde \alpha \frac{\partial \mathcal{G}_E}{\partial q_d}.
	\end{align}
\end{subequations}
Since when one substitutes $q_a = 1 - q_d - \frac{\chi}{K}$ in the energetic term, $\mathcal{G}_E$ at first order in $K$ does not depend on $\chi$, one has that the dependence on $\chi$ can be removed:
\begin{subequations}
	\begin{align}
		q_d = 2(1-q_d) \tilde \alpha \frac{d \mathcal{G}_E}{d q_d}
	\end{align}
\end{subequations}
where the \emph{total derivative with respect to $q_d$} of the energetic term reads
\begin{equation}
	\frac{d \mathcal{G}_E}{d q_d} = \frac{\partial \mathcal{G}_E}{\partial q_d} + \frac{\partial \mathcal{G}_E}{\partial q_a} \frac{d q_a}{d q_d} = \frac{\partial \mathcal{G}_E}{\partial q_d} - \frac{\partial \mathcal{G}_E}{\partial q_a}.
\end{equation}

\color{black}
\section{Numerical implementation of learning algorithms }
\label{sec:appendix_numerics}
This section aims at providing the reader with further details on the numerical experiments. We validate our theoretical observations experimentally through the use of the Langevin Dynamic (LD) algorithm and Gradient Descent (GD). We considered a teacher-student matching where the number of hidden-unit $K$ of the teacher and its associated activation function $\sigma(\cdot)$ are the same as those use for the student, same as the outer activation $f(\cdot)$. The student and teacher outer weight are equal $A_k=A^{\star}_k$, the algorithm will like to find an estimation for $\bW^{\star} \in \mathbb{R}^{K\times N}$.  

To sample from the Gibbs posterior, we employed a discretized version of the Langevin Dynamics (LD) algorithm. At each epoch $e$, the weights are updated according to the rule:
\begin{equation}
	\bW(e+1) = W(e) - \eta \nabla_{\bW} \mathcal{L}(\bW(e)) + \sqrt{2T\eta}  \xi_e,
\end{equation}
where	$\eta$ is the learning rate,	$T = 1/\beta$ is the temperature,	$\xi_e$ is matrix whose entries are sample from a standard Gaussian and $\mathcal{L}(\bW)$ is the empirical loss.

The  full batch GD was performed by taking the temperature to zero ($T=0$).

We used two initialization variants for the algorithm:
\begin{itemize}
	\item	\textit{LD Planted Init}: the student weights were initialized close to the teacher's weights $\bW^\star$. Precisely, we set $\bW(0) = \bW^\star + \kappa  \boldsymbol{\xi}$, where (e.g. $\kappa = 0.7$ for our simulation) controls the noise amplitude and the coefficient of $\boldsymbol{\xi}\in\mathbb{R}^{K\times N}$ are i.i.d sample from standard Gaussian.
	\item	\textit{LD Random Init} and Gradient Descent: the weights $\bW(0)$ were initialized randomly with i.i.d. standard normal entries.
\end{itemize}


In the zero-temperature limit, corresponding to Empirical Risk Minimization (ERM), we set $T = 10^{-4}$ in the Langevin algorithm. This value was sufficiently low to closely approximate the $\beta \to \infty $ limit, and we confirmed that the numerical observables (e.g., overlaps, generalization error) matched well with theoretical predictions.

The learning rate $\eta$ was adapted depending on the sample complexity $\tilde{\alpha} = P/(KN)$; in practice, we selected $\eta \in [10^{-1}, 10^{-4}]$, using smaller values of $\eta$ for larger $\tilde{\alpha}$, and vice versa.\\




\paragraph{\textbf{Regression.}}

For the regression task, we chose the identity function as the outer activation, i.e. $f(\cdot) = I_d$.
The regularized empirical loss associated to this task is, given by:
\begin{equation}
	\mathcal{L}(\bW) =  \sum_{\mu=1}^P \ell\left(y_\mu^\star, \varphi_{\mathbf{A},\bW}(\bx_\mu)\right) + \lambda \|\bW\|_F^2, \label{loss_regression}
\end{equation}
and $\ell(y, \hat{y})=(y- \hat{y})^2$ denotes the mean squared error between prediction and label. The number of hidden units $K=10$ was enough to validate the theory. \\

\paragraph{\textbf{Classification.}}
For the classification task, the outer activation chose was $f(\cdot)=\mathrm{Sign}(\cdot),$
the teacher weights $\bW^\star$ were chosen to lie on the hypersphere $(\mathbb{S}^{N-1})^K$.
The empirical loss used is  \begin{equation}
	\mathcal{L}(W) = \sum_{\mu=1}^P \ell\left(y_\mu^\star, \varphi_{\mathbf{A},\bW}(\bx_\mu)\right),\label{loss_classification}
\end{equation}
where the loss used in our numerics is the Hinge loss $\ell(y, \hat{y})_{\kappa}= \hat{y}\times f(\hat{y})~ max(0,\kappa-f(y)\times f(\hat{y}))$ with $\hat{y},y$ being the outer-preactivation and $\kappa$ the margin (we set it to zero in our simulation).  We set $K=100$ to match the theory.

Since the teacher weight matrix  $\bW^\star$ lies on the sphere, to enforce the spherical prior constraints assumed in the theory for the student, we renormalize each row $\bw_{k}\in \mathbb{R}^{N}$ at every training epoch $e$ as follows: 
\begin{equation}
	\bw_{k}(e) \leftarrow \frac{\bw_{k}(e)}{\|\bw_{k}(e)\|_2}. 
\end{equation}
The spherical constraint could have also being enforce by chosing an appropriate regularization parameter $\lambda$  (as the one we have in \eqref{loss_classification} for \textbf{regression}) which for this case should satisfies $\lambda \beta =1.$\\

\paragraph{\textbf{Bias correction procedure.}}
To ensure label centering, we performed bias correction at each training step. After generating predictions on a dataset, we computed the empirical mean of the predicted labels and subtracted it from each prediction. These centered predictions were then used to compute the training error component of the loss. The empirical test error was also evaluated on these debiased predictions. In the classification setting, the bias corresponds to the mean of the pre-activations of the output layer, and this correction was applied before passing the outputs through the final activation function $f(\cdot)$.\\

\paragraph{\textbf{Generalization error estimation.}}
the numerical evaluation of the Gibbs generalization error $\epsilon_g$ in Eq.~\eqref{equ:generalization_error} under the Langevin learning paradigm was performed as follows. After the algorithm has converge, we sampled student weight configurations $\bW(e)$ at time steps $e = 1, \dots, E,$  ($E$ being the number of student weight samples from LD once it has converge, e.g. we used with $E = 500$ for our numerics). For each configuration $\bW(e)$, we computed an estimate of the generalization error, denoted $\hat{\epsilon}^{e}_g$ , defined as the empirical mean squared error over a test set of size $1000$. The final estimator of the Gibbs generalization error was then obtained by averaging these individual estimates:
\begin{equation}
	\hat{\epsilon}_g =\frac{1}{E} \sum_{e=1}^{E} \hat{\epsilon}_g^{e}.
\end{equation}
After convergence, gradient descent (GD) based learning yields a single estimation that have being used to estimate the generalization, only an average over the test set was perform. We note here that the mean square error with proper factor (a $1/4$ is needed for the classification) serves as test error estimator for the classification and regression task.\\

\paragraph{\textbf{Teacher-student overlap estimation.}}
At an epoch $e$, a student weight matrix $\bW(e)$ is being use to estimate the teacher-student overlap matrix $m=m_d\mathbf{I}_K +(m_a /K)\mathbf{1}_K\mathbf{1}_K^\intercal$ by computing $\hat{m}_e =\bW^{\star}\bW^{\top}(e)/N\in \mathbb{R}^{K\times K}$. These overlaps were then averaged over $E = 500$ (those corresponding to the ones obtained after convergence of the algorithm) configurations to obtain the empirical estimator of the teacher-student overlap:\begin{equation}
	\hat{m} = \frac{1}{E} \sum_{e=1}^{E} \hat{m}_e.
\end{equation}
We numerically evaluate $m_0=m_a+m_d/K$ by computing an empirical average of the diagonal elements of $\hat{m}$. Similarly, an empirical average of the off-diagonal entries of $\hat{m}$ has being perform to numerically estimate $m_1=ma/K.$
All evaluations were conducted at fixed sample complexity $\tilde{\alpha} = P / (KN)$, ensuring consistency across different algorithmic regimes. The permutation symmetry that comes with the use of outer-weight $A^{\star}_k=A_k=1$ was broken by instead of using $A^{\star}_k$ (respectively $A_k$ ) we chose to used $A^{\star}_k\leftarrow A^{\star}_k+\gamma_k$ (respectively $A_k\leftarrow A_k+\gamma_k$) where $\gamma_k \sim \mathcal{N}(0,10^{-5})$, $10^{-5}$ being the standard deviation. This operation allows us to break the symmetry among the hidden neurons while still being close enough to the theory we aim at describing. One could follows similar procedure to compute \textit{Student-Student} overlap $q$  and the \textit{Self-Student} overlap $q_{\rm self}.$\\

\paragraph{\textbf{Computational resources.}}
Our simulations were performed using a single GPU of type NVIDIA A100-PCIE-40GB. Over the course of training, the GPU was active only 0.9\% of the time per epoch, indicating that the GPU accounted for just 0.9\% of the total computation time per epoch, with the remaining 99.1\% executed on the CPU. The GPU was primarily used for the training step, while other operations (such as the computation of observables) were performed on the CPU.

Across all algorithms, a single epoch took approximately 0.7~\text{s}. The number of epochs used for each figure was as follows:
\begin{itemize}
	\item	Figure 2: $2 \times 10^7$ epochs,
	\item	Figure 3: $2.8 \times 10^8$ epochs,
	\item	Figure 4: $1.26 \times 10^6$ epochs.
\end{itemize}
In total, the simulations required $2.406 \times 10^8$ epochs, corresponding to approximately $4678.33$ computing hours. Of this, only $42.11$ GPU hours were effectively used.\\

All the data presented in the paper, as well as the code used to run the different algorithms, can be found in this  \href{https://github.com/GibbsNwemadji/Narrow-one-hidden-layer-networks}{Github repository}.


\end{document}